\documentclass[11pt,a4paper]{article}                           

\usepackage{amsmath}
\usepackage{amssymb}
\usepackage[usenames]{color} 
\usepackage{multirow}
\usepackage{graphicx}
\usepackage{epstopdf}
\usepackage{array}
\usepackage{hhline}
\usepackage{cite}
\usepackage{microtype,colortbl}
\usepackage[bf]{caption}
\usepackage{color}
\usepackage[usenames,dvipsnames]{xcolor}
\usepackage{subcaption}
\usepackage{pstool}
\usepackage{bbold}
\usepackage{tikz}
\usepackage{ytableau}
\usepackage{braket}
\usepackage[T1]{fontenc}

\usepackage{ifpdf}
\ifpdf
  \usepackage[pdftex,
    pdftitle={},
    pdfauthor={},
    pdfsubject={},
    bookmarksopen, bookmarksnumbered, bookmarksopenlevel=2]{hyperref}
\fi
\def\hybrid{\topmargin -20pt    \oddsidemargin 0pt
        \headheight 0pt \headsep 0pt
        \textwidth 6.25in       % A4 paper
        \textheight 9 in       % A4 paper
        \marginparwidth .875in
        \parskip 5pt plus 1pt 
          \jot = 1.5ex
   }
\hybrid
\numberwithin{equation}{section}
\numberwithin{table}{section}

\newcommand{\beq}{\begin{equation}}  \newcommand{\eeq}{\end{equation}}
\newcommand{\bal}{\begin{aligned}}   \newcommand{\eal}{\end{aligned}}
\newcommand{\bea}{\begin{eqnarray}}  \newcommand{\eea}{\end{eqnarray}}

\newcommand{\bmat}{\left(\begin{array}}
\newcommand{\emat}{\end{array}\right)}

\newcommand{\bbC}{\mathbb{C}}
\newcommand{\bbR}{\mathbb{R}}

\newcommand{\nn}{\nonumber}

\newcommand{\cO}{\mathcal{O}}

\newcommand{\cE}{\mathcal{E}}

\newcommand{\cL}{\mathcal{L}}

\newcommand{\cN}{\mathcal{N}}

\newcommand{\cG}{\mathcal{G}}

\newcommand{\cH}{\mathcal{H}}

\newcommand{\cV}{\mathcal{V}}

\newcommand{\cM}{\mathcal M}

\newcommand{\I}{\text{Im}}

\usepackage{cancel}

\newcommand{\be}{\begin{equation}}
\newcommand{\ee}{\end{equation}}

\newcommand{\im}{\mathbf{i}}

\newcommand{\bbZ}{\mathbb{Z}}
\newcommand{\subs}{\subset}

\newcommand{\conj}[1]{\overline{#1}}
\definecolor{Gray}{gray}{0.95}

\begin{document}

\baselineskip=15pt
\parskip 5pt plus 1pt

\vspace*{-1.5cm}
\begin{flushright}    % Publication numbers
  {\small 
  
  }
\end{flushright}

\vspace{2cm}
\begin{center}        % Main title
  {\Huge Moduli Space Holography\\[.3cm] and the Finiteness of Flux Vacua}

\end{center}

\vspace{0.5cm}
\begin{center}        % Authors
{\large  Thomas W.~Grimm}
\end{center}

\vspace{0.15cm}
\begin{center}        % Institutes
\emph{%$^1$ 
Institute for Theoretical Physics \\
Utrecht University, Princetonplein 5, 3584 CE Utrecht, The Netherlands}
 
\end{center}

\vspace{2cm}

%%%%%%%%%%%%%%%%%%%%%%%%%%%%%%%%%%%%%%%%%%%%%%%
%%%%%%%%%%%%%%%%%%%%%%%%%%%%%%%%%%%%%%%%%%%%%%%
%%%%%%%%%%%%%%%%%%%%%%%%%%%%%%%%%%%%%%%%%%%%%%%
%%%%%%%%%%%%%%%%%%%%%%%%%%%%%%%%%%%%%%%%%%%%%%%
%%%%%%%%%%%%%%%%%%%%%%%%%%%%%%%%%%%%%%%%%%%%%%%
%%%%%%%%%%%%%%%%%%%%%%%%%%%%%%%%%%%%%%%%%%%%%%%
%%%%%%%%%%%%%%%%%%%%%%%%%%%%%%%%%%%%%%%%%%%%%%%
%%%%%%%%%%%%%%%%%%%%%%%%%%%%%%%%%%%%%%%%%%%%%%%

\begin{abstract}
\noindent
\baselineskip=15pt
A holographic perspective to study and characterize 
field spaces that arise in string compactifications is suggested. A concrete correspondence is developed by studying
two-dimensional moduli spaces in supersymmetric string compactifications. It is proposed 
that there exist theories on the boundaries of each moduli space, whose crucial data are given by 
a Hilbert space, an $Sl(2,\bbC)$-algebra, and two special operators. This boundary data is motivated by 
asymptotic Hodge theory and the fact that the physical metric on the moduli space of Calabi-Yau manifolds asymptotes near any infinite distance boundary to a Poincar\'e metric with $Sl(2,\bbR)$ isometry.
The crucial part of the bulk theory on the moduli space is a sigma model for group-valued matter fields.  
It is discussed how this might be coupled to a two-dimensional gravity theory.  
The classical bulk-boundary matching is then given by the proof of the famous $Sl(2)$ orbit theorem of Hodge theory, which is 
reformulated in a more physical language. Applying this correspondence to the flux landscape in Calabi-Yau fourfold compactifications it is 
shown that there are no infinite tails of self-dual flux vacua near any co-dimension one boundary. This finiteness 
result is a consequence of the constraints on the near boundary expansion of the bulk solutions that match to the
boundary data. It is also pointed out that there is a striking connection of the finiteness result for supersymmetric 
flux vacua and the Hodge conjecture.  

\end{abstract}

\thispagestyle{empty}
\clearpage

\setcounter{page}{1}

%%%%%%%%%%%%%%%%%%%%%%%%%%%%%%%%%%%%%%%%%%%%%%%
%%%%%%%%%%%%%%%%%%%%%%%%%%%%%%%%%%%%%%%%%%%%%%%
%%%%%%%%%%%                 %%%%%%%%%%%%%%%%%%%
%%%%%%%%%%%  DOCUMENT BODY  %%%%%%%%%%%%%%%%%%%
%%%%%%%%%%%                 %%%%%%%%%%%%%%%%%%%
%%%%%%%%%%%%%%%%%%%%%%%%%%%%%%%%%%%%%%%%%%%%%%%
%%%%%%%%%%%%%%%%%%%%%%%%%%%%%%%%%%%%%%%%%%%%%%%
%%%%%%%%%%%%%%%%%%%%%%%%%%%%%%%%%%%%%%%%%%%%%%%

\newpage

\tableofcontents

%\newpage

%%%%%%%%%%%%%%%%%%%%%%%%%%%%%%%%%%%%%%%%%%%%%%%
\section{Introduction}
\label{sec:intro}
%%%%%%%%%%%%%%%%%%%%%%%%%%%%%%%%%%%%%%%%%%%%%%%

The search for general principles that identify effective theories that can be consistently coupled 
to quantum gravity has recently attracted much attention \cite{Palti:2019pca}. These principles 
have been formulated in a number of quantum gravity or `swampland' conjectures. 
A motivation for this work provides the so-called distance conjecture \cite{Ooguri:2006in}. It deals 
with effective theories with scalar fields and suggests that, if a UV completion 
with gravity exists, it has to admit states with a certain universal behaviour 
when approaching points in field space that are at infinite shortest geodesic distance. 
The profoundness of this conjecture arises from the fact that it links properties of the field space 
to the existence of certain states in the underlying theory near such infinite distance points. 
Even if one might not know the whole spectrum of states, at least a subsector has to follow a 
rather constrained asymptotic behaviour. In this work, we suggest that this asymptotic structure 
is dictated by a holographic principle and the existence of an auxiliary boundary theory living 
at infinite distance boundaries of field space. 

The physical motivation for the holographic dictionary lies largely in the distance conjecture
combined with the observation that the asymptotic field space metric and the relevant towers of states 
follow stringent constraints in all known examples arising from string compactifications  \cite{GPV,Lee:2018urn,Lee:2018spm,Grimm:2018cpv,Corvilain:2018lgw,Font:2019cxq,Lee:2019xtm,Marchesano:2019ifh,Grimm:2019wtx,Lee:2019wij,Baume:2019sry,Enriquez-Rojo:2020pqm,Gendler:2020dfp,Heidenreich:2020ptx}. 
Furthermore, it is intriguing to 
interpret the distance conjecture as describing a mechanism of ensuring that exact global symmetries 
are absent in any gravity-coupled theory with finitely many states \cite{Banks:1988yz,Banks:2010zn}. Adapting the converse perspective,
we claim that at any infinite distance boundary 
in field space a global symmetry arises and that we can attach this data to the boundary. 
Our intuition is then derived 
from the expectation that combining the existence of a global symmetry with all possible positivity constraints, 
e.g.~of the field space metric, will strongly restrict
the asymptotic behaviour of the theory.  That this is indeed the case in supersymmetric string compactifications 
was recently highlighted in \cite{GPV,Grimm:2018cpv,Corvilain:2018lgw,Grimm:2019wtx,Grimm:2019ixq,Gendler:2020dfp}. In fact, in the vector sector of 
$\cN=2$ string compactifications it is the asymptotic global symmetry and the positivity of the physical couplings 
and masses in the asymptotic regime that fixes much of the asymptotic structure. The underlying 
mathematical reason for this observation can be described using asymptotic Hodge theory \cite{AST_1989__179-180__67_0,HodgeTheoryMN49}. 
This broad and abstract theory will allow us to develop the holographic dictionary to a significant 
extend. 

The detailed construction that we will present is motivated by an in-depth understanding of 
the field spaces that arise in string compactifications on Calabi-Yau manifolds.
More precisely, we start our discussion with a study of the asymptotic behaviour of the moduli space 
of geometric deformations that preserve the Calabi-Yau condition and later strip away the 
underlying geometric motivation. 
The geometric settings have been studied intensively in the past and it is well-known that there are two types of deformations of 
Calabi-Yau geometries, the complex structure deformations 
and the K\"ahler structure deformations. Since by mirror symmetry the latter deformations can be realized as 
a special subset of the former \cite{Hori:2003ic}, we will exclusively focus on complex structure deformation space in this work. 
It is central to this work that the complex structure moduli space has boundaries. These 
correspond to choices of complex structures for which the Calabi-Yau manifold degenerates. While 
some of these degeneration points, such as the large complex structure point, have been studied in much 
detail in the past, it is important to stress that there are a plethora of such degenerations and there is no detailed
classification of the possibilities yet (see \cite{Kerr2019HodgeTO,Kerr2020HodgeTO} for recent progress). 
The crucial point is that at the boundaries of moduli space the associated manifold is so singular 
that the usual geometric structures, such as the Hodge norms determining the kinetic terms of fields, 
degenerate and can no longer be applied. Asymptotic Hodge theory shows, however, that there 
is a more abstract structure living \textit{on} the boundary and we claim that this structure can 
be thought of as defining parts of a boundary theory. We will descibe in this 
work which set of boundary data determines the behaviour of the couplings 
in the effective theory and the moduli space 
close to boundary. In this geometric picture we thus find that when 
tuning the Calabi-Yau space to become singular such that a global symmetry emerges in moduli space, 
the structure of the effective theory and the moduli space is largely fixed by  
\textit{global symmetry},  \textit{positivity}, and \textit{holomorphicity}. While our findings are compatible with the 
expectations from the distance conjecture at boundaries that are at infinite distance, we will see that the constrained 
asymptotic behaviour arises 
more generally. 
 
The existence of a holographic description of the Calabi-Yau moduli space can also be motivated 
by noting that the physical metric on moduli space, i.e.~the Weil-Petersson metric which arises in string 
compactifications, always asymptotes to a metric containing the factors of Poincar\'e metric at any infinite distance 
boundary \cite{wang1}. In a real two-dimensional setting, which will be the main focus of this work, this means that the physical metric 
asymptotes to the two-dimensional Poincar\'e metric which is a patch of Euclidean AdS$_2$. This metric 
has an $\mathfrak{sl}(2,\bbR)$ isometry algebra that will non-trivially translate to a global symmetry algebra $\mathfrak{sl}(2,\bbC)$
on the boundary. We will see that this boundary $\mathfrak{sl}(2,\bbC)$ is indeed a result of the emerging global symmetry and 
exists more generally even if the boundary is not at infinite distance. 
The key quantity relevant to evaluate the asymptotic physical metric is the period matrix 
of the Calabi-Yau manifold. It encodes how the Hodge decomposition over the middle cohomology changes when 
moving over the complex structure moduli space. A remarkable result of Schmid \cite{Schmid} states that asymptotically 
this information is captured by a so-called nilpotent orbit, which packages the asymptotic behaviour in a seemingly simple 
polynomial way. For example, in the large complex structure or large volume boundary the nilpotent orbit captures this information in 
the periods remaining after dropping all exponential corrections.\footnote{We stress that the nilpotent orbit exists near every boundary and, in general, indirectly captures exponential corrections that are needed for the positivity of the Hodge norm
in the asymptotic regime.} The nilpotent orbits
will serve as the motivation for the bulk and boundary theories that we 
discuss in this work. 
Eventually, however, the results can be formulated without any reference to 
nilpotent orbits. They arise as solutions to the bulk theory that match the boundary data. We will 
call such solutions `physical' in the following, since they appear in actual geometric compacitifactions. 
Viewed abstractly, however, both the bulk and boundary data can then be formulated without reference
 to a geometric string theory setting. 

In order to construct the bulk action we restrict our attention to real two-dimensional field spaces. In other words, the field 
space of the effective theory will thus be viewed as a two-dimensional worldsheet. Ideally we would like to construct 
a gravity model coupled to a sigma-model on the worldsheet. The matter sector of this theory will be constrained 
by field equations that also arise in asymptotic Hodge theory. It is known from \cite{Schmid,CKS}
that nilpotent orbits in one complex dimension provide solutions to Nahm's equations that satisfy a certain constraint and match a well-defined set of
boundary conditions. An action principle associated to Nahm's equations was discussed long ago in \cite{donaldson1984} and 
we will generalize it to a sigma model action on the worldsheet. A significant generalization of
this action to the multiple variables $tt^*$-system of \cite{Cecotti:1991me} appeared more recently in \cite{Cecotti:2020rjq}.
We also comment on the coupling of two-dimensional gravity to the matter sector. This is similar in spirit to the suggestion by Cecotti 
\cite{Cecotti:2020uek}, who proposes to couple this sigma model to Einstein gravity for higher-dimensional worldsheets.\footnote{The described perspective has been developed independently.} While we will not present a complete action principle, we will successively build 
up a set of field equations. These turn out to admit solutions that are the nilpotent orbits that can arise in Calabi-Yau compactifications,
together with the physical metric on moduli space. Crucially this requires to fix boundary conditions which we propose stem from a 
boundary theory.
  
 To motivate the existence of a boundary theory we will again start with a nilpotent orbit, which we consider 
 as the physical solutions to the bulk theory, and extract the data on the boundary that fixes such solutions.
 In order to do that we will use the famous Sl(2) orbit theorem of Schmid \cite{Schmid} 
and Cattani, Kaplan, and Schmid \cite{CKS}.  
 The set of boundary data will consist of an $\mathfrak{sl}(2,\bbC)$ symmetry algebra acting on 
 a finite-dimensional Hilbert space. The latter can be obtained as complexification of the charge or flux lattice 
 relevant in the effective theory, and corresponds in geometric setting to the middle 
 cohomology group of the Calabi-Yau manifold. The Hilbert space has a special 
 $\mathfrak{sl}(2,\bbC)$-compatible norm that is induced by an operator $Q_\infty$.  
 Geometrically this $Q_\infty$ defines a Hodge decomposition that exists on the boundary of the moduli space
 despite the fact that the corresponding compactification geometry is badly singular \cite{Schmid,CKS}.
 The $\mathfrak{sl}(2,\bbC)$ algebra turns out to be non-trivially related to the global symmetry
 in the asymptotic bulk solutions. We describe that boundary data contains a real nilpotent operator, 
 which we call phase operator, that encodes how the asymptotic global symmetry is rotated into the 
 boundary $\mathfrak{sl}(2,\bbC)$ and how positivity constraints on the bulk solution map to the boundary. 
 The additional information 
 contained in the phase operator turns out to be central to the whole construction.\footnote{In Hodge theory 
 this operator was introduced by Deligne \cite{DeligneRHS,CKS} as a unique rotation of any complex mixed Hodge structure into a mixed Hodge structure split over 
 the real numbers.} Taken together this boundary data will suffice for our construction and serves as evidence for the 
 existence of a boundary theory. Further evidence for the existence of such a boundary theory is provided by 
 a number of conjectures put forward in \cite{GPV}, \cite{Lee:2019wij}, and \cite{Lanza:2020qmt,Lanzatoappear}, each discussing aspects of 
the theories that might emerge at infinite distance boundaries.   
  While we will leave its full construction to future work, we will sometimes refer to 
 the boundary data as describing a boundary theory. 
 
 Reconstructing the bulk solutions matching the boundary data turns out to be highly non-trivial and 
 contained in the proof of the $Sl(2)$ orbit theorem \cite{Schmid,CKS}. Remarkably, the aforementioned 
 boundary data specifies the bulk solution uniquely. To see this we will solve the matter equations of motion 
 with a near boundary expansion and then determine their properties and eventually their dependence 
 on the boundary operators. 
 The constraints on the coefficients arise from the $\mathfrak{sl}(2,\bbC)$
 symmetry and will turn out to be central in the finiteness proof that we discuss in the last part of this work.
 In showing that there is a unique reconstruction of the bulk solution from the boundary data, we discuss 
 how the phase operator becomes of crucial importance. We determine 
 a single matrix equation \eqref{fixall} which provides the unique match \cite{CKS}. It then follows that
 coefficients in any bulk solution matching to the boundary are universal non-commutative polynomials in 
 the $\mathfrak{sl}(2,\bbC)$ generators and the phase operator.

In the final part of this work we will highlight some first non-trivial physical applications of the holographic 
perspective by discussing the finiteness of the flux landscape and the validity of the distance conjecture. 
In particular, we address in detail the longstanding question about the finiteness of 
flux vacua in Type IIB and F-theory flux compactifications \cite{Grana:2005jc,Douglas:2006es}. Formulated in F-theory or 
M-theory language, such compactifications are specified by a Calabi-Yau fourfold with a background 
flux $G_4$. The classical equations of motion then demand that this flux is self-dual in 
a general vacuum, while consistency demands that the flux-square is bounded by a tadpole constraint. 
While in the bulk of the moduli space it easy to argue that there are only finitely many fluxes and 
self-dual loci in moduli space, these could accumulate near its boundaries \cite{Ashok:2003gk,Denef:2004ze}. 
We show  that this does not happen when approaching 
any co-dimension one boundary  \cite{Schnellletter,GrimmSchnell}. The result derives from the described bulk-boundary construction 
and is a consequence of the fact that the near boundary expansion 
of the bulk solution is constraint by the boundary data to forbid infinite tails. To 
prove finiteness for all boundaries will be the aim of \cite{GrimmSchnell}.  
It is interesting to point out that in the supersymmetric case in which the $G_4$ fluxes are restricted to 
be of $(2,2)$-type a famous result \cite{CDK} provides 
a general proof of finiteness near any boundary. The significance of the latter publication arises 
due to the fact that this result can also be obtained by assuming the Hodge conjecture. Restricted 
to the co-dimension one boundaries, the main tool of \cite{CDK} is precisely the $Sl(2)$ orbit theorem underlying 
the correspondence discussed here. This gives further support to the significance of the described structures and 
the power of this formalism.

This article is structured as follows. In section \ref{motivation} we begin by motivating our 
constructions by recalling some facts about asymptotic Hodge theory. In particular, we 
describe how the Hodge decomposition of the cohomology groups of forms behaves near 
the boundary of moduli space and how this behaviour is captured by nilpotent orbits. 
We then discuss the asymptotic form 
of the Weil-Petersson and the Hodge metric on moduli space and show when 
they asymptote to the Poincar\'e metric near the boundaries. In section \ref{bulktheory}  we turn 
to the discussion of the bulk theory on moduli space. We formulate field equations 
and an action principle for group-valued matter fields and discuss aspects of coupling 
this theory to gravity. Important aspects of the boundary theory are then discussed in section \ref{boundary_theory}, 
where it is explained how a set of boundary data is fixed by symmetry and positivity. 
Technically most involved is section \ref{bulk_boundary}, in which we describe how the boundary 
data singles out special sets of bulk solutions and constrains their behaviour. It contains 
some of the key steps of the proof of the $Sl(2)$-orbit theorem reformulated to support the 
holographic perspective.  In the final section \ref{finitenesssec} we then apply these finding 
to address the finiteness of flux vacua on Calabi-Yau fourfolds. We show the finiteness 
of self-dual fluxes near co-dimension one boundaries and comment on the finiteness 
of $(2,2)$-fluxes. We close with some remarks on applying the holographic perspective to 
the distance conjecture. The paper contains one appendix \ref{computedelta} discussing the computation 
of the phase operator.

\section{Motivation using asymptotic Hodge theory} \label{motivation}

In this section we provide the motivation for the construction of the bulk theory and the 
bulk-boundary matching by introducing some results from asymptotic Hodge theory. While many 
of the described facts are true for general K\"ahler manifolds, we will restrict our attention 
to complex $D$-dimensional Calabi-Yau manifolds $Y_D$. In this cases, the geometry 
of the complex structure moduli space $\cM$ of $Y_D$ can be encoded by the moduli dependence 
of the $(D,0)$-form $\Omega$. 
We first introduce the Hodge norm and the Hodge decomposition in 
section \ref{sec:Hodge-norm+decomp} and comment on its relevance in string compactifications. 
In section \ref{Near-boundary} we then restrict our
attention to the near boundary region in $\cM$. 
We explain how the Hodge decomposition near the boundary can always be encoded by a 
 expansion that is polynomial in the moduli and is best described by a so-called nilpotent orbit. In passing we 
argue that this expansion nevertheless encodes `non-perturbative' terms in the 
periods of $\Omega$ at most boundaries. Crucial for developing the bulk theory
is the fact that the nilpotent orbits satisfy a set of differential equations. We introduce 
these equations in section \ref{nilpotent_diffs}, point out their relation to Nahm's equations, 
and discuss an associated action principle. 
Finally, in section \ref{near-boundary-metric}, 
we introduce two metrics on the moduli space $\cM$ and discuss their near boundary expansion. 
The first one is the Weil-Petersson metric and is the physical metric in string compactifications 
on $\cM$. The second one is the Hodge metric and closely related to the Hodge norm. We note that the asymptotic 
form of the Weil-Peterson metrics contains a Poincar\'e metric at all infinite distance boundaries, while this fact is more 
generally true for the Hodge metric. The isometry group $Sl(2,\bbR)$ of the Poincar\'e metric
will translate to part of the symmetry group found in the boundary theory in section \ref{boundary_theory}.

\subsection{Hodge norm and Hodge decomposition in the bulk} \label{sec:Hodge-norm+decomp}

In order to introduce a holographic picture of the moduli space 
we first have to specify which quantities we want to keep track of. Let us denote by $Y_D$ a 
compact Calabi-Yau manifold of complex dimension $D$.
For concreteness we will set our 
focus on the behaviour of the Hodge norm of a $D$-form cohomology class of $Y_D$. Considering
two elements $\alpha,\beta \in H^D(Y_D,\bbC)$, the Hodge norm arises from the inner product
\beq \label{Hodge-inner-prod}
  \int_{Y_D} \bar \alpha \wedge * \beta = \frac{1}{D!}\int_{Y_D} d^{2D} x \sqrt{\text{det} g} \, \bar \alpha_{\mu_1...\mu_D}  \beta^{\mu_1...\mu_D}\ 
\eeq 
and will be denoted by 
\beq \label{Hodge-norm}
 \| \alpha \|^2 =   \int_{Y_D}  \bar \alpha \wedge *   \alpha\ .
\eeq
Note that the inner product   \eqref{Hodge-inner-prod} is induced 
 by the Hodge norm and therefore it often suffices to discuss the latter. 
In addition to the inner product induced by the norm we can also define the wedge-product \footnote{Note that 
in the mathematical literature \cite{Schmid,CKS} this inner product is denoted by $S(\alpha,\beta)= \langle \beta,\alpha \rangle$.}
\beq \label{def-S}
    \langle \alpha,\beta \rangle:=   \int_{Y_D} \alpha \wedge  \beta\ ,
\eeq
which is symmetric for $D$ even and skew-symmetric for $D$ odd. It will be important in the following 
to consider transformations $g$ preserving $\langle\cdot,\cdot\rangle$. The group of such transformations over the real numbers 
will be denoted by $G_{\bbR}$, while the corresponding algebra is denoted by $\mathfrak{g}_{\bbR}$. Hence, 
we have
\beq \label{def-gg}
  g\in G_{\bbR}:\quad \langle g\alpha, \beta\rangle=\langle\alpha, g^{-1} \beta\rangle \ ,  \qquad \quad 
  L \in \mathfrak{g}_{\bbR}:\quad \langle L\alpha, \beta \rangle=-\langle\alpha, L \beta \rangle\ . 
\eeq
As an example, we note that for Calabi-Yau threefolds one has $G_{\bbR}= Sp(2h^{2,1}+2,\bbR)$. 
The complex version of this group and algebra are henceforth denoted by $G_{\bbC},\ \mathfrak{g}_{\bbC}$. 

When computing the effective actions arising from compactifications of string theory the Hodge norm \eqref{Hodge-norm} appears in many instances. 
As a first example, note that in Type IIB string theory  on a Calabi-Yau threefold the Hodge norm determines 
the kinetic terms of the four-dimensional gauge fields, which arise by expanding the R-R four-form $C_4$ into three-forms $H^{3}(Y_3,\bbZ)$.
Picking a symplectic basis $(\alpha_M,\beta^N)$ with $ \langle \alpha_M,\beta^N \rangle = \delta_M^N$ of $H^{3}(Y_3,\bbZ)$ we write 
$C_4 = A^M\wedge \alpha_M - \tilde A_M\wedge \beta^M$. The four-dimensional vectors $A^M$ and $\tilde A_M$ are electric and magnetic $U(1)$ gauge fields in 
the effective theory, respectively. The charged particles in the effective theory arise from D3-branes wrapped on three-cycles in $Y_3$. 
The space $H^{3}(Y_3,\bbZ)$ can be identified with the charge lattice of these states under $(A^M,\tilde A_N)$. The relevance 
of these states in the distance conjecture will be briefly discussed in the very last section \ref{distance_comments}. 
A second example, which will be central to section \ref{finitenesssec}, are F-theory and M-theory compactifications on Calabi-Yau fourfolds. 
In these cases the flux scalar potential  
induced by a background four-form flux $G_4$ in $H^4(Y_4,\bbZ/2)$ is determined by the Hodge norm. The 
lattice  $H^4(Y_4,\bbZ/2)$ corresponds to the flux lattice.

The goal of the following discussion is to keep track of the dependence of the Hodge norm on the complex structure deformations 
of the manifold $Y_D$. For Calabi-Yau manifolds it can be shown that there exists an unobstructed moduli
space $\cM$, the complex structure deformation space. This space is a K\"ahler manifold of complex 
dimension $h^{D-1,1}=\text{dim}\,H^{D-1,1}(Y_D)$. In order to investigate the change of \eqref{Hodge-norm}
along $\cM$, we consider the Hodge decomposition 
\beq \label{Hodge-dec}
   H^{D}(Y_D,\bbC) = H^{D,0} \oplus H^{D-1,1} \oplus ... \oplus H^{1,D-1} \oplus H^{0,D}\ ,
\eeq
where $\overline{H^{p,q}}= H^{q,p}$ and $p+q=D$.
This decomposition has to be determined for the chosen complex structure on $Y_D$ and 
hence varies when moving along $\cM$. 
Using the K\"ahler metric on $Y_D$ to determine the 
Hodge star $*$ one shows that 
\beq \label{Hodge-action}
   * w^{p,q}  = i^{p-q} w^{p,q} \  ,\qquad w^{p,q} \in H^{p,q}\ .
\eeq
Furthermore, one has the relation that 
\beq
    \langle w^{p,q},v^{r,s}\rangle =0\ , \qquad \text{for} \  p\neq s\ , q\neq r\ .
\eeq
This implies that one can evaluate the Hodge norm $\|\alpha \|$, defined in \eqref{Hodge-norm}, if the $(p,q)$-decomposition of 
$\alpha$ has been determined. 

\begin{figure}[h!]
\begin{center}
 \includegraphics[width=0.37\textwidth]{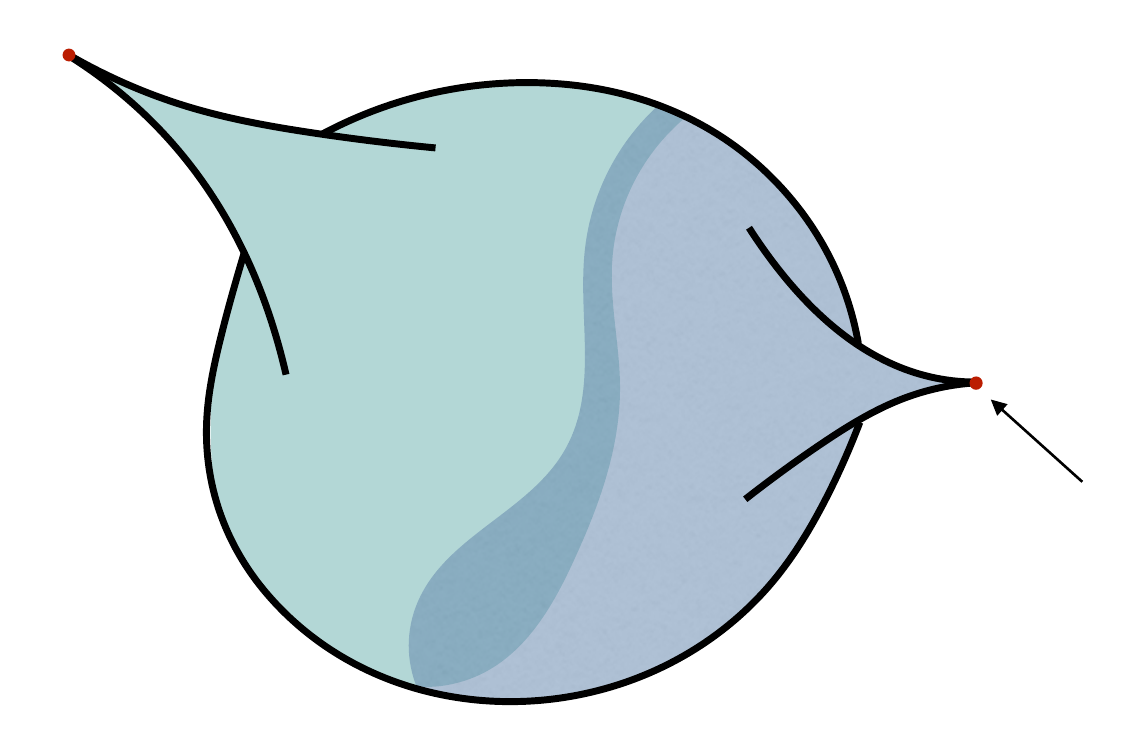}
\end{center}
\begin{picture}(0,0)
\put(95,95){\begin{minipage}{3cm}\small large complex \\
structure patch \end{minipage}}
\put(275,100){\begin{minipage}{3cm}\small conifold \\
 patch \end{minipage}}
 \put(307,55){\small boundary point}
\end{picture}
\vspace*{-1cm}
\caption{Schematic depiction of a complex one-dimensional moduli space. Two overlapping patches are indicated in light green and light blue. These can, for example, contain the large complex structure point and the conifold point of a $Y_3$. \label{fig:patches}}

\end{figure}
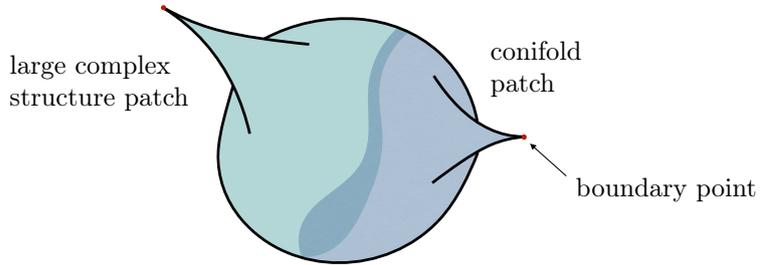

The dependence of \eqref{Hodge-norm} on the coordinates $z^I$ of the moduli space $\cM$ can thus be understood by following 
the $(p,q)$-decomposition along $\cM$. It is actually better to study how 
the spaces 
\beq \label{Fp-filtration}
   F^p = \bigoplus_{r\geq p} H^{r,D-r}\ 
\eeq
change when moving along $\cM$. These spaces vary, at least locally, holomorphically in the complex 
coordinates $z^I$ of $\cM$  \cite{HodgeTheoryMN49}. The original decomposition \eqref{Hodge-dec} is then recovered by $H^{p,q}= F^p \cap \bar F^q$.
The moduli dependence of the $F^p$ is, in general, given by complicated transcendental functions that solve 
partial differential equations known as the Picard-Fuchs equations. Generically the solutions have only a finite radius 
of convergence and, in order to cover the whole moduli space $\cM$ one has to work in patches, 
leading to a picture as in Figure \ref{fig:patches}. 

It is crucial for our considerations to note that the moduli space $\cM$ is, at first, neither smooth nor compact. This is 
due to the fact that, when changing the complex structure, the Calabi-Yau manifold can  become so singular that 
a Hodge decomposition as in \eqref{Hodge-dec} no longer exists. Such singular loci constitute the boundaries 
of the moduli space.  It was shown in \cite{Hironaka,Viehweg} that one can modify the boundary loci, by blowing up possible singularities, 
such that the boundary of $\cM$ can be written as 
\beq \label{boundary}
   \partial \cM = \bigcup_k \Delta_k\ , 
\eeq
where $\Delta_k$ are complex manifolds of complex dimension $h^{D-1,1}-1$ intersecting at normal instance.
In the following sections, we will describe how an extended structure generalizing the decomposition \eqref{Hodge-dec} can 
be defined \textit{on} the boundaries $\cup_k \Delta_k$. Before doing this, we study in more detail the near-boundary 
behaviour of the decomposition \eqref{Hodge-dec}.

\subsection{Near-bounday expansions and nilpotent orbits} \label{Near-boundary}

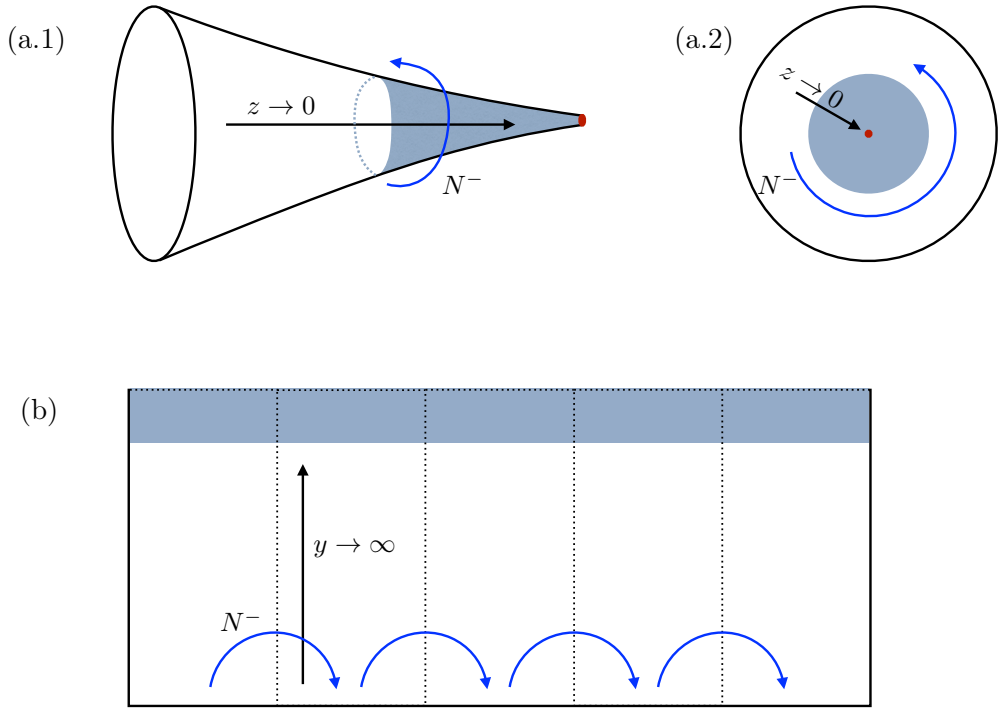
\begin{figure}[h!]
\begin{center}
 \includegraphics[width=0.8\textwidth]{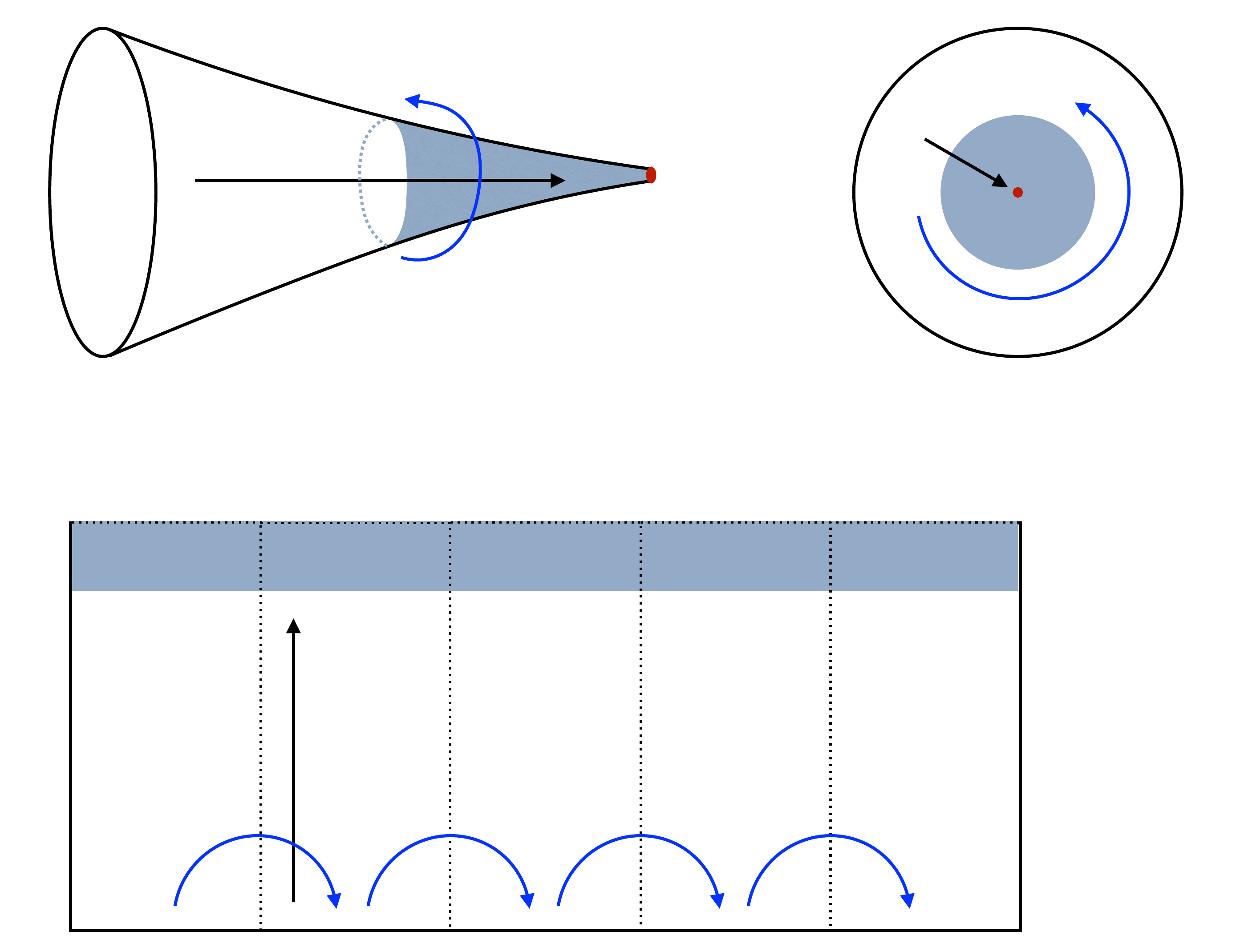}
\end{center}
\begin{picture}(0,0)
\put(20,280){(a.1)}
\put(110,255){\small$z \rightarrow 0$}
\put(183,225){\small $N^-$}
\put(301,225){\small$N^-$}
\put(309,270){\rotatebox{-31}{\small $z \rightarrow 0$}}
\put(135,90){\small $y \rightarrow \infty$}
\put(100,60){\small $N^-$}
\put(270,280){(a.2)}
\put(25,140){(b)}
\end{picture}

\caption{Schematic depiction of the asymptotic region in a complex one-dimensional moduli space. Figures (a.1) and (a.2) show 
the punctured disc parametrized by the complex coordinate $z$. The boundary of interest is the puncture at $z=0$. Figure (b) shows
the universal cover of the punctured disc, the upper half plane parametrized by $t = x+iy$. The boundary of interest is 
now located at $y=\infty$. We also indicate that there can be a log-monodromy matrix $N^-$ appearing in \eqref{nilpotent_orbit}
when encircling the 
puncture or shifting in the upper-half plane. \label{fig:boundary1}}

\end{figure}

In this subsection we discuss behaviour of the decomposition \eqref{Hodge-dec} near any boundary
component $\partial \cM$ of the 
moduli space $\cM$. Recalling that the 
boundary splits into multiple $\Delta_k$, as discussed around \eqref{boundary}, we 
want to consider a local patch containing a co-dimension $n$ boundary.
We thus introduce local coordinates $z^j \equiv e^{2\pi it^j}$, $j=1,...,n$, and $\zeta^\kappa$, such that the boundary 
component is approached in the limit
\beq \label{local-coords}
  z^j \rightarrow 0  \qquad \text{or}\qquad t^j = x^j+iy^j \ \rightarrow \ x^j_0 + i \infty\ .  
\eeq
Suppressing the $\zeta^\kappa$ coordinate directions, the considered configuration can 
be depicted as in figure \ref{fig:boundary1}.

We next recall the first major 
result of asymptotic Hodge theory, which states how the Hodge decomposition behaves for sufficiently large $y^j \gg 1$. 
As stated after \eqref{Fp-filtration} it is convenient to study the vector spaces $F^p$, 
since they vary holomorphically over the moduli space. This means that the $F^p$ are locally only 
depending on the coordinates $z^j$ or $t^j$ introduced in \eqref{local-coords}, but are independent of $\bar z^j$, $\bar t^j$. 
This statement extends to the complex coordinates $\zeta^\kappa$, which we will suppress in the following expressions. 
A main insight of Schmid \cite{Schmid} was that near a boundary $\I\, t^j = y^j \rightarrow \infty$ the $F^p$ always take the form 
\beq \label{pert-nonpert-split}
     F^p(t) \approx F^p_{\rm pol}(t) \ ,
\eeq
where $F^p_{\rm pol}$ is varying, up to possibly an overall rescaling in any direction, as a polynomial in $t$. This `polynomial part'
is given by 
\beq \label{nilpotent_orbit}
   F^p_{\rm pol}(t)   = e^{ t^j N_j} F^p_0\ ,
\eeq
where $N_j$ are constant nilpotent matrices and the $F^p_0$ is independent of $t^j$, but can still depend holomorphically on $\zeta^\kappa$. 
The compatibility between the $N_i$ and the spaces $F^p_0$ can be stated as the condition that $N_i F^p_0 \subset F_0^{p-1}$.
The polynomial piece is known as the \textit{nilpotent orbit}.
Note that \eqref{pert-nonpert-split} is a relation between vector spaces and it is a non-trivial statement that the 
polynomial part alone has $\text{dim}\, F^p = \text{dim}\, F^p_{\rm pol}$. Fixing a direction in the vector spaces 
there are generically exponentially suppressed corrections $\cO(e^{2\pi i t^j})$ to the identification \eqref{pert-nonpert-split}.
These are strongly suppressed in the near-boundary regime $\I \,t^j \gg 1$.  

One can now show that \eqref{nilpotent_orbit} also defines a 
$(p,q)$-decomposition and an associated norm using the analog of \eqref{Hodge-action}. Concretely, we introduce the 
decomposition by 
\beq \label{pol_splitting}
     H^{D}(Y_D, \bbC) = H^{D,0}_{\rm pol} \oplus H^{D-1,1}_{\rm pol} \oplus... \oplus H^{0,D}_{\rm pol}\ ,
\eeq 
by setting $H^{p,q}_{\rm pol} = F^p_{\rm pol} \cap \bar F^q_{\rm pol} $. A norm $\| \cdot\|_{\rm pol}$ is then defined 
as 
\beq \label{Weil-pol}
    \| \omega \|^2_{\rm pol} : = \langle   \bar  \omega, C_{\rm pol}   \omega \rangle \ , \qquad C_{\rm pol} \omega^{p,q} = i^{p-q} \omega^{p,q}\ , 
\eeq
where now $\omega^{p,q} \in H^{p,q}_{\rm pol}$. We also introduce the associated inner product 
\beq  \label{def-innerprod_nil}
   \langle \omega | \nu \rangle_{\rm pol} := \langle \bar \omega,  C_{\rm pol}   \nu \rangle \ ,
\eeq
where we will use bra-ket notation when convenient. 
Note that it is non-trivial that 
the $H^{p,q}_{\rm pol}$ obtained from the polynomial $F^{p}_{\rm pol}$ 
suffice to span the whole space $H^{D}(Y_D,\bbC)$. In contrast, it is \textit{not true}
in general that $F^p_0$ can be used similarly to define a $(p,q)$-decomposition with a well-defined norm. 

%It will be convenient to introduce a basis of $F^{p}_{\rm pol}$, which we denote by $| I \rangle$

At first it seems natural to view $ F^p_{\rm pol} $ as only capturing `perturbative terms', 
while the dropped corrections $\cO(e^{2\pi i t^j})$ in \eqref{pert-nonpert-split} correspond to the non-perturbative corrections. While this interpretation can indeed be made 
more precise in the special limit 
known as the large complex structure limit,\footnote{Mirror symmetry states that complex structure deformations 
are exchanged with K\"ahler structure deformations of a dual Calabi-Yau geometry \cite{Hori:2003ic}. In this dual picture the strings wrapping cycles 
in the dual space do induce perturbative and non-perturbative corrections with the above split.} 
we will discuss in the following that this is \textit{not} generally the case 
at other limits in moduli space. In fact, in most other situations, certain \textit{crucial} non-perturbative corrections are captured 
by $F^p_{\rm pol}(t) $.  
To simplify the discussion let us assume that $\cM$ is one-dimensional, i.e.~we only use the coodinate $t$ to parametrize the Hodge 
decomposition. In order to see which information is captured by $ F^p_{\rm pol} $, we 
give an explicit expansion of the $(D,0)$-form $\Omega$ spanning $F^D = H^{D,0}$, which is a one-dimensional  complex
vector space for a Calabi-Yau manifold. Applying \eqref{pert-nonpert-split} with \eqref{nilpotent_orbit} to $F^D$, we find 
\beq \label{Omega_expand}
   \Omega = e^{t N^-} a_0 + e^{2\pi i t} \tilde a_1 + e^{4 \pi i t} \tilde a_2 + \ldots \ ,
\eeq
where we have fixed the overall normalization of $\Omega$ such that $a_0$ is independent of $t$. The derivative $\partial_t \Omega$ spans, together with $\Omega$ itself the space $F^{D-1}$ and takes the form 
\beq \label{dtOmega}
  \partial_t \Omega = e^{t N^-} N^- a_0 + e^{2\pi i t} \big( 2\pi i  \tilde a_1 + \partial_t \tilde a_1\big) + \ldots \ .
\eeq 
Applying now \eqref{pert-nonpert-split} with \eqref{nilpotent_orbit} to $F^{D-1}$, we see that, as long as $N^- a_0 \neq 0$, 
the first term in \eqref{dtOmega} takes again the form of a nilpotent orbit. However, if one considers a boundary 
with $N^- a_0 = 0$ then the exponential correction in \eqref{Omega_expand} is actually needed, such that 
the whole space $H^{D}(Y_D,\bbC)$ can be obtained from the nilpotent orbit $F^p_{\rm pol}$. In this case 
$ \partial_t \Omega$ is proportional to $e^{2\pi i t}$, but this overall factor can be removed by 
a rescaling since \eqref{nilpotent_orbit} is an equality between vector spaces.  
In other words, the existence of a nilpotent orbit \eqref{nilpotent_orbit} that gives a splitting \eqref{pol_splitting}, 
implies that $\Omega$ should be expanded as 
\beq \label{Omega-expansion}
   \Omega = e^{t N^-} \Big( a_0 + e^{2\pi i t}  a_1 + e^{4 \pi i t}  a_2 + \ldots  \Big)\ ,
\eeq
with the $a_i$, $i>0$ relevant as soon as the vectors $(N^-)^n a_0$ do not suffice to span $H^{D}(Y_D,\bbC)$. This implies that 
at many boundaries exponential corrections in $\Omega$ are \textit{implied} by 
the existence of a nilpotent orbit with the above properties. This matches nicely the recent proposal put forward 
in \cite{Palti:2020qlc} and will be discussed in more generality in \cite{BastianGrimmHeisteeg}.
We also note that the generic presence of exponential corrections has recently been shown in 
\cite{Cecotti:2020rjq} by using other results from Hodge theory.

\subsection{Differential equations and constraints from nilpotent orbits} \label{nilpotent_diffs}

%The study of the variation of the Hodge decomposition \eqref{Hodge-dec} over the moduli space is a mathematically 
%well developed field and captured by the theory of variations of Hodge structures \cite{ReviewHS}. 
With the motivation to develop a holographic perspective, 
it is desirable to formulate the behaviour of  \eqref{Hodge-dec}, and hence the norm 
\eqref{Hodge-norm}, by introducing a set of fields that live on the moduli space $\cM$ and 
admit equations of motions that are inspired by Hodge theory. In the following we will 
motivate such equations of motion by the conditions obeyed by the nilpotent 
orbit $F^p_{\rm pol}$ introduced in section \ref{Near-boundary}. This implies that 
we restrict our attention to the near boundary region. Eventually, the resulting 
equations can be considered without reference to an underlying nilpotent orbit. 

To motivate the equations of motion let 
us again consider a one-dimensional limit \eqref{local-coords}, i.e.~study the nilpotent orbit 
\beq \label{onepar_nilp}
   F^p_{\rm pol}(t) = e^{t N^-} F^p_0\ ,
\eeq
depending on one variable $t$. 
We want to study the moduli dependence 
of \eqref{onepar_nilp} with respect to a fixed Hodge decomposition $H^{p,q}_{\rm ref}$. This decomposition 
could be picked by simply evaluating $F^p_{\rm poly}(t)$ at a fixed $t_0$. However, we will 
make a more educated choice that prepares us for the discussion of the bulk-boundary correspondence
that will follow below. In fact, we will consider a reference Hodge decomposition $H^{p,q}_{\rm ref}$ induced by 
\beq \label{ref-F}
   F^p_{\rm ref} = e^{ i N^-}\, e^{-i\delta}\, F^p_0\ .
\eeq 
This choice corresponds to evaluating \eqref{onepar_nilp} at $t=i$ and multiply the result by a phase matrix $e^{-i\delta}$,
here $\delta$ is a real matrix in $\mathfrak{g}_\bbR$ satisfying $[\delta,N^-]=0$. The matrix $\delta$ can be uniquely associated 
to a given $N^-$, $F^p_0$ and we will call $\delta$ the \textit{phase operator} in the following. We will 
describe in appendix \ref{computedelta} how $\delta$ can be constructed starting from $N^-$, $ F^{p}_0$. 
The reason why we extract a phase in \eqref{ref-F} will become clear below. In a nutshell, we will find that asymptotically 
the moduli space metric admits an isometry group $\mathfrak{sl}(2,\bbR)$, discussed after \eqref{simple-asymptotic-metric}.
The phase operator is extracted in such a way that one can find a real slice $H^{D}(Y_D,\bbR)$ in $H^{D}(Y_D,\bbC)$ on which 
$\mathfrak{sl}(2,\bbR)$ acts faithfully. We 
also introduce the inner product $\langle w| v \rangle_{\rm ref}$ associated to  $H^{p,q}_{\rm ref}$ in analogy 
to \eqref{Weil-pol}, \eqref{def-innerprod_nil}.

Starting from the reference splitting $H^{p,q}_{\rm ref} = F^p_{\rm ref}  \cap \bar F^q_{\rm ref} $ induced by \eqref{ref-F}, we will first 
restrict our attention to the imaginary part $\I\,t=y $ that is taken to be large in \eqref{local-coords}
and set $x = 0$. We introduce a $G_{\bbR}$-valued function $h(y)$ that captures the change in $t=iy$ of 
 $H^{p,q}_{\rm pol}$ with respect to $H^{p,q}_{0}$ by setting
\beq \label{def-h(y)}
   H^{p,q}_{\rm pol}(i y) = h(y) H^{p,q}_{\rm ref}\ ,\qquad \quad h(y) \in G_{\bbR} \ ,
\eeq
or
\beq \label{Fpol-ash}
    F^p_{\rm pol}(iy) = e^{iy N^-} F^p_0 = h(y) F^p_{\rm ref} \ .
\eeq
Recall that $G_{\bbR}$ is the real group keeping the inner product \eqref{def-S} invariant. 
Since $e^{xN^-}$ is also an element of $G_{\bbR}$ one can complete \eqref{def-h(y)} to 
\beq \label{def-hatH}
   H^{p,q}_{\rm pol}(t,\bar t) = e^{x N^-}h(y) H^{p,q}_{\rm ref} \equiv \hat h(x,y) H^{p,q}_{\rm ref}\ ,
\eeq
where we have defined $\hat h(x,y) := e^{x N^-}h(y) \, \in\, G_{\bbR}$.
We therefore get an explicit expression for the $(p,q)$-forms in the asymptotic regime, in terms of 
a reference splitting. While the dependence on $x$ is rather simple,
the dependence on $y$ via $h(y)$ is generally involved. We therefore focus mostly on the $y$-dependence 
in the following, keeping in mind that the $x$-dependence can be restored rather easily. 

We can also choose to encode the splitting $\cH = \bigoplus_{p} H^{p,D-p}_{\rm ref}$ by using an operator $Q$, which 
is independent of $t$.
Such an operator is known as grading element  $Q \in \mathfrak{g}_{\bbC}$, see e.g.~\cite{robles_2016,Kerr2017}, and we will call it 
\textit{charge operator} in the following. It is defined by \footnote{
As a side remark, we note that the action of $Q$ on operators as in \eqref{QcN-matching_first}
can also be formulated more mathematically by using the decomposition of 
the algebra $\mathfrak{g}_\bbC$, defined after \eqref{def-S}, induced by $H^{p,q}_{\rm ref}$. 
Explicitly one splits $\mathfrak{g}_{\bbC} =  \bigoplus_{p \in \bbZ} \mathfrak{g}^{p,-p}_{\rm ref}$ with $\mathfrak{g}^{p,-p}_{\rm ref} = \big\{T \in \mathfrak{g}_\bbC : \, T H^{a,b}_{\rm ref} = H^{a+p,b-p}_{\rm ref} \big\}$, where  $\overline{ \mathfrak{g}^{p,-p}_{\rm ref}}= \mathfrak{g}^{-p,p}_{\rm ref}$ and $[ \mathfrak{g}^{p,-p}_{\rm ref}, \mathfrak{g}^{q,-q}_{\rm ref}] \subset  \mathfrak{g}^{p+q,-(p+q)}_{\rm ref}$. The notation 
$\mathfrak{g}^{p,-p}_{\rm ref}$ indicates that one is dealing with a Hodge structure of weight $p-p=0$. The action the charge operator $Q$ is then induced by the split as $[Q, T] = p \, T$ for $T \in \mathfrak{g}^{p,-p}_{\rm ref}$.}
\beq \label{Q_forms}
      Q |w\rangle_{\rm ref} = \frac{1}{2} (2p-D) |w\rangle_{\rm ref} \qquad \text{for} \quad |w\rangle_{\rm ref} \in H^{p,D-p}_{\rm ref}\ , 
\eeq 
where we are using a bra-ket notation for states in $\cH$. Since $\overline{H_{\rm ref}^{p,q}} = H^{q,p}_{\rm ref}$ we 
conclude that 
\beq  \label{Q0chargeimaginary}
   \bar Q = - Q\ , \qquad Q \in i \mathfrak{g}_{\bbR}\ .
\eeq
Recalling that the inner product $\langle w| v \rangle_{\rm ref}$ is induced 
by a real Weil operator $C_{\rm ref}$ as in \eqref{pol_splitting}, \eqref{Weil-pol}, we can use $Q$ to 
write 
\beq \label{prod-withQ}
   \langle w| v \rangle_{\rm ref} = \langle w,  C_{\rm ref} v\rangle\ , \qquad C_{\rm ref} = i^{2Q}\ . 
\eeq 
The charge operator $Q$ will play a central role in the construction, since it encodes the Hodge decomposition. 
The fact that it is not a real operator implies that its eigenstates with real eigenvalues are complex.

The function $h(y)$ satisfies a set of equations that will constitute the base of the bulk theory
discussed in section \ref{bulktheory}. 
We will formulate these equations in the algebra $\mathfrak{g}_{\bbR}$ associated to $G_{\bbR}$ and thus 
define $  \cN^0(y)\ ,  \cN^-(y)\ \in \mathfrak{g}_{\bbR}$ by setting
\beq \label{cN-def-h}
      \cN^0(y) := - 2 h^{-1}  \partial_y h\ ,  %= (\cN^0)^\dagger\ , 
      \qquad
       \cN^-(y) := h^{-1} N^- h \ ,\qquad
          \cN^+(y) := (\cN^-(y))^\dagger\ , 
\eeq
where we abbreviate $\partial_y = \frac{d}{dy}$ and denote by $\dagger$ the operation to adjoin an operator 
with respect to the inner product 
$\langle w| v \rangle_{\rm ref}$. Note that the adjoint $\cO^\dagger$ to $\cO \in \mathfrak{g}_{\bbC}$ is given by 
\beq 
    \cO^\dagger = - C_{\rm ref}^{-1} \bar \cO C_{\rm ref} \ ,
\eeq
as can be seen from \eqref{def-innerprod_nil} and \eqref{def-gg}.
Given these definitions one shows \cite{Schmid}\footnote{This is shown in Lemma 9.8.}
that any $ \cN^0(y)\ ,  \cN^\pm (y)$ coming from a nilpotent orbit \eqref{onepar_nilp} obeys
the differential equations
\beq \label{Nahm1_cN}
  \boxed{\rule[-.25cm]{.0cm}{.7cm} \quad
\partial_{y} \cN^{\pm} = \pm \tfrac{1}{2} [\cN^{\pm},\cN^0] \ , \qquad  
  \partial_{y} \cN^0 =- [\cN^{+} ,\cN^{-}]\ ,
  \quad }
\eeq
as well as the algebraic relations \footnote{
It might be useful to point out that \cite{CKS} displays these conditions in a different form: $\cN^- - \cN^+$ has charge $q=0$ under ad$_Q$, while $\cN^0$ has 
a $q=1$ and $q=-1$ component under ad$_Q$. Furthermore, they require $\cN^-+\cN^+ = i^{-Q} \cN^0 i^Q$, and $\cN^+ = - i^{-2Q} \cN^- i^{2Q}$.
These conditions are equivalent to the conditions stated here.}
\beq \label{QcN-matching_first}
   \boxed{\rule[-.25cm]{.0cm}{.7cm} \quad  \big[Q,\cN^{0}\big]= i  (\cN^+ + \cN^-) \ , \qquad     \big[Q,\cN^{\pm}\big] =-\tfrac{i}{2} \cN^0 \ .\quad}
\eeq
Note that from \eqref{prod-withQ} and \eqref{QcN-matching_first}  we infer
\beq \label{QN-dagger}
  Q^\dagger = Q \ ,\qquad (\cN^0)^\dagger = \cN^0\ , \qquad \cN^+ = (\cN^-)^\dagger\ ,
\eeq 
where we have chosen the last equality as definition of $\cN^+$ in \eqref{cN-def-h}. Note that 
the equations \eqref{Nahm1_cN}, \eqref{QcN-matching_first} are both constraints on the function 
$h(y)$ and make no reference anymore to the nilpotent orbit.

It is not hard to check that the differential   
equations \eqref{Nahm1_cN} are equivalent to Nahm's equations  $\partial_y T_i =-[T_j,T_k]$, for every cyclic permutation of $i,j,k$. 
We can then use that in \cite{donaldson1984} it was argued that Nahm's equations  \eqref{Nahm1_cN} can be analyzed by using an 
action principle for the fields $h(y)$. The one-dimensional action reads 
\bea \label{Nahm-action}
   S_{\rm Nahm}(h) &=& \frac{1}{2} \int  \Big(  \text{Tr} \big|(h^{-1} \partial_y h)^\dagger + h^{-1} \partial_y h \big|^2 +   2 \text{Tr} \big|h^{-1} N^- h \big|^2  \Big) dy\ ,\\
       &=&  \frac{1}{2} \int  \Big( \frac{1}{4}\text{Tr} |(\cN^0)^\dagger+\cN^0|^2 + 2 \text{Tr} |\cN^-|^2  \Big) dy\ , \nn 
\eea
where $|A|^2 = A^\dagger A$, with the dagger and the trace is evaluated in the norm $\langle v|w \rangle_{\rm ref}$. 
To see this, we check that the two equations to the left in \eqref{Nahm1_cN} are automatically satisfied with the 
Ansatz \eqref{cN-def-h}, while the last one is obtained from \eqref{Nahm-action}. The latter statement follows if
we vary the action by using $\delta \cN^- = - [h^{-1}\delta h , \cN^-] $ and $\delta \cN^0 = - [h^{-1}\delta h , \cN^0] - 2 \frac{d}{dy} (h^{-1} \delta h)$.
Note that we can equivalently formulate the conditions \eqref{Nahm1_cN}, \eqref{QcN-matching_first} and the action \eqref{Nahm-action} as 
constraints on $\hat h(x,y)$. We will see in section~\ref{bulktheory} that the construction naturally generalizes to an 
action for $\hat h(x,y)$ that can be coupled to two-dimensional gravity. 
%We suggest in section \ref{bulktheory} to interpret $\hat h(x,y)$ as a bulk field on the moduli 
%space with \eqref{Nahm1_cN}, \eqref{QcN-matching_first} constraining its dynamics. 

Before turning to the discussion of moduli space metric, let us close with a number of remarks. 
It is useful to note that the equations \eqref{Nahm1_cN} can be rewritten in many different forms that highlight certain aspects of the construction.
Firstly, we can use \eqref{QcN-matching_first} to obtain a differential equation in Lax form,
\beq \label{Lax11}
   \partial_y\cN^- = - i [\cN^-, [Q, \cN^- ]] \ ,
\eeq
which highlights the fact that there is an underlying integrable structure. 
Secondly, we can define the complex operators 
\beq \label{def-cL+-1}
 \cL_{\pm 1}(y) := \tfrac{1}{2} \big( \cN^+ + \cN^- \mp i \cN^0  \big) \ , \quad \cL_0(y) := i\big(\cN^- -\cN^+) \ ,
\eeq
which satisfy $\cL^\dagger_0 = \cL_0$ and $\cL_{1}^\dagger = \cL_{-1}$. The index $\alpha=(-1,0,1)$ indicates 
 the charge of these operators, 
\beq \label{QonLalpha}
      [Q,\cL_\alpha] = \alpha \cL_\alpha \ .
\eeq
The equations \eqref{Nahm1_cN} can then be written in the form of the $tt^*$ equations \cite{Cecotti:1991me}. Setting 
$  D \equiv \partial_t - \tfrac{i}{4}\cL_0$ and $C \equiv \tfrac{1}{2}\cL_{- 1 }$,
%one has $\bar D = \partial_{\bar t}+ \frac{i}{4}\cL_0$, $\bar C = -\frac{i}{2}\cL_{-1}$, where we recall that $t=x+iy$, $\partial_t = \frac{1}{2}(\partial_x - i \partial_y)$. 
the conditions \eqref{Nahm1_cN} and \eqref{QcN-matching_first} are then equivalent to
\bea \label{DC-system}
    && [D,C]= [D,\bar C] = 0 \ , \qquad [D,\bar D] = - [C,\bar C]\ , \\
    && [Q,C] = -C,\qquad [Q,\bar C] = \bar C \ , \qquad [D,Q]=[\bar D,Q] = 0 \ . \nn 
\eea
It is long known that these equations emerge from the variations of Hodge structures \cite{Cecotti:1991me}. This connection 
also provides the link between our work and the recent papers \cite{Cecotti:2020rjq,Cecotti:2020uek}. 

\subsection{Metric on the moduli space and its near boundary expansion} \label{near-boundary-metric}

There are at least two natural metrics, denoted by $g^{\text{\tiny WP}}$ and $g^{\text{\tiny H}}$, that can be defined on the moduli space $\cM$.
In Calabi-Yau compactifications the physically most relevant metric is 
the so-called \textit{Weil-Petersson metric} $g^{\text{\tiny WP}}$. It determines the kinetic terms of the complex coordinates
$z^K$, when interpreting them as scalar fields in the low energy effective action. In the study 
of the distance conjecture in Calabi-Yau compactifications $g^{\text{\tiny WP}}$ is therefore relevant to distinguish 
finite and infinite distance geodesics. As was shown in \cite{wang1} and exploited in \cite{GPV,Grimm:2018cpv}, infinite 
distances only can occur when approaching a boundary with an associated non-vanishing $N^-$ 
that satisfies $N^- a_0 \neq 0$, with $a_0$ appearing in \eqref{Omega_expand}. We will see that 
the Weil-Petersson metric near such points takes a characteristic asymptotic form. In fact, we
will argue that at each such limit an $\mathfrak{sl}(2,\bbR)$ isometry algebra emerges. This feature 
turns out to be generally true for $N^- \neq 0$ if one considers the \textit{Hodge metric} $g^{\text{\tiny H}}$ on moduli space, which we will
introduce below. Later on, in section \ref{boundary_theory}, we will see that this $\mathfrak{sl}(2,\bbR)$ is key in determining
the symmetry of the boundary theory. 

To begin with, we recall that the Weil-Petersson metric is a K\"ahler metric and 
can be derived via $g^{\text{\tiny WP}}_{I \bar J} = \partial_{z^I} \partial_{\bar z^J}K$ from a K\"ahler potential 
\beq  \label{K-WP}
   K= - \log i^{D} \int_{Y_D}  \Omega \wedge \bar  \Omega = - \log \| \Omega(z) \|^2 \ ,
\eeq
where $\Omega(z)$ is the $(D,0)$-form on $Y_D$ spanning $H^{D,0}$ and varies 
holomorphically in the $z^I$. This metric can be explicitly evaluated by 
introducing a basis of $(D-1,1)$-forms as derivatives of $\Omega$. These 
are defined by evaluating $\partial_{z^I} \Omega$ and projecting  the result
to the $(D-1,1)$ part, $\big[\partial_{z^I} \Omega\big]^{(D-1,1)}  \equiv \nabla_i \Omega $. 
We have denoted by $ \nabla_i \Omega$ the resulting $(D-1,1)$-forms.\footnote{It can be checked 
that $\nabla_i$ is the K\"ahler-covariant derivative in this case.} Taking derivatives 
of \eqref{K-WP} one thus derives the Weil-Petersson 
metric
\beq \label{gWP}
   g^{\text{\tiny WP}}_{I \bar J}  =  \frac{\int_{Y_D}   \nabla_I \Omega \wedge \overline{ \nabla_J \Omega}}{\int_{Y_D} \Omega \wedge \bar \Omega}\ . 
\eeq

Let us now give a first, rough, evaluation of this metric near the boundary of the moduli space
by using the expansions introduced in section \ref{Near-boundary}. For simplicity we will 
only focus on a single coordinate $t = x+iy$ in the regime $y \gg 1$ and drop all $\zeta^\kappa$-dependence 
labelling the position on the boundary. Inserting \eqref{Omega-expansion} into the 
K\"ahler potential \eqref{K-WP} we compute the line element 
\beq \label{asymp_WP}
 ds^2_{\text{\tiny WP}}= \frac{1}{y^2}  \Big(\hat d + \gamma(y) \Big) (dy^2 + dx^2)\ ,
 \eeq
 where $\gamma(y)$ vanishes as $y \rightarrow \infty$.
The integer $\hat d \in \{0,...,D\}$ is the highest power of $N^-$ that does not annihilate $a_0$ introduced in \eqref{Omega-expansion}, 
 i.e.~$(N^-)^{\hat d} a_0 \neq 0$ while $(N^-)^{\hat d+1} a_0 = 0$. Therefore, as long as $\hat d>0$, i.e.~$N^- a_0 \neq 0$ 
 the metric asymptotes to the Poincar\'e metric, which is also describing a patch of Euclidian Anti-de Sitter space. Being motivated 
 to establish a holographic perspective one might thus want to either treat the case $\hat d=0$ separately, or consider a more suitable 
 metric. In fact, the case $\hat d = 0$ describes precisely the situation in which the boundary is at \textit{finite} distance, i.e.~that there 
 exists a path to the boundary of finite length in the Weil-Petersson metric. In light of the distance conjecture \cite{Ooguri:2006in}, which 
 discusses infinite distance boundaries, we thus indeed expect that the finite distance case $\hat d = 0$ is special. The details 
 on how the distinction of the cases $\hat d=0$ and $\hat d>0$ plays out in studying the states relevant to the distance conjecture  
 was explained in detail in \cite{GPV,Grimm:2018cpv,Corvilain:2018lgw}. From the perspective of the underlying 
 structure, however, the case $\hat d=0$ is not particularly special as we will see by 
 looking at another well-known metric, the Hodge metric on $\cM$, 
 in which this distinction disappears.

While the Weil-Petersson metric is directly physically relevant in Calabi-Yau compactifications, the
the so-called Hodge metric $g^{\text{\tiny H}}$ is central in Hodge theory. Its properties have been studied in numerous 
mathematical works (see e.g.~\cite{Lu99,Lu01,lu_sun_2004,LuFang05,Douglas:2006zj,Lu:2009aw,PearlsteinPeters}). In fact, it was also used in the physical study of  
Calabi-Yau compactifications with background fluxes in~\cite{Douglas:2006zj,Lu:2009aw}. 
 Crucially, it turns out to have `nicer' properties when considering its 
curvature tensors, which is mainly due to its universal asymptotic behaviour. The Hodge metric is defined by 
\beq \label{gH}
  g^{\text{\tiny H}}_{I \bar J}  =  \sum_{p=0}^D \cG^{\alpha_p \bar \beta_p} \int_{Y_D}   \nabla_I \chi_{\alpha_p} \wedge * \overline{\nabla_J \chi_{\beta_p}}     \ ,    
\eeq
where $\chi_{\alpha_p}$ is a basis of $(D-p,p)$-forms and $ \nabla_I \chi_{\alpha_p}$ is the derivative $\partial_{z^I} \chi_{\alpha_p}$ projected 
onto the $(D-p-1,p+1)$-component. The matrix $G^{\alpha_p \bar \alpha_p}$ is the inverse of the metric 
\beq \label{Gpp}
  \cG_{\alpha_p \bar \alpha_p} = \int_{Y_D}   \chi_{\alpha_p} \wedge * \overline{  \chi_{\alpha_p}}\ .
\eeq
Note that the sum defining $g^{\text{\tiny H}}_{I \bar J}$ contains terms that are equal to $g^{\text{\tiny WP}}_{I \bar J}$. In fact, one immediately sees 
that the summand $p=0$ in \eqref{gH} is exactly the Weil Peterson metric \eqref{gWP}, since the only $(D,0)$ is $\Omega$ 
and the metric \eqref{Gpp} yields a single term proportional to $\int \Omega \wedge \bar \Omega$ in this direction. 
In addition, the metric $g^{\text{\tiny WP}}_{I \bar J}$ also appears from the summand $p=D$, since in this case one finds only 
the $(0,D)$-form $\bar \Omega$. This leads us to conclude that $g^{\text{\tiny H}}_{I \bar J} = 2 g^{\text{\tiny WP}}_{I \bar J}+\ldots$, 
with the omitted terms being positive definite expressions in the curvature tensors of $g^{\text{\tiny WP}}_{I \bar J}$. A direct computation
reveals that one can explicitly relate the Weil-Petersson and Hodge metric for Calabi-Yau manifolds of arbitrary dimension $D$. 
For example, it was found in \cite{Bershadsky:1993cx,lu2005hodge, lu_sun_2004} that  
\bea \label{WP-Hodge-Relation}
 %  \text{K3}: \qquad g^{\text{\tiny H}}_{I \bar J} &=& 2 g^{\text{\tiny WP}}_{I \bar J} \ , \nn \\
  D=3: \qquad g^{\text{\tiny H}}_{I \bar J} &=& (h^{2,1} + 3) g^{\text{\tiny WP}}_{I \bar J}+ R^{\text{\tiny WP}}_{I \bar J}\\
   D=4: \qquad g^{\text{\tiny H}}_{I \bar J} &=& 2(h^{3,1} + 2) g^{\text{\tiny WP}}_{I \bar J}+ 2R^{\text{\tiny WP}}_{I \bar J}\ , \nn 
\eea
where $R^{\text{\tiny WP}}_{I \bar J}$ is the Ricci tensor computed in the Weil-Peterson metric. 
A key observation is that the metric $g^{\text{\tiny H}}_{I \bar J} $ has a nice asymptotic behaviour. In fact, 
we will see that its asymptotic form always splits off a part that is a
Poincar\'e metric as long as one has $N^-_i \neq 0$, for at least one $N_i$. 

Near the boundary of the moduli space we can use the nilpotent orbit to derive 
the metric $g_{I \bar J}$. Let us, as above, denote by $ \chi_{\alpha_p}$ a basis of $(D-p,p)$-forms, which are now 
in the decomposition \eqref{pol_splitting}, and denote by $ \nabla_I \chi_{\alpha_p} $ the $z^I$-derivative projected to 
the $(D-p-1,p+1)$-piece. We now use the notation \eqref{def-innerprod_nil} to write the metric \eqref{gH} as
\beq \label{nilpgH}
  g_{I \bar J}  =  \sum_{p=0}^D \langle   \chi_{\alpha_p}  | \chi_{\beta_p} \rangle_{\rm pol}^{-1} \langle  \nabla_I \chi_{\alpha_p}  |  \nabla_J \chi_{\beta_p} \rangle_{\rm pol}     \ ,    
\eeq
where we have used the notation established in \eqref{def-innerprod_nil}.

To give a first  study the asymptotic behaviour of  the Hodge metric $ g^{\text{\tiny H}}_{I \bar J}$ 
we again focus on a single coordinate $t = x+iy$ in the regime $y \gg 1$ and drop all $\zeta^\kappa$-dependence.
To evaluate the leading metric we use \eqref{Omega-expansion} and its successive derivatives with respect to $t$.
The line element for the metric $g_{t\bar t}$ now takes the form 
\beq \label{simple-asymptotic-metric}
ds^2_\text{\tiny H} = \frac{1}{y^2}  \Big( c^{(0)} + \hat \gamma(y) \Big) (dy^2 + dx^2)\ ,
\eeq
where now $ c^{(0)} = \frac{1}{12} \sum_i  d_i ( d_i + 1)( d_i + 2)$ with a sum over the irreducible 
$\mathfrak{sl}(2,\bbR)$-representations with highest weight $ d_i$ in the boundary theory introduced in section \ref{boundary_theory}, 
see \cite{GrimmHeisteegMonee} for details. Crucially, we realize that as long as $N^-\neq 0$
we have $c^{(0)}>0$. Note that this implies that the metric indeed becomes the Poincar\'e metric 
$ \frac{c^{(0)}}{y^2}  (dy^2 + dx^2)$ in the limit $y \gg 1$. It is not hard to check that this metric 
admits an $Sl(2,\bbR)$ isometry group. Furthermore, recall that $Sl(2,\bbR) \cong SO(2,1)$ is the global conformal 
group in one dimension. It might therefore be tempting to associate a conformal quantum mechanical 
system to this setting. In fact, we will see in section \ref{boundary_theory} that there is indeed a type of `boundary theory'
associated to each asymptotic limit.  In fact, we will see that the asymptotic metric
 \eqref{simple-asymptotic-metric} admits an expansion 
\beq \label{simple-asymptotic-metric_exp}
 ds^2_{\text{\tiny H}} = \frac{1}{y^2}  \left( c^{(0)} +\frac{c^{(1)}}{y}  +\frac{c^{(2)}}{y^2} + \ldots \right) (dy^2 + dx^2)\ ,
\eeq
and find that the coefficients $c^{(i)}$ are determined non-trivially by a set of boundary data. 
This strategy can also be applied to obtain the 
expansion of the Weil-Petersson metric. In the case of having an infinite distance boundary the 
`boundary theory' fixes the the asymptotic expansion in powers of $1/y$ at $y\rightarrow \infty$.
Note, however, that in contrast to standard AdS$_2$ holography, 
we are \textit{not} considering the conformal boundary at $y \rightarrow 0$.

\section{The bulk theory on the moduli space} \label{bulktheory}

In this section we discuss aspects of the classical bulk theory living on the moduli space. 
The aim is to find field equations and an 
action principle for a metric and a matrix-valued field, such that the classical 
solutions include the `physical' nilpotent orbits \eqref{nilpotent_orbit}.
In other words, we use the statements of Hodge theory reviewed in section \ref{motivation} and 
the existence of a nilpotent orbit as a motivation for a bulk theory. It will then become 
clear that the field equations of this theory only yield back a nilpotent orbit, if certain 
boundary conditions are imposed. These are provided by the boundary theory introduced 
in section \ref{boundary_theory}.
%which then determine the 
%near boundary behaviour of the Hodge metric and Hodge decomposition. 
One of the goals of our construction is to capture the information about the asymptotic 
geometry of moduli spaces without reference to Hodge theory. 
As noted before, there is a natural action principle associated to one-parameter nilpotent orbits \cite{donaldson1984}. 
Recently, a significant generalization has been discussed in \cite{Cecotti:2020rjq}. Cecotti also suggested 
in \cite{Cecotti:2020uek} that an Einstein-Hilbert term with negative cosmological 
constant can be coupled. Since such a coupling is trivial in two-dimensions, i.e.~in the settings most 
relevant to this work, we suggest in section \ref{bulk-action-gravity} to deviate from \cite{Cecotti:2020rjq}. 
It should be stressed, however, that many 
aspects of the argument in section \ref{matteraction} and \ref{bulk-action-gravity} are similar to \cite{Cecotti:2020rjq,Cecotti:2020uek} and have been observed independently as part of this project. The fact that there is a bulk gravity action on moduli space 
fits rather naturally to the holographic perspective suggested here. 

\subsection{Bulk action for matter fields}\label{matteraction}

In order to construct a bulk theory compatible with the Hodge theory analysis of section \ref{motivation},
we first discuss the relevant field equations without reference to a nilpotent orbit. We then aim 
to find an action on the moduli space encoding these equations. We will focus on 
a real two-dimensional moduli space and later comment on possible higher-dimensional 
generalizations. 

 Instead of fixing $\cM$ to be the moduli space of a Calabi-Yau manifold, we consider $\cM$ abstractly as being
the real two-dimensional world-sheet of a bulk sigma model. We denote the local world-sheet coordinates by 
$\sigma^1,\sigma^2$. This sigma-model has matter fields $\hat h(\sigma)$ that take values 
in the group $G_\bbR$, which is the target space of the bulk sigma model. 
The group $G_\bbR$ acts on a finite-dimensional complex Hilbert space $\cH$ and 
preserves some bilinear form $\langle v,w \rangle$. We denote the fixed inner product on $\cH$ by 
$ \langle v |w \rangle_{\rm ref}$. A non-trivial requirement is that the inner product 
can be written using a grading element $Q \in i \mathfrak{g}_\bbR$, which we call the charge operator.  
By definition, this operator is a semisimple algebra element which splits $\mathfrak{g}_\bbC$ into eigenspaces with 
integer eigenvalues and obeys $\bar Q = - Q$. Given any operator $\cO \in \mathfrak{g}_\bbC$, we can then make
a decomposition $\cO = \sum_l \cO_l$ such that
\beq
     [Q, \cO_l] = l \cO_l \ , \qquad l = -D,..., D\ .  
\eeq 
This grading element gives an essentially equivalent way to formulate a Hodge decomposition and their infinitesimal 
variations.\footnote{See \cite{robles_2016,Kerr2017} for the mathematical details.} Motivated by the Hodge theory construction we 
now require as in section \ref{motivation} that 
\beq
    \langle v |w \rangle_{\rm ref} =  \langle v, i^{2Q}w \rangle\ . 
\eeq
Note that we infer by using \eqref{def-gg} that $Q^\dagger = Q$, with the adjoint taken with respect to $ \langle v |w \rangle_{\rm ref}$.

Given this structure we can now discuss the field equations for the matter fields $\hat h$.
Let us first restate the second equation in \eqref{Nahm1_cN} using $\hat h$, which 
amount to writing
\beq \label{h-Nahm}
  2  \partial_{y}  \big(\hat h^{-1}  \partial_y \hat h \big)=\big[(\hat h^{-1} \partial_x \hat h)^\dagger, \hat h^{-1} \partial_x \hat h \big]\ .
\eeq
Furthermore, we recall that the equations \eqref{QcN-matching_first} can be rewritten using $\hat h$ and take the form 
 \beq \label{Q-restr_eom}
   \boxed{\rule[-.2cm]{.0cm}{.7cm} \quad  -2\big[Q,\hat h^{-1} \partial_{y} \hat h\big]= i \Big( (\hat h^{-1} \partial_{x} \hat h)^\dagger +  \hat h^{-1} \partial_{x} \hat h \Big) \ , \qquad     \big[Q,\hat h^{-1} \partial_{x} \hat h \big] =i\hat h^{-1} \partial_{y} \hat h \ . \quad }
 \eeq
 We note that this equation implies 
 \beq \label{self-dual}
    (\hat h^{-1} \partial_{y} \hat h)^\dagger = \hat h^{-1} \partial_{y}\hat h\ . 
 \eeq
Clearly, the equations \eqref{h-Nahm}, \eqref{Q-restr_eom} are not democratic in $x,y$, which is due to the fact that in the nilpotent orbit solutions the coordinate $x$ appears only through 
the exponential $e^{x N^-}$ as seen in \eqref{def-hatH}. This simple dependence yields the conditions  
\beq \label{hathderiv}
   \partial_{x}  \big(\hat h^{-1}  \partial_y \hat h \big) = \partial_{x}  \big(\hat h^{-1}  \partial_x \hat h \big) = 0\ . 
\eeq
We can also realize these constraints by requiring the existence of a continuous symmetry 
\beq \label{h-rotation}
   \hat h(x+c,y) = e^{c N^-} \hat h(x,y)\ ,
\eeq
where $c$ is any real constant and $ N^-$ is some real matrix in $\mathfrak{g}_{\bbR}$. 
To obtain an unconstrained set of equations of motion democratic in $\sigma^1\equiv x, \sigma^2\equiv y$ we can combine \eqref{hathderiv} with \eqref{h-Nahm} and \eqref{self-dual} to obtain 
\beq \label{full-eom}
   \sum_\alpha \partial_{\sigma^\alpha}  \big(\hat h^{-1}  \partial_{\sigma^\alpha} \hat h + (\hat h^{-1}  \partial_{\sigma^\alpha} \hat h)^\dagger \big) - \big[(\hat h^{-1} \partial_{\sigma^\alpha} \hat h)^\dagger, \hat h^{-1} \partial_{\sigma^\alpha} \hat h \big] = 0\ .
\eeq
In the following we will view these equations as equations of motion for $\hat h$.
That this is plausible can be further motivated in various ways. Firstly, we could have considered in section \ref{motivation} 
the full variation of Hodge structure as done in \cite{Cecotti:2020rjq}. Secondly, we can obtain the equations of motion \eqref{full-eom}
from an action that arises as natural generalization of \eqref{Nahm-action} as we will see in the following.

To obtain a bulk matter action principle, we start with the one-dimensional action  \eqref{Nahm-action}. To 
obtain a two-dimensional action for the matter fields $\hat h(\sigma)$, we first replace $ h^{-1} \partial_y h= \hat h^{-1} \partial_y \hat h$
and $ h^{-1} N^- h = \hat h^{-1} \partial_x \hat h$ by temporarily assuming the simple $x$-dependence $\hat h = e^{x N^-} h$. 
Integrating over $\sigma^1\equiv x$ the action \eqref{Nahm-action} then generalizes to an action $S(\hat h)$ 
as
\beq
   S(\hat h) = \frac{1}{2} \int d^2 \sigma  \Big(  \text{Tr} | (\hat h^{-1} \partial_{\sigma^2} \hat h)^\dagger + \hat h^{-1} \partial_{\sigma^2} \hat h |^2 +   2 \text{Tr} | \hat h^{-1} \partial_{\sigma^1} \hat h |^2  \Big)\ ,
\eeq
where $|A|^2 = A^\dagger A$ and $d^2 \sigma= d\sigma^1 d \sigma^2$. We can check that adding the terms 
Tr$( \hat h^{-1} \partial_{\sigma^1} \hat h)^2$ and Tr$( (\hat h^{-1} \partial_{\sigma^1} \hat h)^\dagger)^2$ to the action does not change the equations of motion. This allows us to write 
\beq \label{action-with-gen_1d_flatgauge}
    \boxed{\rule[-.25cm]{.0cm}{.9cm} \quad S_{\rm mat}(\hat h) =  \frac{1}{2} \int  d^2\sigma\,  \sum_{\alpha} \text{Tr} | (\hat h^{-1} \partial_{\sigma^\alpha} \hat h)^\dagger + \hat h^{-1} \partial_{\sigma^\alpha} \hat h |^2  \ .\quad }
\eeq
That the equations of motion resulting from this action are indeed given by  \eqref{full-eom} can be checked by using    
\beq
   \delta (\hat h^{-1} \partial_{\sigma^\alpha} \hat h) = - [\hat h^{-1}\delta \hat h , \hat h^{-1} \partial_{\sigma^\alpha} \hat h] + \frac{d}{d\sigma^\alpha} (\hat h^{-1} \delta \hat h)  \ .
   % \qquad \delta \cN^0 = - [\hat h^{-1}\delta \hat h , \cN^0] - 2 \frac{d}{d\sigma^2} (\hat h^{-1} \delta \hat h)
\eeq
Let us note that the action \eqref{action-with-gen_1d_flatgauge} agrees with the one found in \cite{Cecotti:2020rjq}, which was shown to yield the $tt^*$ equations. We also see that the generalization to higher-dimensional field spaces of \eqref{action-with-gen_1d_flatgauge} and \eqref{Q-restr_eom}
appears to be straightforward. This is partly deceiving, since further constraints arise from imposing commutativity of 
certain derivatives $\hat h^{-1} \partial_{\sigma^\alpha} \hat h$. When looking at near boundary solutions a very non-trivial structure 
emerges \cite{CKS} that we will not further analyze in this work. Our study of solutions to 
\eqref{full-eom} and \eqref{Q-restr_eom} near the boundary does apply, however, to higher-dimensional moduli spaces where $\hat h$ depends on two coordinates $\sigma^1,\sigma^2$ near the boundary and a number of directions that remain in the bulk and are suppressed in the notation.

It is important to note that the action \eqref{action-with-gen_1d_flatgauge} has to be supplemented by \eqref{Q-restr_eom}. 
From the above discussion 
it should be clear that the existence of $Q$ and the constraint \eqref{Q-restr_eom} is central to the construction
and it would be desirable to find an action principle that also yields the latter constraint. In the bulk-boundary matching 
of section \ref{bulk_boundary}, however, we will only need the field equations \eqref{h-Nahm} and \eqref{Q-restr_eom} and therefore 
leave the construction of a complete action principle to future work. 
Furthermore, it is crucial to impose a set of boundary conditions 
to obtain solutions to these field equations that correspond to nilpotent orbits as 
we will see in section \ref{bulk_boundary}. Firstly, one has to require the 
symmetry \eqref{h-rotation} which constrains the $x$-dependence of $\hat h$. 
Secondly, note that a solution to \eqref{h-Nahm} with \eqref{Q-restr_eom}  gives also 
a solution to Nahm's equations \eqref{Nahm1_cN} when setting
\beq  \label{def-cN-hath}
\cN^0 := - 2 \hat h^{-1}  \partial_y \hat h\ ,  %= (\cN^0)^\dagger\ , 
      \qquad
       \cN^-  :=  \hat h^{-1}  \partial_x \hat h \ ,\qquad
          \cN^+  := (\cN^-(y))^\dagger\ .
\eeq
 The solutions to Nahm's equations give rise at the poles 
of $\cN^0,\cN^{\pm}$ to a special triples of operators that commute as generators of $\mathfrak{sl}(2,\bbR) \cong \mathfrak{su}(2)$ 
\cite{hitchin1983}.
In the situation at hand, we are interested in solutions near $y = \infty$ and we will consider 
general bulk solutions of the form 
\beq \label{boundary11}
  \cN^0  = \frac{\tilde N^{0}}{y} + \cO\big(y^{-\frac{3}{2}} \big)\ , \qquad \cN^{\pm}  = \frac{\tilde N^{\pm }}{y} + \cO\big(y^{-\frac{3}{2}} \big)\ .
\eeq
The coefficients of the slowest decreasing term of $\cN^0,\cN^{\pm}$ are triples ($\tilde N^0,\tilde N^{\pm}$), which are 
generators of an algebra $\mathfrak{sl}(2,\bbR)$ as we will discuss below. 
One can think of the ($\tilde N^0,\tilde N^{\pm}$) as setting part of the boundary conditions for the solution  $\cN^{0},\cN^{\pm}$. 
Furthermore, to implement the positivity of the norms over the moduli space, we also need to make sure that there 
is a well-defined reference structure $|v\rangle_{\rm ref}$, which we will argue should exist on the boundary.
Therefore, to single out solutions that correspond to physical situations, i.e.~to a valid $F^p_{\rm pol}$, we will then 
turn this into a requirement  for a specification of the physical boundary conditions. In the next section \ref{boundary_theory}
we suggest that these boundary conditions arise from a certain `boundary theory'.

\subsection{On the coupling to gravity} \label{bulk-action-gravity}

In the following we would like to briefly comment on the possibility 
to couple the matter action of section \ref{matteraction} to gravity. As before, we will 
mostly focus on 
a one-dimensional moduli space, but later comment on possible generalizations. 
Several aspects of the following discussion have recently also appeared in \cite{Cecotti:2020uek}.
We will make contact with the results given there and highlight where our construction differs. 

From a holographic perspective it would be natural if there is actually a gravity theory on the moduli space with a dynamical metric $g_{\alpha \beta}$.
For a real two-dimensional $\cM$, however, we recall that Einstein gravity is non-dynamical and the Einstein-Hilbert term reduces to a constant. 
In fact, we can think of \eqref{action-with-gen_1d_flatgauge}
as a two-dimensional string world-sheet action gauge-fixed to the trivial metric $\delta_{\alpha \beta}$. 
The coupling to a general metric $g_{\alpha \beta}$ yields then the action 
\beq \label{fullmatter_action}
   S_{\rm mat}(g,\hat h) = \int_{\cM} d^2 \sigma \sqrt{g} \ \cL_{\rm mat} \ , 
 \eeq
 with
 \beq
      \cL_{\rm mat} =  \frac{1}{2}  g^{\alpha \beta}\, \text{Tr} \big[(\hat h^{-1} \partial_{\sigma^\alpha} \hat h)^\dagger +  \hat h^{-1} \partial_{\sigma^\alpha} \hat h ] \big] \big[(\hat h^{-1} \partial_{\sigma^\beta} \hat h)^\dagger +  \hat h^{-1} \partial_{\sigma^\beta} \hat h ] \big] \ .
 \eeq
There are now two sets of equations of motion: (1) the equations of motion of $\hat h$, and (2) the equations of motion for $g_{\alpha \beta}$. 
The latter simply correspond to the statement that the energy-momentum tensor vanishes, 
\beq \label{Tvanishes}
     T_{\alpha \beta}^{\rm mat} =0\ .
\eeq 
Let us now compute $T_{\alpha \beta}^{\rm mat}$ and 
check when it vanishes on-shell for a nilpotent orbit solution $\hat h$.

In evaluating the energy momentum tensor $T_{\alpha \beta}^{\rm mat}$ we use 
the important observation that in complex coordinates $t = \sigma^1+ i \sigma^2$ one finds that 
the integrand in \eqref{action-with-gen_1d_flatgauge} can be expressed by using the Hodge metric \eqref{nilpgH}
when evaluated on a solution $\hat h$ corresponding to a nilpotent orbit. 
To see this, we use a basis of  $(D-p,p)$-forms $\hat h | v\rangle_{\rm ref}$ in the expression for the Hodge metric \eqref{nilpgH}.
If $\hat h | v\rangle_{\rm ref}$ is of type  $(D-p,p)$ then also $| v\rangle_{\rm ref}$ is of type $(D-p,p)$, but now in the reference 
Hodge decomposition \eqref{def-hatH}.
Since $\cL_{-1}$ obeys 
 \eqref{QonLalpha} we conclude that $\cL_{-1} | v\rangle_{\rm ref} $ is a $(D-p-1,p+1)$-form as can be checked by recalling \eqref{Q_forms}. 
Hence, we realize that $\cL_{-1}$ acts as $ \nabla_t$ appearing in the Hodge metric  \eqref{nilpgH}
and allows us to rewrite the line element of $g^{\text{\tiny H}}$.
Concretely, using $\text{Tr}(\cL_{-1}^2)=-\text{Tr}([Q,\cL_{-1}] \cL_{-1})=0$, the line element of the Hodge metric can be written as
\bea
g_{t\bar t}^{\text{\tiny H}}(\hat h) dt d\bar t &=& \text{Tr} | \cL_{-1} dt |^2  =  \frac{1}{2}\text{Tr}|\cN^+ + \cN^-|^2  (d\sigma^1)^2 + 
\frac{1}{2}\text{Tr} | \cN^0|^2 (d\sigma^2)^2  \\
 &=&  \frac{1}{2} \sum_{\alpha} \text{Tr} | (\hat h^{-1} \partial_{\sigma^\alpha} \hat h)^\dagger + \hat h^{-1} \partial_{\sigma^\alpha} \hat h |^2 (d\sigma^\alpha)^2 \ , \nn 
\eea
where we have used \eqref{def-cL+-1} in the second equality and \eqref{QN-dagger}, \eqref{def-cN-hath} in the third.
Therefore, the Lagrangian density $\cL_{\rm mat}$ and the energy momentum tensor $T_{\alpha \beta}^{\rm mat}$ can be evaluated on a nilpotent orbit solution as
\bea \label{Tmatter_H}
    \cL_{\rm mat} = \frac{1}{2} g^{\alpha \beta} g^{\text{\tiny{H}}}_{\alpha \beta} \ , \qquad  
    T_{\alpha \beta}^{\rm mat} = g^{\text{\tiny H}}_{\alpha \beta}-\frac{1}{2} g_{\alpha \beta}\, g^{\gamma \delta} g^{\text{\tiny H}}_{\gamma \delta} \ , 
\eea 
where we have expressed $g^{\text{\tiny H}}_{\alpha \beta}$ in real coordinates, keeping in mind that this metric is K\"ahler and 
hence obeys $g^{\text{\tiny H}}_{tt} = g^{\text{\tiny H}}_{\bar t \bar t}=0$. It is now easy to see that $T_{\alpha \beta}^{\rm mat} = 0$
for any choice of metric $g_{\alpha \beta}$ on $\cM$ that satisfies $g_{tt} = g_{\bar t \bar t}=0$. 
In particular, we can consider $g_{\alpha \beta}$ to be the K\"ahler metrics
$g_{\alpha \beta} = g^{\text{\tiny WP}}_{\alpha \beta}$ or $g_{\alpha \beta} = g^{\text{\tiny H}}_{\alpha \beta}$ and 
satisfy the complete set of equations of motion of \eqref{fullmatter_action}. The latter fact was 
also shown in general dimension in \cite{Cecotti:2020rjq,Cecotti:2020uek}.

While this result is encouraging 
it appears that one should look for a more 
sophisticated two-dimensional gravity theory. Ideally one 
would like to construct a two-dimensional theory, such that 
$g^{\text{\tiny WP}}_{t\bar t}$ with the expansion \eqref{asymp_WP} is a solution at infinite distance points. Starting 
again with the nilpotent orbit solutions for the matter theory and imposing \eqref{Tvanishes}, we realize from the above discussion 
that there is only a single real function, namely $g_{t \bar t}$, in  
the metric $g_{\alpha \beta} $ to be fixed. This can be done by imposing one real equation relating the 
matter part of the theory to the pure gravity part. In particular, we will now check that imposing
\beq \label{relatewithmatter}
     \boxed{\rule[-.15cm]{.0cm}{.6cm} \quad R - 2 \Lambda = 2\kappa^2 \cL_{\rm mat}\quad }
\eeq
does indeed ensure that $g_{t\bar t} =g^{\text{\tiny WP}}_{t\bar t}$, if we fix $\Lambda$ and $\kappa$ appropriately. 
Considering $\cL_{\rm mat}$ on a nilpotent orbit solution $\hat h$, we use \eqref{Tmatter_H} to show that \eqref{relatewithmatter} 
can be written as
\beq
    g^{\alpha \beta} (R_{\alpha \beta} - 4 \Lambda g_{\alpha \beta}) = \kappa^2 g^{\alpha \beta} g_{\alpha \beta}^{\text{\tiny H}}\ .
\eeq
The real equation fixes $g_{t\bar t}$ in case we impose $T_{\alpha \beta}^{\rm mat} = 0$. To 
ensure the match with $g_{t\bar t}=g^{\text{\tiny WP}}_{t\bar t}$, we now fix $\kappa,\Lambda$ by 
assuming $g_{\alpha \beta} = g^{\text{\tiny WP}}_{\alpha \beta}$ 
and exploiting the relation of the Hodge metric $g^{\text{\tiny H}}_{\alpha \beta}$ with $g^{\text{\tiny WP}}_{\alpha \beta}$. 
Recall the that we have given the explicit relations in \eqref{WP-Hodge-Relation} for Calabi-Yau threefolds ($D=3$) and Calabi-Yau 
fourfolds ($D=4$) and refer to \cite{Lu:2009aw} for the general discussion. For these two cases one finds 
\beq \label{fix-Dkappa}
   D=3:\quad \Lambda=-1\ , \ \kappa^2=1\ , \qquad  D=4:\quad \Lambda=-\frac{3}{4}\ , \ \kappa^2=\frac{1}{2}\ . 
\eeq

It would be desirable to formulate an action principle imposing the field equations \eqref{Q-restr_eom}, \eqref{full-eom}, \eqref{Tvanishes}, \eqref{relatewithmatter} on $\hat h$ and $g_{\alpha \beta}$.  
 A natural choice appears to be the coupling of \eqref{fullmatter_action} to the Jackiw-Teitelboim gravity action \cite{Teitelboim:1983ux,Jackiw:1984je}.
Introducing an auxiliary field $\Phi(\sigma)$ the full action then reads
\beq \label{full-bulk-action}
     S(\hat h,g,\Phi)= - \frac{1}{2 \kappa^2}  \int_{\cM} d^2 \sigma \sqrt{g} \Phi (R -2 \Lambda) + \int_{\cM} d^2 \sigma \sqrt{g} \Phi \cL_{\rm mat}\ .
\eeq 
The field $\Phi$ acts as a Lagrange multiplier and enforces the condition \eqref{relatewithmatter}.
We stress, however, that this theory is likely not complete and deserves further study. In particular, we find 
that the coupling to $\Phi$ also modifies the matter equations of motion and we recall that we need  
to impose the constraint \eqref{Q-restr_eom} on the matter fields. Furthermore, we have not discussed 
possible boundary terms in the action \eqref{full-bulk-action}. It would be desirable to identify the 
correct gravity model, if it exists at all, from the plethora of two-dimensional possibilities \cite{Grumiller:2002nm,Grumiller:2006rc}.

Let us close by noting that for higher-dimensional moduli spaces a coupling to gravity with a 
Einstein-Hilbert action with a cosmological constant is possible \cite{Cecotti:2020uek}. Also in these 
cases one can determine $\kappa$ and $\Lambda$ depending on $D$ and the dimensionality of $\cM$, 
such that the `physical' metric $g^{\text{\tiny WP}}_{\alpha \beta}$ is a solution to the Einstein equations. In light 
of the above observations and with a focus on a holographic perspective 
it would be nice to see if this is indeed the correct coupling to gravity. 
In the remainder of the paper we will mostly restrict to the two-dimensional situation and, since we are only talking about 
classical aspects of the theory, it will suffice to work with the field equations \eqref{full-eom} and \eqref{Q-restr_eom} directly.

\section{The Sl(2,$\mathbf{\bbC}$) boundary theory} \label{boundary_theory}

In this section we discuss the structure that arises at the boundary $\partial \cM$ of the 
moduli space $\cM$ with a focus on a $\text{dim}\cM-1$ dimensional component. Recalling that the 
boundary splits into multiple such components $\Delta_k$, as discussed around \eqref{boundary}, we thus 
want to describe the boundary theory for a fixed component, say $\Delta_0$, and describe how in 
a local patch around it the Hodge norm and the Hodge decomposition can be determined in the bulk. 
We introduce local coordinates $z \equiv e^{2\pi it}$ and $\zeta^\kappa$, such that the boundary component is approached in the limit
\beq \label{one-parameter-limit}
  z\rightarrow 0  \qquad \text{or}\qquad t = x+iy \ \rightarrow \ x_0 + i \infty\ .  
\eeq
Suppressing the $\zeta^\kappa$ coordinate directions, the considered configuration can 
be depicted as in figure \ref{fig:boundary1}.
The cases of higher co-dimension is significantly more involved and goes beyond the 
scope of this work.

Let us begin by briefly motivating how the relevant data on the boundary is extracted. 
As for the bulk theory the motivation comes from Hodge theory, or more precisely 
the existence of a nilpotent orbit introduced in section \ref{Near-boundary}. The non-trivial 
task is then to extract the relevant data in the limit $y \rightarrow \infty$. Roughly stated, one 
performs a clever expansion of $F^p_{\rm pol}$ in $t$ around the point $y=\infty$
and analyses the information carried by the various terms near the boundary of $\cM$. 
One then finds that an intriguing algebraic structure with $\mathfrak{sl}(2,\mathbf{\bbC})$
symmetry emerges and we propose that this should be the symmetry algebra of a boundary theory. 
To understand how the nilpotent orbit defines the boundary data 
the reader may later consult section \ref{bulk_boundary} where we 
match the bulk and boundary theories. Given a nilpotent 
orbit the boundary data can always be extracted as we will describe in section \ref{sec-uniqueness}. 

In this section we describe the for us relevant structures  
of the boundary theory abstractly without describing how the data is derived from 
Hodge theory. More precisely, we will discuss the following boundary data in detail: 
\begin{align} \label{boundary_data}
    &\text{(1) Hilbert space:}&& \cH\ \text{and $\mathfrak{g}$-invariant}\ \langle v , w \rangle \ ,& \nn  \\
   &\text{(2) boundary charge decomposition:}&& Q_\infty\ \text{with}\  \langle v |w \rangle_\infty = \langle \bar v , e^{\pi i Q_\infty} w \rangle\ , &\\
    &\text{(3)}\ \mathfrak{sl}(2,\bbC)\text{-algebra in $\mathfrak{g}$:} && (L_{-1},L_0,L_{+1})\ ,&  \nn \\
    &\text{(4) phase operator in $\mathfrak{g}$:} && \hat \delta\ .  & \nn 
\end{align}
Our discussion will make very little reference to the underlying Hodge theory and 
will be formulated using operators rather than $(p,q)$-splittings.\footnote{In the mathematics literature such an approach was given by Robles \cite{robles_2016}.} It is important to 
stress that we will not attempt to construct the full boundary theory, but rather note 
that the data \eqref{boundary_data} is sufficient for the purposes of this work.
We will will sometimes refer loosely to the data \eqref{boundary_data} as being the 
`boundary theory'.

As an aside, let us not that an actual boundary theory could be obtained by using the data \eqref{boundary_data} to define a conformal quantum mechanics model.
To introduce a time dependence, we could consider the boundary to still 
depend on a coordinate $\tau =\frac{1}{2\pi i} \log x_0$ that is unaffected when taking the limit \eqref{one-parameter-limit} 
and view $\tau$ as Euclidean time. On the quotient space with coordinates $z$, we have $\tau \cong \tau +1$. 
One might want to visualize this cutting out the singularity of the disk (a.2) in figure \ref{fig:boundary1}. 
Working on the disk we can build local operators $ \cO_q^{[d]} (\tau)$ of charge $q$ and weight $d$ by  
setting 
\beq
    \cO_q^{[d]} (\tau)= \sum_{l=-d}^d e^{2 \pi i l \tau} \cO^{(d,l)}_q \ ,
\eeq
where $q,l,d$ are the eigenvalues under $Q_\infty$, $L_0$, and the $\mathfrak{sl}(2,\bbC)$-Casimir, respectively. 
This decomposition of operators will be discussed in detail in section \ref{sec:boundary_operators} below. 
As for the quantum mechanical model with $\mathfrak{sl}(2,\bbR)$-symmetry suggested in \cite{deAlfaro:1976vlx}, 
it is not immediate how to interpret this theory as a CFT$_1$. 
There is no distinguished vacuum state and the formulation of a state-operator-map is obscured. 
Interestingly, both issues have been addressed in \cite{Chamon:2011xk} for the model of \cite{deAlfaro:1976vlx}. 
The proposed solutions appear to be equally important in our setting. 
Despite these similarities we believe that the actual boundary theory should be more involved 
and we hope to address a detailed formulation in the future.

\subsection{Boundary charge operator and $\mathfrak{sl}(2,\bbC)$-algebra}

To start with we note that this boundary theory consists of a finite 
dimensional Hilbert space of states $\cH$. This Hilbert space is obtained 
by the complexification of a lattice $\cL$ as $\cH= \bbC \otimes \cL$. Note 
that as a vector space $\cH$ is isomorphic to the space introduced in section \ref{bulktheory}. 
We will also consider its real version $\cH_\bbR = \bbR \otimes \cL$.
In geometric settings, such as in Calabi-Yau compactifications, 
we identify $\cL = H^D(Y_D,\bbZ)$, $\cH=H^D(Y_D,\bbC)$, and $\cH_\bbR=H^D(Y_D,\bbR)$.\footnote{More precisely, we would actually have 
to consider the primitive parts of $H^D(Y_D)$.} 
We require that the norm of $\cH$ is induced by a
charge operator $Q_\infty$ and that it is compatible with an $\mathfrak{sl}(2,\bbC)$-algebra. We will introduce 
these objects in the following. 

\noindent
\textbf{The charge operator $\mathbf{Q_\infty}$ and inner product.}
The charge operator $Q_\infty$ can be viewed as defining a boundary Hodge decomposition
\beq \label{boundary-decomposition_sing}
  \cH  = H^{D,0}_\infty \oplus H^{D-1,1}_\infty \oplus... \oplus H^{0,D}_\infty\ ,
\eeq
where $ H^{q,p}_\infty = \overline{ H^{p,q}_\infty}$ and $p+q=D$.\footnote{This decomposition can still vary with the change in the coordinates 
$\zeta^\kappa$ not taken to a 
limit, but is, of course, independent of $t$.} The decomposition \eqref{boundary-decomposition_sing} 
can equally be encoded by specifying a charge operator $Q_\infty \in \mathfrak{g}_\bbC$ acting as 
\beq \label{Qinfty_onstates}
      Q_\infty |w\rangle = \frac{1}{2} (2p-D) |w\rangle \qquad \text{for} \quad |w\rangle \in H^{p,D-p}_\infty\ , 
\eeq 
where we are using a bra-ket notation for states. Note that one infers from the properties of \eqref{boundary-decomposition_sing}
that 
\beq  \label{chargeimaginary}
   \bar Q_\infty = - Q_\infty\ .
\eeq
We also require that the decomposition 
\eqref{boundary-decomposition_sing} allows us to define a boundary Hodge operator $C_\infty$ by setting 
\beq
     C_\infty |w\rangle = i^{2p-D} |w\rangle  \quad   \text{for}\quad  |w\rangle \in H^{p,D-p}_\infty\ ,
\eeq
which defines a norm $\| \cdot \|_\infty $. Formally one can thus relate $C_\infty = i^{2 Q_\infty} = e^{\pi i Q_\infty}$, with 
$Q_\infty$ acting as in \eqref{Qinfty_onstates}.
Concretely, we define the inner product and the norm by 
\beq \label{infinity_norm}
    \langle v | w \rangle_\infty  := \langle \bar v , C_\infty w \rangle  \ , \qquad \| v \|^2_\infty =  \langle v |  v \rangle\ .
\eeq
It is a non-trivial fact that such a charge operator $Q_\infty$ with associated 
non-degenerate norm exists on every boundary component 
of a complex structure moduli space \cite{CKS}. To come to this 
conclusion the reader can consult section \ref{bulk_boundary} and read the explanation  with the assumption 
that an underlying nilpotent orbit exists. 

\noindent
\textbf{The boundary algebra $\mathfrak{sl}(2,\bbR)$.}  We next introduce an
operator algebra representing $\mathfrak{sl}(2,\bbR)$ on $\cH$. This can be 
motivated by the fact that the Hodge metric introduced in section \ref{near-boundary-metric} always asymptotes 
to the Poincar\'e metric, which has an $\mathfrak{sl}(2,\bbR)$ isometry algebra. 
We thus define real generators  $N^0,N^\pm \in \mathfrak{g}_\bbR$ that
satisfy the  angular momentum algebra \footnote{Note that, interpreting  $\mathfrak{sl}(2,\bbR)$ as the global conformal group in one dimensions, one 
can identify $H = \frac{1}{2}N^0$ as the global Hamiltonian, $K=N^-$ as the special conformal transformations, and $P=-N^+$ as translations. 
Below in \eqref{N0-dagger} we will see that that with respect to the inner product \eqref{infinity_norm} we have $P^\dagger = -K$.}
\beq \label{sl2R-algebra}
   [ N^0,N^\pm] = \pm 2 N^\pm\ ,\qquad [N^+, N^-] = N^0\ . 
\eeq
We want to ensure that this algebra is compatible with the split \eqref{boundary-decomposition_sing} 
and hence introduce commutation relations with the charge operator $Q_\infty$. 
This operator is imaginary and hence 
we are naturally lead to formulate compatibility conditions over the 
complex numbers. We thus require 
\beq \label{QN-comp}
    \big[Q_\infty,N^{0}\big]= i  (N^+ + N^-) \ , \qquad     \big[Q_\infty,N^{\pm}\big] =-\tfrac{i}{2} N^0 \ .
\eeq
Since $C_\infty = i^{2Q_\infty}$ these compatibility conditions imply that 
\beq \label{N0-dagger}
   (N^0)^\dagger = N^0\ , \qquad  (N^{+})^\dagger = N^-
\eeq 
where $\cO^\dagger = - C^{-1}_\infty \bar \cO C_\infty$ is the adjoint operator to $\cO$ with 
respect to the inner product $ \langle v | w \rangle_\infty$ introduced in \eqref{infinity_norm}.
It is interesting to stress that there are at least two major differences to the conformal quantum mechanics model
discussed, for example, in \cite{deAlfaro:1976vlx,Chamon:2011xk}. Firstly, we are considering finite-dimensional 
representations of $\mathfrak{sl}(2,\bbR)$ which are thus not unitary. Secondly, the central importance 
of the operator $Q_\infty$ naturally leads us to consider a complex operator algebra. 
In fact, we will see next that the boundary symmetry is better described by the algebra $\mathfrak{sl}(2,\bbC)$ rather
than $\mathfrak{sl}(2,\bbR)$.

\noindent
\textbf{The boundary algebra $\mathfrak{sl}(2,\bbC)$.} 
Due to the fact that the split \eqref{boundary-decomposition_sing} is over the complex numbers, 
we are lead to introduce the generators $L_\alpha \in \mathfrak{g}_\bbC$ representing an $\mathfrak{sl}(2,\bbC)$ action on $\cH$. These are 
defined in terms of the $N^0,N^\pm$ as
\beq \label{def-Ls}
    L_{\pm 1} := \tfrac{1}{2} \big(N^+ + N^- \mp i N^0  \big) \ , \quad L_0 :=  i\big(N^- - N^+) \ .
\eeq
Using \eqref{sl2R-algebra} we see that the $L_\alpha$ satisfy the commutation relations 
\beq \label{commutationrelationsL}
     [L_0,L_{\pm 1}] = \pm 2\,L_{\pm 1}\ , \qquad [L_{1},L_{-1}]=L_0\ .
 \eeq
Furthermore, we have that 
 \beq \label{barL}
    \bar L_0= - L_0\ , \qquad \bar L_1 = L_{-1}\ .
\eeq
In addition, we note from \eqref{QN-comp} that the compatibility with the split \eqref{boundary-decomposition_sing} amounts 
to the commutation relations with $Q_\infty$ given by  
 %\footnote{To define $Q_\infty$,
%we decompose the algebra $\mathfrak{g}_\bbC$ introduced after \eqref{def-S} by setting 
%\beq \label{gC-decomposition_infty}
%   \mathfrak{g}_{\bbC} =  \bigoplus_{p \in \bbZ} \mathfrak{g}^{p }_\infty \ , \qquad  \mathfrak{g}^{p }_\infty = \big\{T \in \mathfrak{g}_\bbC : \ [Q_\infty,T]= %p\, T \big\}\ ,
%\eeq
%which corresponds to a spit with $T\in \mathfrak{g}^{p }_\infty $ if $T H^{a,b}_\infty = H^{a+p,b-p}_\infty$ as in \eqref{gC-decomposition-H_0}. Note the one has $\overline{\mathfrak{  g}^{p }_\infty}  =  \mathfrak{g}^{-p }_\infty$ and  $[\mathfrak{g}^{p }_\infty , \mathfrak{g}^{p }_\infty ]  \subset \mathfrak{g}^{p+q}_\infty$ as above. 
%}
\beq \label{L-restrictions}
     [Q_\infty, L_\alpha ]  = \alpha \, L_\alpha \ ,
\eeq
which clarifies the meaning of the subscript on $L_\alpha$.\footnote{In mathematics a triple with $L_0$ having even charge and 
$L_{\pm 1}$ having odd charge while satisfying \eqref{commutationrelationsL}
and \eqref{barL} is also known as \DH okovi\'c-Konstant-Sekiguchi (DKS) triple and plays an important role in the DKS correspondence.}
Note that \eqref{barL} together with \eqref{L-restrictions} implies that \footnote{Recall 
that $\mathfrak{g}_\bbC$ is the algebra of elements preserving $\langle\cdot , \cdot \rangle$ given in \eqref{def-S}, e.g.~one has $\langle L \alpha,\beta\rangle = - \langle \alpha,L \beta \rangle$.} 
\beq \label{Lboundary_dagger}
     L_{\alpha}^\dagger = L_{-\alpha}\ ,
\eeq
where $L^\dagger_\alpha$ is the adjoint operator with 
respect to the inner product $ \langle v | w \rangle_\infty$.

The conditions on $L_\alpha$ ensure the compatibility of the $\mathfrak{sl}(2,\bbC)$ representation with the 
boundary decomposition \eqref{boundary-decomposition_sing}. The statement can rephrased 
by noting the $\mathfrak{sl}(2,\bbC)$ itself has a natural charge split with 
eigenspaces $\mathfrak{sl}(2,\bbC)^{0},\ \mathfrak{sl}(2,\bbC)^{1}\ ,\ \mathfrak{sl}(2,\bbC)^{-1}$ 
of charge $-1,0,1$, respectively. These charge eigenspaces are 
the one-dimensional complex spaces spanned by $\mathbf{l}_0 = i\big( \mathbf{n}^- -  \mathbf{n}^+)$, and 
$\mathbf{l}_{\pm 1} =  \tfrac{1}{2} \big(\mathbf{n}^+ +  \mathbf{n}^- \mp i  \mathbf{n}^z\big) $, respectively. Here one is using the 
standard $\mathfrak{sl}(2,\bbR)$ generators
\beq
   \mathbf{n}^- = \left( \begin{array}{cc} 0 & 1 \\ 
   0 & 0 \end{array} \right)\ ,\qquad    \mathbf{n}^+ = \left( \begin{array}{cc} 0 & 0 \\ 
   1 & 0 \end{array} \right)\ , \qquad    \mathbf{n}^z = \left( \begin{array}{cc} -1 & 0 \\ 
   0 & 1 \end{array} \right)\ .
\eeq 
The equation \eqref{L-restrictions} then 
corresponds to the requirement that this split of $\mathfrak{sl}(2,\bbC)$ is represented in the  
split of $\mathfrak{g}$ induced by $Q_\infty$. In fact, it requires that there exists a map $\varphi: Sl(2,\bbC) \rightarrow G_{\bbC}$ such that 
$ \varphi_*  \mathbf{l}_\alpha = L_\alpha$.
%\beq
 %   \varphi_*  \mathbf{n}^\pm = N^\pm\ , \qquad  \varphi_*  \mathbf{n}^z = N^0\ . 
%\eeq
In mathematical terms such a representation is known as a horizontal $Sl(2)$ with respect to
the splitting \eqref{boundary-decomposition_sing}.

\subsection{Boundary eigenstates} \label{boundaryeigenstates}
 
 With the existence of the $\mathfrak{sl}(2,\bbC)$ algebra on the boundary, 
we can define another canonical splitting of $\cH$. 
This splitting can be performed over the 
real numbers or complex numbers, i.e.~by considering $\cH_\bbR$ or $\cH$. 

\noindent
\textbf{The real boundary splitting.} 
For the real space $H^{D}(Y_D, \bbR)$ we introduce an eigenbasis labelled by two quantum numbers $(l,d)$ for the $\mathfrak{su}(2)$-algebra \eqref{sl2R-algebra}.
The label $l \in \{ -   d , ...,   d \}$ corresponds to the eigenvalue of $N^0$, while $d \in \{0,...,D\}$ is related to the eigenvalues $d(d+2)$ of 
the Casimir operator 
\beq \label{Casimir}
    N^2 = 2 N^+ N^-+2 N^- N^+ +(N^0)^2\ .
\eeq
Denoting the eigenstates by $ |d,l \rangle $ we thus have  
\begin{align} \label{bra-basis}
   && N^2 |d,l \rangle  &\ =\ d(d+2) |d,l \rangle\ ,&&   &&\hspace*{-1.7cm} d=0,...,D\ , && \nn \\
   &&N^0 |d,l  \rangle  &\ =\ l |d,l \rangle \ ,&&    &&\hspace*{-1.7cm} l = -d,...,d\ ,&&
\end{align}
and note that 
there can be many states with the same labels, but we will not distinguish them with an extra index.
The set of highest weight states is then given by 
\beq
  \text{highest weight states $ | d,d \rangle$:}\quad  N^+ | d,d \rangle= 0 \ , 
\eeq
with all other states being generated by acting with $N^-$. 
We thus have the split 
\beq \label{V-decomposition}
\cH_\bbR = \bigoplus_d \bigoplus_{l \in \{ -   d , ...,   d \} }  V_l^d \equiv \bigoplus_{\ell  \in \cE} V_{\ell}\ ,
\eeq
where $ V_l^d$ are the real vector spaces spanned by the eigenstates $| d,l\rangle$ introduced in \eqref{bra-basis}. We have also introduced the 
set $\cE$ of all possible values $\ell=(d,l)$ in order to simplify the expressions.

The splitting is orthogonal 
in the boundary inner product introduced in \eqref{infinity_norm}. Consider $| d,l\rangle \in V_l^d$ and $| d',l'\rangle \in V_{l'}^{d'}$. Then one has the 
identities 
\bea
    \langle d,l | d',l' \rangle &=& 0 \ \qquad \text{for} \quad d\neq d', l\neq l'\ , \label{orthogonality} \\
     \langle d,l | d,l \rangle &  >& 0 \ . \label{positivity}
\eea
Note that in general one has several states $| d,l \rangle$ with the same labels. 
The positivity expression \eqref{positivity} then should be read as the statement that the matrix 
formed from these states is positive definite. This is compatible with the statement 
that $\|\cdot \|_\infty$ introduced in \eqref{infinity_norm} is actually a norm.  The properties \eqref{orthogonality} and 
\eqref{positivity} are non-trivial and important in many applications. A consequence 
of \eqref{infinity_norm} is that the operators $N^2,N^0$ are self-adjoint with respect to the inner product \eqref{infinity_norm}. 

\noindent
\textbf{The complex boundary splitting.} 
Given the decomposition \eqref{V-decomposition} one can also determine a decomposition of 
the complex vector space $H^{D}(Y_D, \bbC)$ into $L^2$, $L_0$-eigenspaces, where $L^2= 2 L_{1} L_{-1}+2 L_{-1} L_{1} +(L_0)^2$
is the Casimir operator.
Let us denote the eigenvectors by $|d,l \rangle\!\rangle$, i.e.~we require 
\beq
   L_0 \, |d,l \rangle\!\rangle = l \, |d,l \rangle\!\rangle\ , \qquad L^2 \, |d,l \rangle\!\rangle = d\, |d,l \rangle\!\rangle\ . 
\eeq
The important observation that we want to use in the following is 
the fact that there exists a transformation $\rho$, defined as 
\beq \label{def-rho}
   \rho = \text{exp} \tfrac{i\pi}{4}(N^+ + N^-)  = \text{exp} \tfrac{i\pi}{4}(L_{1} + L_{-1}) \ ,
\eeq
which relates the real and complex versions of $\mathfrak{sl}(2)$ as
\beq \label{L-N-relation}
 L_0 = \rho\, N^0 \, \rho^{-1}  \ , \qquad L_{\pm 1} =  \rho \, N^{\pm}\, \rho^{-1} \ . 
\eeq
This can be checked by using \eqref{sl2R-algebra}, \eqref{def-Ls}, and the fact that the adjoint representations 
satisfy $\text{Ad}_{e^X} = e^{\text{ad}_X}$. 
Therefore, we can relate the eigenstates of $N^2$, $N^0$ introduced in \eqref{bra-basis} to the eigenstates 
 $|d,l \rangle\!\rangle$ via
 \beq
      |d,l \rangle\!\rangle = \rho  |d,l \rangle \ , 
 \eeq
 where we recall that there can be several basis elements with the same $(d,l)$, but we have suppressed the
 further index labelling them.
 Hence we find that the complex cohomology admits a decomposition 
\beq \label{V-decomposition-c}
\cH = \bigoplus_d \bigoplus_{l \in \{ -   d , ...,   d \} }  \cV_l^d \ . 
\eeq
where $ \cV_l^d$ is the complex vector space spanned by the $ |d,l \rangle\!\rangle $.

\noindent
\textbf{Weights and charges of states.} Having decomposed the states 
according as eigenstates of $L_0$, $L^2$, we also can add a label indicating 
the charge under $Q_\infty$. Noting that  \eqref{L-restrictions} implies $[Q_\infty,L_0]=[Q_\infty,L^2]=0$ it 
is possible to find simultaneous eigenstates of $L_0,L^2$ and $Q_\infty$ that 
form an \textit{orthogonal} basis. We 
denote these by $|d,l ; q \rangle\!\rangle$ with 
\begin{align} \label{states_threelables}
   && L^2 |d,l ; q \rangle\!\rangle  &\ =\ d(d+2) |d,l ; q \rangle\!\rangle\ ,&&   &&\hspace*{-1.7cm} d=0,...,D\ , && \nn \\
   &&L_0 |d,l ; q \rangle\!\rangle  &\ =\ l |d,l ; q \rangle\!\rangle \ ,&&    &&\hspace*{-1.7cm} l = -d,...,d\ ,&&\\ 
   && Q_\infty |d,l ; q \rangle\!\rangle  &\ = \ \frac{1}{2} (2q-D)  |d,l ; q \rangle\!\rangle \ , &&   &&\hspace*{-1.7cm} q = 0,...,D\ ,&&   \nn
\end{align}
with $d,\ l,\ q$ being integer valued. 
In accordance with the CFT language, we call $l$ the \textit{weight} of a state, while $q$ is the \textit{charge} of the state.

\subsection{Boundary operators and the phase operator} \label{sec:boundary_operators}
 
In the following we will discuss the operators of the boundary theory acting 
on the finite-dimensional Hilbert space $\cH$. Analog to the 
 state decomposition of subsection \ref{boundaryeigenstates} we will then 
 introduce a decomposition of operators. 
 In addition, we will  introduce a special operator 
$\hat \delta$, the phase operator, completing the data that has to be specified for the boundary theory. 
 
\noindent
\textbf{Boundary splitting of operators.} 
The operators $L^2,L_0$, and $Q_\infty$ introduced in the previous subsections can also be used 
to split the space of operators on $\cH$. Concretely, any operator $\cO$ acting on $\cH$ admits 
an expansion \footnote{Note that this splitting corresponds to the so-called Deligne splitting of $\mathfrak{g}_\bbC$
and one uses the notation $\cO_{r,s}$, more concretely one has $\sum_d \cO_{p}^{(d,p+q)}=\cO_{p,q}$.}
\beq \label{weight-charge-expansion-operator1}
     \cO = \sum_{d,l,q \in \bbZ} \cO^{(d,l)}_q \ , 
\eeq
with 
\begin{align} \label{weight-charge-expansion-operator2}
  && (\text{ad} L)^2 \cO^{(d,l)}_q &\ =\ d(d+2)\, \cO^{(d,l)}_q\ , &&  &&\hspace*{-2cm} d=0,...,D & \nn \\
  &&\big[L_0,\cO^{(d,l)}_q\big] &\ =\  l\, \cO^{(d,l)}_q\ , &&  && \hspace*{-2cm} l = -d,...,d &\\
   &&\big[Q_\infty ,\cO^{(d,l)}_q \big] &\ =\ q\, \cO^{(d,l)}_q   \ , &&  && \hspace*{-2cm} q = -D,...,D\ . &\nn  
\end{align}
where we have used the shorthand notation (ad$L)^2$ to denote
\beq
   (\text{ad} L)^2 \cO :=2\big[L_{1},\big[L_{-1},\cO \big]\big]+2\big[L_{-1},\big[L_{1},\cO \big]\big]+\big[L_0,\big[L_0,\cO \big]\big]\ .
\eeq
We will call $l$ the \textit{weight} of the operator $\cO^{(d,l)}_q$ that descents from a \textit{highest weight} $d$, while $q$ denotes its \textit{charge}. 
Accordingly, from \eqref{L-restrictions}, \eqref{commutationrelationsL} the operators $(L_{-1},L_0,L_{1})$ have charges $(-1,0,1)$ and weights $(-2,0,2)$ and highest weights $(2,0,2)$, respectively.  
Note that the adjoint operator $\cO^\dagger = - C^{-1}_\infty \bar \cO C_\infty$ with respect to the inner product \eqref{infinity_norm} admits also 
a decomposition into $L^2,L_0$, and $Q_\infty$ components. Due to the fact that $C_\infty = i^{2Q_\infty}$
we find that 
\beq \label{adjoint-decomp}
     \big(\cO^\dagger\big)^{(d,l)}_q  =  \big(\cO_{-q}^{(d,-l)} \big)^\dagger\ .
\eeq

It will be convenient to sometimes not perform the $L^2$-decomposition. We then suppress the 
index $d$ in \eqref{weight-charge-expansion-operator1} and write 
\beq \label{weightonly}
  \cO^{(l)}_q \equiv  \sum_{d \in \bbZ} \cO^{(d,l)}_q\ .
\eeq
The space of operators with charge less than $q$ and weight less than $p$ will be denoted by \footnote{Note that 
in \cite{CKS} these spaces were denoted by $L^{p,q} \equiv \Lambda^{(p+q)}_p$. } 
\beq \label{space_chargeless}
     \Lambda^{(p)}_q  = \text{span} \big\{ \cO_n^{(s)} \ ,  n\leq q\ , s \leq p ,\ r=0,...,D\big\}\ .
\eeq
Note that these spaces are `filtered' with $  \Lambda^{(p)}_q \subset \Lambda^{(\tilde p)}_{\tilde q}$ for $q\leq \tilde q$ and $p \leq \tilde p$
and that $[L_{-1}, \Lambda^{(p)}_q] \subset \Lambda^{(p-2)}_{q-1}$ as can be checked using \eqref{commutationrelationsL} and \eqref{L-restrictions}. 

\noindent
\textbf{The phase operator.} The remaining information about the boundary theory is encoded in a phase operator $\hat \delta$. 
This operator satisfies the commutation relation 
\beq \label{phase-commutes}
    [\hat \delta,L_{-1}] = 0 \ .
\eeq
It is crucial, however, that in general, $[\hat \delta,L_{1}] \neq 0$, $[\hat \delta,L_{0}] \neq 0$ and $[\hat \delta,Q_\infty] \neq 0$. 
To specify the failure of commuting with $L_0,Q_\infty$ we perform an expansion as in \eqref{weight-charge-expansion-operator1}
with \eqref{weightonly} by writing 
\beq
   \hat \delta = \sum_{l,q \in \bbZ} \hat \delta_{q}^{(l)}\ .
\eeq
$\hat \delta$ does not commute with $L_0,Q_\infty$ if it has components other then $\hat \delta_{0}^{(0)}$. 
To constrain $\hat \delta$ further we impose conditions on its components. Firstly, we require that it stems 
from real operator $\delta \in \mathfrak{g}_\bbR$ by the transformation 
\beq
     \hat \delta = \rho \, \delta\, \rho^{-1}\ , \qquad \overline{\hat \delta} = \rho^{-2} \hat \delta \rho^2\ .
\eeq
with $\rho$ given in \eqref{def-rho}. Since $\rho$ is complex, $\hat \delta$ is not real. 
Furthermore, one has to 
ensure that $\hat \delta$ is only build out of $\hat \delta_{q}^{(p)}$ with charge less than $-1$ and weight 
less than $-2$. This amounts to the statement that  $ \hat \delta = \sum_{q \leq -1,\, p\leq -2}   \delta^{(p)}_q$ or, by 
using \eqref{space_chargeless} that 
\beq \label{hatdelta_inL}
     \hat \delta\ \in\ \Lambda^{(-2)}_{-1} \ .
\eeq
Note that we will see later in section \ref{bulk_boundary} that $\hat \delta$ is related to the $\delta$ introduced in \eqref{ref-F} via $\hat \delta = \rho^{-1} \delta \rho$, when matching bulk and boundary data. In fact, the properties \eqref{phase-commutes}, \eqref{hatdelta_inL} of $\hat \delta$ 
are motivated from the properties of charge operators associated to nilpotent orbits as described in appendix \ref{computedelta}.

Let us indicate the importance of the phase operator $\hat \delta$. It encodes a deformation of the 
inner product which is compatible with all the structures and is required to match a general bulk solution. 
In fact, one could also define another norm on the boundary that depends on $\delta$ by 
replacing $C_\infty$ in \eqref{infinity_norm}. In this case, the operation 
of taking the adjoint $ \cO^\dagger$ is no longer compatible with 
the eigendecomposition as in \eqref{adjoint-decomp}. Interestingly, this is reminiscent of the discussion 
of phases in the principle series representations of $Sl(2,\bbR)$ given in \cite{Anninos:2019oka}.

\subsection{Classification of boundary theories} \label{classification}

It is interesting to point out that the data that we have just specified
can be used to classify allowed boundary theories. Let us consider a geometric setting with $\cH= H^D(Y_D,\bbC)$. A first non-trivial fact that one has to use in such 
a classification is the identity 
\beq \label{H-identical}
   \text{dim}\, H^{p,q}  = \text{dim}\, H^{p,q}_\infty\ . 
\eeq
This implies, if we restrict ourselves to Calabi-Yau manifolds, that the space $H^{D,0}_\infty$ is 
one-dimensional and the dimension of $H^{D-1,1}_\infty$ counts the total number of 
deformations spanning the moduli space $\cM$. The basic idea is to combine the information 
captured by the split \eqref{V-decomposition-c} into $\mathfrak{sl}(2,\bbC)$-eigenspaces 
with the $(p,q)$-decomposition \eqref{boundary-decomposition_sing} on the boundary. 
This can be done by using that $N^- F^i_{\infty} \subset F^{i-1}_\infty$ for $F^i_\infty$ 
defined analog to \eqref{Fp-filtration}. 

Let us exemplify this first for Calabi-Yau threefolds $Y_3$ \cite{Kerr2017}, which has 
been used in physical applications in \cite{GPV,Grimm:2018cpv}. In the threefold 
case the highest power of $N^-$ acting on $H^{3,0}_\infty$ can be three, such that we can introduce 
four principle cases 
\beq
   (N^-)^{\hat d}  H^{3,0}_\infty \neq 0 ,\quad   (N^-)^{\hat d+1}  H^{3,0}_\infty = 0 \  \quad \left\{ \begin{array}{cccc} \hat d=0 & \hat d=1 & \hat d=2 & \hat d=3 \\
   \text{I} & \text{II} & \text{III} & \text{IV}\\  \end{array}\right. \ .
\eeq 
This information can be refined further by counting the number $n_{2,1}$ of elements in $H^{2,1}_\infty$ that are \textit{not} annihilated by $N^-$. 
One records this information by a subindex leading to in total $4h^{2,1}$ types \cite{Kerr2017}
\beq \label{sing_types}
   \text{I}_{n_{2,1}}\ , \quad \text{II}_{n_{2,1}-2}\ ,  \quad \text{III}_{n_{2,1}-4} \ ,  \quad \text{IV}_{n_{2,1}-2}\ . 
\eeq
Using the decomposition \eqref{V-decomposition-c} one can then infer that in the various cases the possible 
minimal and maximal value of $n_{2,1}$ is restricted (see \cite{Kerr2017,GPV,Grimm:2018cpv} for details). 
Remarkably, this classification can be mapped, using mirror symmetry, to a classification of 
limits in the K\"ahler moduli space \cite{Grimm:2018cpv,Corvilain:2018lgw}. This lead recently \cite{Grimm:2019bey} to the suggestion to classify Calabi-Yau threefolds 
into graphs formed by the types \eqref{sing_types}. 

For Calabi-Yau fourfolds the classification of boundary theories proceeds in an analogous way. 
In this case, however, we are dealing with five principle cases, labelled by I, II, III, IV, V in \cite{Grimm:2019ixq}, 
since the highest possible power $\hat d$ of $N^-$ with a non-trivial action on $H^{4,0}_\infty$ is four. 
Furthermore, there are now two sub-indices to each principle case that indicate how many of the $(3,1)$- and 
$(2,2)$-forms degenerate near the boundary. This data was used in \cite{Grimm:2019ixq} to study asymptotic 
flux compactifications. Some important subtleties in the classification of such theories for fourfolds 
have been pointed out in \cite{Kerr2017}. 

It is crucial to stress that the classification does not capture the information in $\hat \delta$. While it would be very interesting  
to include this information, this has not been done so far. In fact, it is tempting to conjecture that 
the boundary theory is, in fact, a theory that dynamically determines the values of 
$\hat \delta$ and enforces the properties described in section \ref{sec:boundary_operators}.

\section{Bulk-boundary correspondence} \label{bulk_boundary}

In this section we have a detailed look at the matching of the 
boundary data introduced in section \ref{boundary_theory} with 
solutions to the bulk theory discussed in section \ref{bulktheory}. 
The aim is to determine general solutions  to the bulk equations and 
describe how the matching with the boundary data gives a  
restriction to physically viable nilpotent orbits $F^p_{\rm pol}$.
Mathematically this matching is known as the 
correspondence between nilpotent orbits and limiting mixed Hodge structures \cite{Schmid,CKS}. 
The following discussion essentially summarizes some of the main steps in the 
proof of the Sl(2)-obrit theorem and thus follow to large extend the seminal papers of Schmid \cite{Schmid} 
and Cattani, Kaplan, Schmid \cite{CKS}. However, we 
will adapt a more physical language and combine some of the steps in a somewhat 
different manner. In particular, we aim to make the eigenspace decompositions of
operators \eqref{weight-charge-expansion-operator2} manifest in the complete analysis. 
From a physics perspective the following approach of determining asymptotic solutions 
is a crucial part of the AdS/CFT duality and relevant, for example, for holographic 
renormalization \cite{Skenderis:2002wp} and bulk reconstruction \cite{DeJonckheere:2017qkk,Harlow:2018fse}.

Let us recall 
here, for the convenience of the reader, that the bulk theory of section \ref{bulktheory} is defined 
by a Hilbert space $\cH$ with a reference inner product $\langle v|w \rangle_{\rm ref} = \langle v , i^{2Q} w\rangle$, which 
is determined in terms of a reference charge operator $Q$. The bulk matter fields are given by a 
matrix valued function $\hat h \in \mathfrak{g}_\bbR$ varying over the field space. In 
a one-parameter asymptotic limit $t = x+iy$, with $y \rightarrow \infty$ being the boundary, one generally has 
a non-trivial $y$-dependence, while the $x$-dependence is fixed by 
symmetry requirement \eqref{h-rotation} to be $\hat h(t,\bar t) = e^{x N^-} h(y)$. In fact, a crucial information in the boundary data 
is the nilpotent matrix $N^-$, which encodes the transformation \eqref{h-rotation} of $\hat h$ under the shift $x \rightarrow x+1$.
The $y$ dependence in a bulk solution $\hat h(t,\bar t)$ are constrained by the equations of motion of the action \eqref{action-with-gen_1d_flatgauge}.  
We have shown that bulk solutions $h(y)$ can be used to define three
operators $\cN^0(y),\cN^{\pm}(y) \in \mathfrak{g}_\bbR$ as in \eqref{def-cN-hath}. These satisfy the 
bulk differential equations \eqref{h-Nahm}, i.e.~we have
\beq \label{Nahm1_later}
 (\text{C}1):\qquad \partial_{y} \cN^{\pm} = \pm \tfrac{1}{2} [\cN^{\pm},\cN^0] \ , \qquad  
  \partial_{y} \cN^0 =- [\cN^{+} ,\cN^{-}]\ .
\eeq
To extract a physical $h(y)$ from these operators, we also needed to require that the reference 
charge operator $Q$ acts on these solutions of \eqref{Nahm1_later} as
\beq \label{QcN-matching}
   (\text{C}2): \qquad \big[Q,\cN^{0}\big]= i  (\cN^+ + \cN^-) \ , \qquad     \big[Q,\cN^{\pm}\big] =-\tfrac{i}{2} \cN^0 \ , 
\eeq
as already given in \eqref{Q-restr_eom}. Note that these conditions imply
$ (\cN^+)^{\dagger}= \cN^- $ and $(\cN^0)^\dagger = \cN^0$, 
with respect to the inner product $\langle v|w \rangle_{\rm ref}$ on $\cH$ that is induced by 
 $Q$. 
 %This is compatible with the fact that the $\cN^0(y),\cN^{\pm}(y)$ are defined in terms 
 %of $h(y)$ as in \eqref{cN-def-h}. 
 
 In this section we will determine  the general form of a solution to the bulk theory near $y=\infty$ and 
 show how it is uniquely fixed by the boundary data. In particular, 
we will see that a general bulk solution takes the form 
\bea \label{simple-h-expansion}
    h(y) &=&  g(\infty) \Big( 1 + \frac{g_1}{y} + \frac{g_2}{y^2} + \ldots \Big)   y^{-\frac{1}{2}\tilde N^0} \ , \\
    h(y)^{-1} &=&  y^{\frac{1}{2}\tilde N^0}  \Big( 1 + \frac{f_1}{y} + \frac{f_2}{y^2} + \ldots \Big)  g(\infty)^{-1}   \ , \nn 
\eea
where $ \tilde N^0 \in \mathfrak{g}_{\bbR}$, $g(\infty) \in G_{\bbR}$, and $g_k,\, f_k$ are operators on $\cH$.
The matrices $g_k,\, f_k,\, g(\infty)$ and $\tilde N^0$ are functions of the boundary data and can 
be determined, at least in principle, explicitly. The key requirement 
to fix the bulk data $g_k$, $g(\infty)$ and boundary data is given by  
\beq \label{fixall}
   \boxed{ \rule[-.25cm]{.0cm}{.8cm} \quad e^{i \delta} = g(\infty) \Big( 1+  \sum_{k>0} \frac{1}{k!} (-i)^k (\text{ad} N^-)^k g_k \Big)\ ,  \quad }
\eeq
where we recall that $\delta$ and $N^-$ are part of the boundary information. 
The matrix $\tilde N^0$ is fixed by the boundary data as 
\beq \label{tildeN0fixing}
  \tilde N^0 = g(\infty)^{-1} N^0 g(\infty)  \ . 
\eeq
It is a non-trivial 
fact that \eqref{fixall} together with the field equations \eqref{Nahm1_later}, \eqref{QcN-matching} for $h(y)$ fixes the solution uniquely. 
This section is intended to explain this in detail.  
 
Before turning to this detailed discussion, let us note that even without giving an 
explicit expression for the $g_k,\, f_k$ and $g(\infty)$ in terms of the boundary data, we can show that these coefficients always admit 
several special properties that allow us to control the general behaviour of a bulk solution matching to the 
boundary data. In particular, one finds that 
\beq \label{gn-property}
   \text{ad}_{N^-}^{k+1}  g_k =\text{ad}_{N^-}^{k+1}  f_k = 0 \ ,\qquad \hat g_k,\hat f_k \in \bigoplus_{n \leq k-1,\, q} \Lambda^{(n)}_q\ ,
\eeq 
where we have set $\hat g_k = \rho g(\infty)\, g_k\, g(\infty)^{-1} \rho^{-1}$, with $\rho$ introduced in \eqref{def-rho}, and recall the definition \eqref{space_chargeless} of $\Lambda^{(p)}_q$. We will also argue that when picking $g(\infty)$ to match the boundary data via \eqref{fixall} it can be written as 
\beq \label{ginfty-properties}
    g(\infty) = e^{\zeta} \ , \qquad [N^-, \zeta ] = 0\ , \quad \hat \zeta \in \Lambda^{(-2)}_{-1}\ ,
\eeq
where $\hat \zeta = \rho \zeta \rho^{-1}$. The properties of $\zeta$ resemble those of the phase operator $\delta$ given in \eqref{phase-commutes} 
and \eqref{hatdelta_inL}. We will see that $\hat \zeta$ can indeed be expressed as a universal non-commutative polynomial in the components $\hat \delta^{(p)}_q$.
Taken together \eqref{simple-h-expansion}, \eqref{gn-property}, and \eqref{ginfty-properties} comprise the statements of the celebrated Sl(2) orbit theorem \cite{Schmid,CKS}. 
 
%We can now define complex 
% combinations 
%\beq \label{def-ccLs_later}
%    \tilde \cL_{\pm 1}(y) := \tfrac{1}{2} \big(\cN^0 \pm i \cN^+ \pm i \cN^- \big) \ , \quad \tilde \cL_0(y) := i\big(\cN^- -\cN^+) \ ,
%\eeq
% These satisfy a condition 
% \beq \label{cLdagger=L}
%   [Q,\tilde \cL_\alpha]= \alpha \tilde \cL_\alpha\ , \qquad  \tilde\cL^{\dagger}_{\alpha} =\tilde \cL_{-\alpha} \ ,
%\eeq
%with respect to the inner product $\langle v|w \rangle_0$ on $\cH$.

%As discussed in \eqref{def-cL+-1}, one can use the $(\cN^0,\cN^\pm)$ to define complex combinations $\cL_\alpha$ that 
%obey $[Q_\infty, \cL_\alpha]= \alpha \cL_\alpha$ \textit{point-wise for every} $y$. 
%In other words complex fields then satisfy the conditions analog to \eqref{L-restrictions} and \eqref{Lboundary_dagger}, but 
%now holding  
%While this might appear natural 
%at first, we will see later that the crucial information in the solutions $\cN^0(y),\cN^{\pm}(y)$ is the change of the 
%original $Q$-splitting with $y$ and is better captured by complex operators $\cL^0(y),\cL^\pm(y)$ introduced below in 
%\eqref{def-cLs-rho}. 

\subsection{Bulk theory solutions as series expansions } \label{series_expansions-bL}

The Nahm equations \eqref{Nahm1_later} have been studied intensively in the literature. When considering $\cN^0\ ,\cN^\pm$ 
to vary holomorphically in a complex parameter with a simple pole, 
a solution to \eqref{Nahm1_later} has residues that form a representation of $ \mathfrak{su}(2)$.
Therefore, solutions to \eqref{Nahm1_later}  naturally lead to triples, such as $N^0$, $N^\pm$ introduced in section \ref{boundary_theory}.
In the situation at hand, we are interested in solutions in a real parameter $y$ near $y = \infty$. 
The general ansatz for a solution then takes the form 
\beq \label{boundary1}
   \cN^0  = \frac{\tilde N^{0}}{y} + \cO\big(y^{-\frac{3}{2}} \big)\ ,\qquad \cN^\pm  = \frac{\tilde N^{\pm}}{y} + \cO\big(y^{-\frac{3}{2}} \big)\ . %\qquad \cL_{\pm 1}  = \frac{L_{\pm1}}{y} + \cO\big(y^{-\frac{3}{2}} \big)\ ,
\eeq
Stated differently, we introduce $\tilde N^0$, $\tilde N^{\pm}$ satisfying a $\mathfrak{sl}(2,\bbR)$ 
algebra 
\beq \label{sl2R-tilde}
 [ \tilde N^0, \tilde N^\pm] = \pm 2 \tilde N^\pm\ ,\qquad [\tilde N^+, \tilde N^-] = \tilde N^0\ ,
\eeq
to determine the coefficients of the slowest decreasing term of $\cN^{0},\cN^{\pm}$.
Note that the leading $1/y$ terms alone also solve \eqref{Nahm1_later}, however, we will need to consider terms sub-leading in the limit 
$y\rightarrow \infty$ to construct a general bulk solution. 

The leading coefficients $\tilde N^{0},\tilde N^{\pm}$
will be matched with the boundary $\mathfrak{sl}(2,\bbR)$ given by $N^0,N^\pm$. 
%Since $N^-$ is related to 
%the symmetry transformation of $\hat h$, it is natural to set $N^- = \tilde N^-$. 
Note that in general the symmetries do not need to directly 
and we parameterize this freedom with an element $\zeta \in \mathfrak{g}_{\bbR}$, i.e.~we consider the identification
%elements $\tilde N^+$, $\tilde N^0$, do not need to directly match with the boundary $\mathfrak{sl}(2,\bbR)$, 
%but rather be rotated by an element $\zeta \in \mathfrak{g}_{\bbR}$, $[\zeta,N^-]=0$ such that
\beq \label{N-rot}
    N^0 = e^{\zeta} \tilde N^0 e^{-\zeta} \ ,\qquad     N^+ = e^{\zeta} \tilde N^+ e^{-\zeta}\ , \qquad N^- =e^{\zeta} \tilde N^- e^{-\zeta}\ .   
\eeq 
Clearly, such rotations preserve the algebra \eqref{sl2R-tilde}. Picking different $\zeta$ corresponds 
to picking different reference inner products $\langle v|w \rangle_{\rm ref}$. To determine 
a bulk solution matching given boundary data, we choose to use the reference inner product
\beq \label{Q-rot}
    Q \equiv e^{-\zeta} Q_\infty e^{\zeta}\ , \qquad \langle v|w \rangle_{\rm ref}\equiv \langle e^{-\zeta} v| e^{-\zeta} w \rangle_{\infty} \ , 
\eeq 
with $Q_\infty$ and $\langle v|w \rangle_\infty$ defined on the boundary in \eqref{Qinfty_onstates} and \eqref{infinity_norm}, respectively.
Equivalently, recalling the definition of $Q$ and $Q_\infty$, we can write
\beq \label{Fref-Finf}
     F^{p}_{\rm ref} = e^{-\zeta} F_{\infty}^p\ .
\eeq
In the following we will describe how the full solution $\cN^{0},\cN^{\pm}$ is fixed by the data of the boundary theory. In particular, 
we will see that translated back into the bulk fields $h(y)$, the solutions have the form \eqref{simple-h-expansion}. 
The element $\zeta \in \mathfrak{g}_{\bbR}$ appearing in \eqref{N-rot} and \eqref{Q-rot} sets the overall transformation $g(\infty)$
and we will set $g(\infty) = e^{\zeta}$ later on. 
Indeed this matches the fact that $h(y)$ depends on the choice of reference basis $|w \rangle_{\rm ref}$ 
as seen in \eqref{def-h(y)}.

To perform the analysis for this section it will turn out to be convenient to work with complex operators, analog to the relation between 
$N^0,N^\pm$ and $L_\alpha$ given in \eqref{L-N-relation}, and also rotate by the real element $\zeta$ as in \eqref{N-rot}, \eqref{Q-rot} 
to get a direct match with the boundary data. We thus define 
\beq \label{def-cLs-rho}
   \mathbf{L}^{\pm}(y) := \rho\, e^{\zeta} \cN^{\pm}(y)e^{-\zeta}\, \rho^{-1}\ , \qquad \mathbf{L}^{0}(y) := \rho\, e^{\zeta} \cN^{0}(y)e^{-\zeta}\, \rho^{-1}\ ,
   % \cL_{\pm 1} = \tfrac{1}{2} \big(\cN^0 \pm i (\cN^+ + \cN^-) \big) \ , \quad \cL_0 = i\big(\cN^- - \cN^+) \ ,
\eeq
where 
\beq \label{rho-repeat}
    \rho = \text{exp} \tfrac{ \pi i}{4}(N^+ + N^-)= \text{exp} \tfrac{ \pi i}{4}(L_{1} + L_{-1})
\eeq 
as already defined in \eqref{def-rho}. With these redefinitions the differential equations \eqref{Nahm1_later} are trivially rewritten as 
\beq \label{Nahm2_later}
 %\boxed{ \rule[-.25cm]{.0cm}{.7cm}\quad 
 (\text{C}1'):\qquad  \partial_{y} \mathbf{L}^{\pm} = \pm \tfrac{1}{2} [\mathbf{L}^{\pm},\mathbf{L}^0] \ , \qquad  
  \partial_{y} \mathbf{L}^0 =- [\mathbf{L}^{+} ,\mathbf{L}^{-}]\ .
  %\quad }
\eeq
Furthermore, we insert \eqref{Q-rot} into the conditions \eqref{QcN-matching} and derive that
\bea \label{charges_cL}
  (\text{C}2'):\qquad    \big[2Q_\infty -  L_0, \mathbf{L}^{0}\big]&=& 2 i  ( \mathbf{L}^+ +  \mathbf{L}^-) +i \big[  L_1, \mathbf{L}^0\big]- i[L_{-1},  \mathbf{L}^{0}\big] \ ,  \nn \\
  \big[2Q_\infty  -  L_0 , \mathbf{L}^{\pm}\big]   &=& -i  \mathbf{L}^0 + i \big[ L_1, \mathbf{L}^\pm\big] - i[L_{-1},  \mathbf{L}^{\pm}\big] \ , 
\eea
where we have explicitly evaluated $\rho\, Q_\infty \, \rho^{-1} = Q_\infty  - \tfrac{1}{2} L_0 - \tfrac{i}{2}  L_1 + \tfrac{i}{2}L_{-1}$ 
by using \eqref{L-restrictions}. Consistent 
with \eqref{boundary1} these conditions are indeed satisfied when replacing $ \mathbf{L}^\pm$ with $L_{\pm 1}/y $ and 
$ \mathbf{L}^0$ with $L_{0}/y $. Finely, using the $ \mathbf{L}^0, \mathbf{L}^{\pm 1}$, we want to 
make sure that they stem indeed from a real function. One trivial way to implement this 
is to impose 
\beq \label{cL-real}
     (\text{C}3'):\qquad  \bar {\mathbf{L}}^{\pm} = \rho^{-2}  \mathbf{L}^\pm \rho^2\ , \qquad    \bar {\mathbf{L}}^{0} = \rho^{-2}  \mathbf{L}^0 \rho^2\ .
\eeq
In summary, we realize that the conditions $(\text{C}1),(\text{C}2)$ for the real $\cN^{0},\cN^{\pm}$ are now rewritten as
$(\text{C}1'),(\text{C}2'),(\text{C}3')$ for the complex $\mathbf{L}^0,\mathbf{L}^{\pm}$.

We next have a closer look at the full expansions of $\mathbf{L}^0,\mathbf{L}^{\pm}$. We will consider the expansions of the 
form 
\beq \label{cL-expansion}
   \mathbf{L}^0 (y) = \sum_{n\geq 0} \frac{ L^0_n}{y^{1+n/2}}\ ,\qquad \mathbf{L}^\pm (y) = \sum_{n\geq 0} \frac{ L^\pm_n}{y^{1+n/2}}\ .
\eeq
In accordance with \eqref{boundary1}, \eqref{N-rot}, and \eqref{L-N-relation}, we fix the slowest decreasing 
term by the boundary conditions
\beq \label{leading-coefficient}
    L^0_0 \equiv L_0 \ , \qquad L^\pm_0 \equiv L_{\pm 1}\ .
\eeq
The goal will be to study the properties of the coefficients $L^{0}_n$, $L^{\pm}_n$. In order to do that we split them 
into a eigen-decomposition under  the commuting operators $L^2$, $L_0$, and $Q_\infty$ as in \eqref{weight-charge-expansion-operator1}. We thus 
write 
\beq 
   L_n^\bullet =\sum_q  \sum_{r>0} \sum_{-r\leq s \leq r} (L^\bullet_n)^{(r,s)}_q\ , \qquad \bullet \in \{0,+,- \}\ ,
\eeq
with $r,s,q$ being the highest weight, the weight, and the charge of $(L^\bullet_n)^{(s,r)}_q$ as defined in \eqref{weight-charge-expansion-operator2}.

The differential equation \eqref{Nahm2_later} then lead to the iterative equations 
\bea \label{iterativeeq}
   (n-s) (L^+_n)^{(r,s+2)}_{q+1} &=& - \Big[L_1,(L^0_n)^{(r,s)}_{q} \Big] + \sum_{0<k<n} \Big[(L^0_k),(L^{+}_{n-k}) \Big]^{(r,s+2)}_{q+1}\ , \nn \\
   (n+s) (L^-_n)^{(r,s-2)}_{q-1} &=&\ \Big[L_{-1},(L^0_n)^{(r,s)}_{q} \Big] + \sum_{0<k<n} \Big[(L^0_k),(L^{-}_{n-k}) \Big]^{(r,s-2)}_{q-1}\ , \\
   (n+2)(L^0_n)_q^{(r,s)}&=&2\Big[ L_1, (L^-_n)_{q-1}^{(r,s-2)}\Big] - 2\Big[ L_{-1}, (L^+_n)_{q+1}^{(r,s+2)}\Big]+ 2 \sum_{0<k<n} \Big[(L^+_k),(L^-_{n-k}) \Big]^{(r,s)}_q\ . \nn
\eea
In addition we can also expand the condition \eqref{charges_cL} in eigen-components to yield
\bea \label{charges_cL_comp}
     i (2q - s)(L^{0}_n)^{(r,s)}_q &=& -2  (L^{+}_n)^{(r,s)}_q - 2 (L^{-}_n)^{(r,s)}_q -\Big[  L_1,(L^{0}_n)^{(r,s-2)}_{q-1} \Big]+\Big[L_{-1}, (L^{0}_n)^{(r,s+2)}_{q+1}\Big] \ ,  \nn \\
  i (2q  -  s )(L^{\pm}_n)^{(r,s)}_q &=& (L^{0}_n)^{(r,s)}_q - \Big[ L_1,(L^{\pm}_n)^{(r,s-2)}_{q-1} \Big] +\big[L_{-1}, (L^{\pm}_n)^{(r,s+2)}_{q+1} \Big] \ . 
\eea
Note that the weights and charges in the various terms in \eqref{iterativeeq}  and \eqref{charges_cL_comp} are in accordance with the fact that $L_{\pm 1}$ 
change the weight by $\pm 2$ and the charge by $\pm1$. Finally, the condition \eqref{cL-real} decomposes 
as
\beq \label{cL-real_comp}
    \overline{(L^{\bullet}_n)^{(r,s)}_{q}} = \rho^{-2} (L^{\bullet}_n)_{s-q}^{(r,s)} \rho^2\ ,
\eeq
where we have used that $\bar Q_\infty = - Q_\infty,\ \bar L_0 = - L_0$, as given in \eqref{chargeimaginary}, \eqref{barL}.
Furthermore, we explicitly computed $\rho^2 Q_\infty \rho^{-2} = Q_\infty - L_0$, $\rho^{2} L_0 \rho^{-2} =- L_0$ and 
applied that $L^2$ is real and commutes with $\rho^2$.  
The main challenge is to extract the constraints on the coefficients $(L^\bullet_n)^{(s,r)}_q$ imposed 
by \eqref{iterativeeq}, \eqref{charges_cL_comp}, and \eqref{cL-real_comp}.

Combining the equations \eqref{iterativeeq} one 
then shows that the $(L_n)^{(r,s)}$ satisfy \cite{Schmid,CKS}
\bea \label{N(r,s)-restrictions}
 (L^\bullet_n )^{(r,s)}_q &=&0\quad \text{unless}\quad |s| \leq r \leq n,\ q\leq n-1 \ ,\nn  \\
  (L^\bullet_n )^{(r,s)}_q &=&0\quad \text{unless} \quad (r,s,n\ \text{even) or}\ (r,s,n\ \text{odd})\ ,  \\
   %N^0_{n}(n,n) = N^0_{n}(n,-n)= N^\pm_{n}(n,\mp n) = N^\pm_{n}(n,\pm(2-n)) = 0\ ,\qquad \text{for}\ n>0 \ . 
   %\big[L_n\big]^{(n,n-\alpha n - |\alpha| n)}_\alpha  = [L_n]^{(n, 2\alpha - n-\alpha n + |\alpha| n)}_\alpha  =0\ ,\qquad \text{for}\ n>0 \ . \nn
  (L^0_n)^{(n,n)}_q  &=&  (L^0_n)^{(n,-n)}_q  = (L^{\pm}_n)^{(n,\mp n)}_q  = (L^\pm_n)^{(n, \pm(2 - n))}_q  =0\ ,\qquad \text{for}\ n>0 \ , \nn
\eea
where one considers either the upper sign or the lower sign in each quantity. 
Furthermore, it turns out that all information in the expansions \eqref{cL-expansion} satisfying the stated constraints 
is in the leading coefficients $(L^\bullet_n)^{(n,s)}_q$. Their algebra can be extracted  
from \eqref{iterativeeq} by setting $r=n$. It is rather non-trivial to show that the sums over $0<k<n$ 
in \eqref{iterativeeq} vanish in this case \cite{CKS}.\footnote{This follows from Proposition 6.17 of \cite{CKS}, which uses yet another 
presentation of the information in \eqref{Nahm1_later}, \eqref{QcN-matching}. }
One concludes that 
\begin{itemize}
\item[1.] The leading coefficients $(L^\bullet_n)^{(n,s)}_q$ satisfy the algebra  
\bea \label{top-component}
 (n-s) (L^+_n)^{(n,s+2)}_{q+1} &=& - \Big[L_1,(L^0_n)^{(n,s)}_q \Big] \ , \nn \\ 
 (n+s) (L^-_n)^{(n,s-2)}_{q-1} &=&  \ \ \Big[L_{-1},(L^0_n)^{(n,s)}_q \Big] \ ,  \\ 
  (n+2)(L^0_n)^{(n,s)}_q&=&2\Big[ L_1, (L^-_n)_{q-1}^{(n,s-2)}\Big] - 2\Big[ L_{-1}, (L^+_n)_{q+1}^{(n,s+2)}\Big]\ . \nn
\eea
Combining this expression with \eqref{charges_cL_comp} the $(L^\bullet_n)^{(n,s)}_q$ also obey the constraint 
   \bea \label{charges_cL_comp_leading}
     i (2q - s)(L^{0}_n)^{(n,s)}_q &=&  (n-s) (L^+_n)^{(n,s)}_{q}+(n+s) (L^-_n)^{(n,s)}_{q} \ .  
 % i (2q  -  s )(L^{\pm}_n)^{(r,s)}_q &=& (L^{0}_n)^{(r,s)}_q - \big[ L_1,(L^{\pm}_n)^{(r,s-2)}_{q-1} \big] + [L_{-1}, (L^{\pm}_n)^{(r,s+2)}_{q+1} \big] \ , 
\eea
%\bea \label{top-component}
% (n-s+2) (L^+_n)^{(n,s)}_{q} &=& - \Big[L_1,(L^0_n)^{(n,s-2)}_{q-1} \Big] \ , \nn \\ 
% (n+s+2) (L^-_n)^{(n,s)}_{q} &=&  \ \ \Big[L_{-1},(L^0_n)^{(n,s+2)}_{q+1} \Big] \ ,  \\ 
 % (n+2)(L^0_n)^{(n,s)}_q&=&2\Big[ L_1, (L^-_n)_{q-1}^{(n,s-2)}\Big] - 2\Big[ L_{-1}, (L^+_n)_{q+1}^{(n,s+2)}\Big]\ . \nn
%\eea
\item[2.] The leading coefficients $(L^\bullet_n)^{(n,s)}_q$ determine the solution $ \mathbf{L}^0(y), \mathbf{L}^{\pm 1}(y)$ satisfying \eqref{Nahm2_later}, \eqref{charges_cL},  
\eqref{cL-real}, uniquely.
\end{itemize}

The next step is to find a solution of \eqref{top-component} fixing the leading coefficients $(L^\bullet_n)^{(n,s)}_q$. 
Remarkably, such solutions can be found to depend on one operator $\hat \eta = \rho \eta \rho^{-1}$ with $\eta \in \mathfrak{g}_\bbR$, 
that satisfies   
\beq \label{hateta_exp}
  \hat \eta = \sum_{q\geq 1 ,n\geq 2} (L^-_n)^{(n,-n)}_{-q} \  \in\ \Lambda^{(-2)}_{-1} \ : \quad [L_{-1} , \hat \eta ] = 0\ .
\eeq 
One can check by straightforward computation that the equations \eqref{top-component} are satisfied by 
% are giving by $n\geq 2$, $s\geq 1$, $1\leq q \leq n+1$:
\bea  \label{L-ansatz} 
   (L^0_n)^{(n,2s-n)}_{s-q} &=& 2\,  a_{q}^{n,s} (\text{ad}\, L_1)^s \hat \eta^{(-n)}_{-q}\ , \nn \\
   (L^+_n)^{(n,2s-n+2)}_{s-q+1} &=& - (n-s)^{-1} a_{q}^{n,s} (\text{ad}\, L_1)^{s+1} \hat \eta^{(-n)}_{-q}\ , \\
    (L^-_n)^{(n,2s-n-2)}_{s-q-1} &=&  (n-s+1) a_{q}^{n,s}   (\text{ad}\, L_1)^{s-1} \hat \eta^{(-n)}_{-q}\ , \nn
\eea
where $n\geq 2$, $s\geq 1$, and $1\leq q \leq n-1$.
Note that the complex coefficients $a_{q}^{n,s}$ are unfixed by  \eqref{top-component}, since these coefficients appear in each 
term with the same $s,q,n$ indices. To fix $a_q^{n,s}$ we first use \eqref{charges_cL_comp_leading} which yields the equation 
\beq \label{a-equation}
i (n-2q) a^{n,s}_q + (n-s) (n-s+1)^{-1} a^{n,s-1}_q - s(n-s) a^{n,s+1}_q = 0\ .
\eeq
Furthermore, we also need to impose \eqref{cL-real_comp} and normalize $\eta^{(-n)}_{-q}$ such that  \eqref{hateta_exp} matches with \eqref{L-ansatz}. The former condition equates $ \overline{ a_{q}^{n,s}} $ with $a_{n-q}^{n,s}$ if one uses 
the fact that $\hat \eta= \rho \eta \rho^{-1}$ and $L_1= \rho N^+ \rho^{-1}$ stem from real operators. The normalization condition 
fixes $a^{n,1}_q$. We thus also have to impose the two constraints
\beq \label{bara=a}
   \overline{ a_{q}^{n,s}} = a_{n-q}^{n,s}\ ,  \qquad a^{n,1}_q = \frac{1}{n} \ .
\eeq
%We normalize $\eta^{(-n)}_{-q}$ such that  \eqref{hateta_exp} matches with \eqref{L-ansatz} which implies $a^{n,1}_q = 1/n$.  
All together the requirements \eqref{a-equation}, \eqref{bara=a} are solved by 
\beq \label{a-solution}
  a_{q}^{n,s} =  i^{s-1}\frac{(n-s)!}{n!}  b^{s-1}_{q-1,n-q-1}\ , 
\eeq
with integers $b^k_{p,q},\ k,q,p\geq 0$ defined by $(1-x)^p (1+x)^q = \sum_{k} b^{k}_{p,q} x^k$.
We check that indeed $b^{k}_{p,q} =(-1)^{k} b^k_{q,p}$ as required for \eqref{bara=a}.

To summarize, we have found that any bulk solution satisfying \eqref{Nahm2_later}, \eqref{charges_cL}, \eqref{cL-real} together with 
the boundary conditions \eqref{leading-coefficient}, is uniquely specified by a single real operator $\eta$ or $\hat \eta = \rho \eta \rho^{-1}$. 
Conversely, it is easy to extract the operator $\eta$ for a given solution via \eqref{hateta_exp}. 
As a next step, we will translate the solution for $ \mathbf{L}^0, \mathbf{L}^{\pm 1}$ into a solution for the bulk fields 
$h(y)$.

\subsection{Asymptotic expansions of the matter fields} \label{expand-matter}

Recall that the matter fields of the bulk theory \eqref{full-bulk-action}, \eqref{action-with-gen_1d_flatgauge} are the matrix-valued functions 
$\hat h(x,y)$, which asymptotically take the form $\hat h(x,y) = e^{x N^-} h(y)$. Note that any 
solution for $\cN^0(y),\cN^\pm(y)$ satisfying \eqref{Nahm1_later}, \eqref{QcN-matching} can be parametrized by an $h(y)$ by 
considering 
\beq \label{cNansatz}
   \cN^0(y) = - 2 h^{-1} \partial_y h\ , \qquad \cN^-(y) = h^{-1} \tilde N^- h \ ,
\eeq
where $\tilde N^-$ is a real matrix that can be fixed using the boundary conditions to be $N^-$.\footnote{One generally 
checks that \eqref{cNansatz} satisfies \eqref{Lax11}, which is obtained from 
\eqref{Nahm1_later} and  \eqref{QcN-matching}.}
We now define a new function $g(y)$ by setting
\beq \label{gy-def}
     h(y) \equiv g(y)\, y^{-\frac{1}{2} \tilde N^0}\ ,
\eeq
where $\tilde N^0$ is the leading coefficient in the $y$-expansion of $\cN^0$ as seen in \eqref{boundary1}.
With this definition we find that $\cN^0$ takes the form 
\beq \label{cN0-ing}
     \cN^0(y) = - 2 y^{\frac{1}{2} \tilde N^0} \big[ g^{-1} D_y g \big] y^{-\frac{1}{2} \tilde N^0}\ ,
\eeq
where $D_y = \partial_y - \frac{\tilde N^0}{2 y}$. 

We are now in the position to make contact to the explicit series expansions of section \ref{series_expansions-bL}. In 
order to do that we need to implement the relation \eqref{def-cLs-rho} between $\cN^0$ and $\mathbf{L}^0$, i.e.~we 
rotate all quantities by $\rho$, defined in \eqref{rho-repeat}, and $e^{\zeta}$. From \eqref{N-rot} and \eqref{L-N-relation} we 
infer that $L^0 = \rho\, e^{\zeta}\tilde N^0 e^{-\zeta}\, \rho^{-1}$. We thus can write \eqref{cN0-ing} as 
\beq \label{L0=hatg}
     \mathbf{L}^0(y) = - 2 y^{\frac{1}{2} L^0} \big[ \hat g^{-1} \hat D_y \hat g \big] y^{-\frac{1}{2} L^0}\ ,\qquad \hat g(y) = \rho \, g(y)\, e^{-\zeta}\, \rho^{-1}\ .
\eeq
Let us stress that $\hat g$ is not real due to the factors of $\rho$.
We next use the explicit expansion \eqref{cL-expansion} for $\mathbf{L}^0(y)$ and determine 
the expansion for $\hat g$.
Using \eqref{N(r,s)-restrictions} we find that  
\beq \label{ginvg}
   \hat g^{-1} \partial_y \hat g = \sum_{k\geq 2}  \frac{B_k}{y^{k}}\ ,
\eeq
with  
\beq  \label{Bn-def}
    B_{k} = - \frac{1}{2} \sum_{s\leq k-2}\  \sum_{r\leq 2k-2-s} \ \sum_{q\leq k-2} \,
   (L^0_{2k-2-s})^{(r,s)}_{q }\ .
\eeq
Note that this form of $B_k$ implies that 
\beq \label{BinLam}
    B_{k}\  \in \ \bigoplus_{p,q \leq k-2} \Lambda^{(p)}_q\ ,  
\eeq
with $ \Lambda^{(p)}_q$ defined in \eqref{space_chargeless}.
It is important to stress that the expansion \eqref{ginvg} starts with a $y^{-2}$ term, so in 
effect the redefinition \eqref{gy-def} ensures that the $y^{-1}$-term is not present in $\hat g^{-1} \partial_y \hat g$
and hence not in $g^{-1} \partial_y g$. We can put these statements together, combine it with an argument for convergence \cite{Schmid,CKS}, and
infer that $g(y)$ and $g(y)^{-1}$ admit Taylor expansions at $y=\infty$ of the form \footnote{The existence of these Taylor expansions constitute the first part of Schmid's Sl(2)-orbit theorem.}
\bea \label{gy-expansions}
     g(y) &=& g(\infty) \Big( 1 + \frac{g_1}{y} + \frac{g_2}{y^2} + \ldots \Big)\ ,\\
     g(y)^{-1} &=&  \Big( 1 + \frac{f_1}{y} + \frac{f_2}{y^2} + \ldots \Big)g(\infty)^{-1}\ . \nn 
\eea
Note that $g(\infty)$, which is the value of $g(y)$ in the limit $y\rightarrow \infty$, drops out from \eqref{ginvg} and hence is not fixed in terms of the $ (L^0_{n})^{(r,s)}_q$.
We have already indicated in \eqref{ginfty-properties} that we can choose the overall transformation $e^\zeta = g(\infty) $, which implies that 
$\hat g$, $\hat g^{-1}$ then has the expansion 
\beq \label{gy-expansions-2}
    \hat g(y) =  \Big( 1 + \frac{\hat g_1}{y} + \frac{\hat g_2}{y^2} + \ldots \Big)\ ,\qquad  \hat g(y)^{-1} =  \Big( 1 + \frac{\hat f_1}{y} + \frac{\hat f_2}{y^2} + \ldots \Big)\ ,
\eeq
with 
\beq \label{def-hatg-hatf}
    \hat g_i = \rho g(\infty) g_i g(\infty)^{-1} \rho^{-1}\ ,\qquad  \hat f_i = \rho g(\infty) f_i g(\infty)^{-1} \rho^{-1}\ .
\eeq 
This choice indeed normalizes the asymptotic expansion, which now 
depends entirely on the $(L^\bullet_{n})^{(r,s)}_q$, which in turn are specified by $\hat \eta$ as discussed after \eqref{hateta_exp}.

To gain a deeper understanding of the properties of the coefficients $\hat g_i$ in \eqref{gy-expansions-2}
we note that one can invert \eqref{ginvg} to write 
\beq \label{hatg-B}
    \hat g_k = P_{k} (B_2,...,B_{k+1})\ ,
\eeq
for a set of \textit{universal} non-commutatitve polynomials $P_k$. These polynomials $P_{k} (B_2,...,B_{k+1})$
are iteratively defines by 
\beq \label{def-Ppoly}
   P_0 = 1\ , \qquad P_{k} = - \frac{1}{k} \sum_{j=1}^k P_{k-j} B_{j+1}\ . 
\eeq
One checks that using \eqref{hatg-B} with \eqref{def-Ppoly}, the differential equation $\partial_y \hat g =  \hat g \cdot \sum_{n\geq 2}  B_n y^{-n}$
is satisfied. Let us now recall that the $B_n$ are fixed in terms of the $(L^0_{n})^{(r,s)}_q$ via \eqref{Bn-def}. The latter are then 
determined from a $\hat \eta$ via \eqref{L-ansatz} for 
the highest term and then recursively via \eqref{iterativeeq}. Taken these facts together one 
finds that the $\hat g_k$ (and the $\hat f_k$) are universal non-commutative polynomials in $L_{+1}$ and $\hat \eta^{(n)}_p$.
Any appearance of $L_0$, $L_{-1}$ in the recursive evaluate can be eliminated by using the fact that ad$_{L_{0}} =[L_0,\cdot]$ acts with 
integer eigenvalues on expressions involving $\hat \eta^{(p)}_q$, $L_1$, and we have $[L_{-1},\hat \eta]=0$. 
In fact, one shows that the $\hat g_k$, $\hat f_k$ are homogeneous of degree $k$ in ad$_{L_{+1}} = [L_{+1},\cdot]$. The latter
is easy to see for the leading term \eqref{L-ansatz}, and can be extend to all terms recursively \cite{CKS}. 
It now also follows that 
\beq \label{gf-anhi}
   \text{ad}_{L_{-1}}^{n+1} \hat g_n = 0 \ , \qquad \text{ad}_{L_{-1}}^{n+1} \hat f_n = 0  \ .
\eeq
Transformed back to $g_i$ we thus gets the condition \eqref{gn-property}, and $\text{ad}_{N^{-}}^{n+1} f_n = 0$. Furthermore, 
we conclude from \eqref{BinLam} and \eqref{hatg-B} that 
\beq  \label{gf-in_Lam}
\hat g_k,\ \hat f_k \in \bigoplus_{p \leq k-1,\, q} \Lambda^{(p)}_q\ .
\eeq
The properties \eqref{gf-anhi} and \eqref{gf-in_Lam} appear to be abstract, but have significant implications in concrete applications
as we discuss in section \ref{finitenesssec}.  

 %It remains to justify that the $g(\infty)$ and $g_k$ are, in fact, fixed entirely by \eqref{fixall}. In other words 
 %we will see that the boundary data fixes the bulk solution \eqref{simple-h-expansion} uniquely.  

\subsection{Uniqueness of the near boundary solution}  \label{sec-uniqueness}

Having determined the expansions \eqref{gy-expansions}
we next discuss a prescription how to fix the solution uniquely. 
In order to do this we first argue that we can bring any bulk 
solution into a form that is reminiscent of the form of a 
nilpotent orbit. Comparing this expression 
with the original nilpotent orbit \eqref{onepar_nilp} allows us 
to show that the boundary data and \eqref{fixall} are 
sufficient to entirely fix the bulk solution. 
The equation \eqref{fixall} relates the boundary data to the coefficients 
in the bulk solution and fixes it uniquely.  

To begin with we aim to bring the bulk solution $\hat h(x,y)$ in a form reminiscent of a nilpotent orbit \eqref{onepar_nilp}. 
Hence, we rewrite the information contained in a solution 
$\cN^0 = -2 \hat h^{-1} \partial_y \hat h$, $\cN^{-} = \hat h^{-1} \partial_x \hat h$. We first derive that 
\beq \label{der1}
  h^{-1} e^{i y \tilde N^-} \frac{d}{dy} \Big(e^{-i y \tilde N^-} h\Big) = -i \cN^- - \frac{1}{2} \cN^0 = -\frac{1}{2} \cL_0 -i\cL_1\ , 
\eeq
where in the second equality we have used the definition \eqref{def-cL+-1} of $\cL_\alpha$.
We can now employ the properties of $\cL_\alpha$ when acting on the reference structure $F_{\rm ref}^p$. The $F_{\rm ref}^p$ are complex vector spaces spanned by 
states with $Q$-charges being larger or equal to 
$p - \frac{1}{2}D$, i.e.~we have $F_{\rm ref}^p =  \bigoplus_{r\geq p} H^{r,D-r}_{\rm ref}$ with \eqref{Q_forms}.
Since $[Q,\cL_\alpha]=\alpha \cL_\alpha$, we see that $\cL_1,\cL_0$ preserve or increase the charge and hence 
conclude that $\cL_0 F^p_{\rm ref} \subset F^p_{\rm ref}$ and 
$\cL_1 F^p_{\rm ref} \subset F^p_{\rm ref}$ showing that $\cL_0$, $\cL_1$ preserve the vector spaces $F^p_{\rm ref}$. 
We now read the expression \eqref{der1} as a relation between a group element $e^{-i y \tilde N^-} h$ and an algebra element $\frac{1}{2} \cL_0 -\cL_1$,
we conclude that $e^{-i y \tilde N^-} h(y) = \kappa f(y)$, where $\kappa, f(y) \in G_{\bbC}$ with $f(y)$ preserving $F^{p}_{\rm ref}$ and $\kappa$ being constant. 
Combining these last two facts we can write 
\beq \label{hF=tildeF}
    h(y) F^p_{\rm ref} = e^{i y \tilde N^-} \tilde F^p_0\ , \qquad \tilde F^p_0 = \kappa F^p_{\rm ref}\ .
\eeq
Clearly, this expression is exactly of the form \eqref{Fpol-ash}, the equation which served as a definition of $h(y)$ when starting 
with a nilpotent orbit. Here we do not make this a priori assumption on $\hat h(x,y)$ and only 
demand that it solves the field equations \eqref{h-Nahm}, \eqref{Q-restr_eom} and has the symmetry \eqref{h-rotation}. 

The question is now to identify the conditions on $\hat h(x,y)$ such that 
$\tilde F^p \equiv e^{i y \tilde N^-} \tilde F^p_0$ is indeed a nilpotent orbit. This requires to enforce that $\tilde N^-$ is nilpotent with $\tilde N^- \tilde F^p_0 \subset \tilde F^{p-1}_0 $ and that the Hodge decomposition 
$\tilde H^{p,q} = \tilde F^p \cap \overline{\tilde F^q}$ satisfies $\overline{\tilde H^{p,q}} = \tilde H^{q,p}$ and induces a well-defined norm. 
The properties of $\tilde  H^{p,q} $ are inherited from the properties of $F^p_{\rm ref}$ and we will see 
below how $F^p_{\rm ref}$ can be matched with the boundary data. While the properties of 
$\tilde N^-$, such as its nilpotency, are inherited from $N^-$ when matched as in \eqref{N-rot}. We conclude that indeed we can 
determine a nilpotent orbit from a bulk solution $h(y)$, with associated $\tilde N^-$, $F^p_{\rm ref}$.

To complete the discussion we notice from \eqref{cNansatz} that $\tilde N^- =  h(y) \cN^-(y) h^{-1}(y)$. 
Inserting the explicit expansion of $h(y)$ given by \eqref{gy-def}, \eqref{gy-expansions} and the expansion of $\cN^-$ given in \eqref{boundary1}, 
we extract the constant term yielding 
\beq
    \tilde N^- = g(\infty) \tilde N^{-} g(\infty)^{-1}\ .     
\eeq
Hence we find that $ [ \tilde N^{-} , g(\infty)] = 0$. Compatible with this condition, we now pick 
\beq \label{fixginf}
   g(\infty) = e^{\zeta}\ , \qquad  [\zeta , N^-] = 0\ , \quad \hat \zeta \in \Lambda^{(-2)}_{-1}\ ,  
\eeq 
which implies that $\tilde N^- = N^-$. This ensures that, when re-introducing the coordinate $x$ by completing $t=x+iy$, 
that the nilpotent orbit derived from $h(y)$, $F^p_{\rm ref}$ transforms with the symmetry $N^-$ associated to the boundary
as in \eqref{h-rotation}. 
Indeed, we can then complete \eqref{hF=tildeF} to $\hat h(x,y) F^p_{\rm ref} = e^{t N^-} \tilde F^p_0$.

It remains to address  how the boundary data fix $\zeta$, which defines $g(\infty)$ via \eqref{fixginf}, and $\eta$, which defines $g_{i}$
as discussed in sections \ref{series_expansions-bL}, \ref{expand-matter}. The central statement is that 
for a given $N^-,\delta \in \mathfrak{g}_{\bbR}$, with $[\delta,N^-]=0$, $\hat \delta \in \Lambda^{(-2)}_{-1}$
there is a unique choice of $\zeta,\eta$ such that \eqref{fixall} is satisfied. 
Let us begin by motivating \eqref{fixall} by comparing the original nilpotent orbit \eqref{onepar_nilp}
to the orbit \eqref{hF=tildeF}. The orbit \eqref{onepar_nilp} was used to introduce a special 
$F_{\rm ref}^p= e^{i N^-} e^{-i \delta} F^p_0$ in \eqref{ref-F}.
Requiring $F_0^p = \tilde F_0^p$ in \eqref{hF=tildeF} we can then find the equality
\beq \label{eq1}
   e^{i \delta}e^{-i N^-} F_{\rm ref}^p = e^{- i y  N^-} h(y) F^p_{\rm ref} \ .
\eeq
We next turn this into an equality of vector spaces obtained 
from $F_\infty^p$. A key step is to realize that $F_\infty^p$ are vector spaces that are 
preserved by $L_0$, $L_1$, following an argument analog to the one after \eqref{der1}. 
We can then use the identity 
\beq
     \rho = e^{i L_{-1}} e^{\frac{i}{2} L_1} e^{\frac{1}{\sqrt{2}} L_0}\ , 
\eeq
which implies together with \eqref{L-N-relation} that on $F^p_\infty$ we have 
\beq
  F^p_\infty =  e^{i N^-} \rho^{-1} F^p_\infty \ . 
\eeq 
Furthermore, using the same reasoning, namely that $L_0$ preserves $F_\infty^p$, 
we have the identity $F_\infty^p = y^{\frac{1}{2} L_0} F_\infty^p$. 
Hence, using $F^p_{\rm ref}= e^{-\zeta} F^p_\infty$ given in \eqref{Fref-Finf} we can thus rewrite \eqref{eq1} as
\bea
   e^{i \delta} e^{-\zeta} \rho^{-1} F_{\infty}^p &=& e^{- i y  N^-} h(y) e^{-\zeta}e^{i N^-} \rho^{-1} y^{\frac{1}{2} L_0} F^p_{\infty} \nn \\
   & =&   e^{- i y  N^-} g(y) y^{-\frac{1}{2} \tilde N^0} e^{-\zeta}e^{i N^-} y^{-\frac{1}{2} N^0} \rho^{-1} F^p_{\infty} \\
    & =&   e^{- i y  N^-} g(y) e^{-\zeta} y^{-\frac{1}{2} N^0} e^{i N^-} y^{\frac{1}{2} N^0} \rho^{-1} F^p_{\infty} \nn \\
    & =&   e^{- i y  N^-} g(y)   e^{i y N^-} e^{-\zeta} \rho^{-1} F^p_{\infty}    \ . \nn 
\eea 
In the fourth identity we have used $ y^{-\frac{1}{2} N^0} e^{i N^-} y^{\frac{1}{2} N^0} = e^{i y N^-}  $ and 
that $[\zeta,N^-]=0$. Inserting the expansion \eqref{gy-expansions} of $g(y)$, with $g(\infty)= e^{\zeta}$, we can now evaluate 
\beq \label{coefficientmatch}
   e^{- i y  N^-} g(y)   e^{i y N^-} e^{-\zeta} \rho^{-1} F^p_{\infty} = e^{\zeta} \sum_{k,l\geq 0} \frac{(-i)^k}{k!}y^{k-l}\, (\text{ad} N^-)^k g_l e^{-\zeta} \rho^{-1} F^p_{\infty}  \ ,
\eeq
where we have introduced $g_0 =1$. 
Due to \eqref{gn-property} each term in the sum has non-positive powers and we can thus evaluate \eqref{coefficientmatch} in the limit $y\rightarrow \infty$. 
This leads to the identity 
\beq
e^{i \delta} e^{-\zeta} \rho^{-1} F_{\infty}^p  = e^{\zeta} \sum_{k\geq 0} \frac{(-i)^k}{k!}\, (\text{ad} N^-)^k g_k e^{-\zeta} \rho^{-1} F^p_{\infty} \ .
\eeq
Comparing coefficients we realize that a sufficient condition for this 
vector space identity to be satisfied is 
\beq
  e^{i \delta}   = e^{\zeta} \sum_{k\geq 0 } \frac{(-i)^k}{k!}\,(\text{ad} N^-)^k g_k \ . 
\eeq
This is the condition \eqref{fixall} announced before and relates the boundary data $\delta,N^-$ with the coefficients 
in a general bulk solution \eqref{simple-h-expansion}. 

Let us now show that indeed the condition \eqref{fixall} is sufficient to fix the bulk solution completely when 
given the set of boundary data specified in section \ref{boundary_theory}. In order to do this we first transform \eqref{fixall}
to 
\beq
  e^{i \hat \delta}  e^{-\hat \zeta} = \sum_{k\geq 0 } \frac{(-i)^k}{k!}\,(\text{ad} L_{-1})^k \hat g_k    \ ,
\eeq
where the $\hat \delta = \rho \delta \rho^{-1} $, $\hat \zeta= \rho \zeta \rho^{-1}$, and $\hat g_i = \rho e^{\zeta} g_i e^{-\zeta} \rho^{-1}$ as above. 
Recall from \eqref{hatg-B} that $\hat g_k = P_k(B_2,...,B_{k+1})$ can be expressed as a function of the coefficients $B_l$ appearing in \eqref{ginvg} 
with $P_k$ being specific non-commutative polynomials introduced in \eqref{def-Ppoly}. Using the Leibniz rule we 
can rewrite this expression as 
\beq  \label{hatdz-P}
    \boxed{ \rule[-.25cm]{.0cm}{.85cm} \quad e^{i \hat \delta}  e^{-\hat \zeta} = 1+\sum_{k\geq 1 } P_k(C_2,...,C_{k+1}) \ , \quad}
\eeq
where $C_{k+1} := \frac{(-i)^k}{k!} (\text{ad} L_{-1})^k B_{k+1}$.
We now aim to find an explicit expression for $C_{k+1}$ in terms of $\hat \eta$. Using the definition \eqref{Bn-def} 
of $B_k$ we first show \footnote{See Lemma 6.32 of \cite{CKS}.}
\beq
  (\text{ad}  L_{-1})^{k-1} B_{k} = - \frac{1}{2} \sum_{l\leq k}\ \sum_{q\leq k-2} \,
   (\text{ad}  L_{-1})^{k-1}  (L^0_{l})^{(l,2k-2-l)}_{q }\ ,\qquad (\text{ad}  L_{-1})^{k} B_{k} = 0\ , 
\eeq 
where we have set $l = 2k-2-s$ in \eqref{Bn-def} and used the fact that $\text{ad} L_{-1}$ lowers the weight of an operator by $2$. 
Furthermore, we derive by using \eqref{L-ansatz} with \eqref{a-solution} together with the $\mathfrak{sl}(2)$-algebra 
that 
\beq
    (\text{ad}  L_{-1})^{k}  (L^0_{l})^{(l,2k-l)}_{k-q} = 2 i^{k-1} k! \ b^{k-1}_{q-1,l-q-1}  \  \hat \eta^{(-l)}_{-q}\ .
\eeq
We are now in the position to evaluate 
\beq \label{Caseta}
  \boxed{ \rule[-.25cm]{.0cm}{.8cm}  \quad C_{k+1}  = i  \sum_{l\geq k+1} \sum_{q\geq 1}\ b^{k-1}_{q-1,l-q-1}  \  \hat \eta^{(-l)}_{-q}\ .\quad }
\eeq
The formula  \eqref{hatdz-P} with \eqref{Caseta} gives us an explicit expression relating $\hat \delta,\hat \zeta$ and $\hat \eta$. In 
fact we will argue next that it allows to determine $\hat \zeta$ and $\hat \eta$ as a function of $\hat \delta$. 

To show that \eqref{hatdz-P} with \eqref{Caseta} can be used to fix $\hat \zeta$ and $\hat \eta$ in terms of $\hat \delta$, we
note that all three operators actually stem from real counterparts $\zeta,\eta,$ and $\delta$. To use this reality condition 
we note that it can be written as 
\beq 
      \bar \cO = \rho^{-2} \cO \rho^2\ , \qquad   \overline{\cO^{(s)}_{q}} = \rho^{-2} \cO_{s-q}^{(s)} \rho^2 \ , \qquad \cO \in \{\zeta,\eta,\delta \}\ ,  
\eeq
as we have already noted in \eqref{cL-real}, \eqref{cL-real_comp} for other operators. 
Since the identity \eqref{hatdz-P} with \eqref{Caseta} is a polynomial in the components of $\hat \zeta$, $\hat \eta$, and $\hat \delta$ 
the transformation involving $\rho$ simply drops on both sides. Hence, we can also
replace in \eqref{hatdz-P}, \eqref{Caseta}:
\beq
   i \rightarrow - i \ , \quad \hat \zeta_{q}^{(s)} \rightarrow \hat \zeta_{s-q}^{(s)}\ , \quad 
  \hat \eta_{q}^{(s)} \rightarrow \hat \eta_{s-q}^{(s)}\ , \quad \hat \delta_{q}^{(s)} \rightarrow \hat \delta_{s-q}^{(s)}\ ,
\eeq
and find an equally valid equation. 
Combined with the original expression we can then either eliminate $\hat \eta$ or $\hat \zeta$ and determine $\hat \zeta$, $\hat \eta$ as a function of 
the components of $\hat \delta$.
For example, the first terms are 
\begin{align}
    &&\hat \zeta_{-1}^{(-2)}=\hat \zeta_{-2}^{(-4)}=0\ , \qquad \hat \zeta_{-1}^{(-3)} =- \frac{i}{2} \hat \delta_{-1}^{(-3)} \ , \qquad \hat \zeta_{-1}^{(-4)} =  -\frac{3i}{4} \hat \delta_{-1}^{(-4)}\ ,& &\nn \\
   &&  \hat \zeta_{-2}^{(-5)} = -\frac{3i}{8}\hat \delta_{-2}^{(-5)} -\frac{1}{8}\Big[\hat\delta_{-1}^{(-2)},\hat\delta_{-1}^{(-3)} \Big]\ , 
   \qquad \hat \zeta_{-3}^{(-6)} = - \frac{1}{8} \Big[\hat\delta_{-1}^{(-2)},\hat\delta_{-2}^{(-4)} \Big]\ . &&
\end{align}
These relations suffice to treat the Calabi-Yau threefold case, but one can expand \eqref{hatdz-P} further 
to determine the relations relevant for any $Y_D$. Note that these relations are abstractly valid 
and do not make use of the Calabi-Yau condition for $Y_D$. 

Note that it remains to show that $\hat g_{k}$ is also fixed by the boundary data. In fact, we know from 
\eqref{hatg-B} with \eqref{BinLam}, \eqref{L-ansatz} that its leading coefficients are fixed by $\hat \eta$, which itself is fixed by $\hat \delta$. The subleading 
coefficient are then determined by the iterative equations \eqref{iterativeeq} and hence also involve ad$L_{1}$.\footnote{The dependence on 
$L_0$, $L_{-1}$ can be eliminated using the evaluating $L_0$ on the weight eigencomponents and $[L_{-1}, \hat \eta]=0$.} In fact, 
we stress that in obtaining $\hat \delta$ from $\delta$ we also need $L_1$. 
We have thus argued that we can evaluate 
\beq
  \hat \zeta = \hat \zeta(\hat \delta)\ ,  \qquad \hat g_k =  \hat g_k(\hat \delta, \text{ad} L_1)\ . 
\eeq
Note that in order to evaluate these relations it is crucial to perform the split of the operators 
$\hat \delta$, $\hat \zeta$ into weight and charge eigencomponents, which are determined by the boundary $\mathfrak{sl}(2,\bbC)$-operator $L_0$ 
and $Q_\infty$. 
Hence, it is possible to evaluate using the described steps the functional dependence
\beq
   \zeta = \zeta(\delta,L_\alpha,Q_\infty)\ , \qquad g_{k}= g_k(\delta,L_\alpha,Q_\infty)\ . 
\eeq
We conclude that all information about the boundary theory is needed to fix the bulk solution \eqref{simple-h-expansion}
with $g(\infty)= e^{\zeta}$ and $\tilde N^0 = e^{-\zeta} \rho^{-1} L_0 \rho e^{\zeta}$. Conversely, one 
can use a bulk solution corresponding to a nilpotent orbit to determine the boundary data. 

\section{The finiteness of the flux landscape and the distance conjecture} \label{finitenesssec}

In this final section we will discuss two interesting applications 
of the holographic perspective developed in this work. In preparation 
of the physics applications we first introduce in section~\ref{sec:growththeorem} 
a powerful consequence of the detailed understanding 
of the near boundary expansion of the matter fields \eqref{simple-h-expansion}.
More precisely, we will return to the analysis of the asymptotic form of the 
Hodge norm \eqref{Hodge-norm} and argue that the leading growth of any fixed
element $F \in \cH$ is determined by its weight decomposition under the boundary 
$\mathfrak{sl}(2,\bbR)$ \cite{Schmid}. This fact will be useful when 
studying flux compactifications and the distance conjecture. 
In section \ref{Finiteness_proof} we then sketch the main aspects of the proof \cite{Schnellletter,GrimmSchnell} that there are no 
infinite tails of flux vacua in the Type IIB or F-theory landscape near any co-dimension one 
boundary. The argument will be formulated for self-dual $G_4$ fluxes on a Calabi-Yau fourfold. 
We will also comment on the situation 
in which the fluxes are of Hodge type $(2,2)$, and note that a general proof is know for this more restrictive case.
Finally, in section \ref{distance_comments} we will return to the discussion of the distance conjecture, which 
was the initial motivation for this work. We briefly comment on how the $\mathfrak{sl}(2)$ structure on the 
boundary might be viewed as generalizing the original duality motivation \cite{Ooguri:2006in} for the conjecture. 

\subsection{Leading behaviour of the Hodge norm}  \label{sec:growththeorem}

We want to understand how the Hodge norm $\|F\|^2 = \int F \wedge *F$ behaves as a function of the moduli as 
mentioned already in the motivation of our constructions around \eqref{Hodge-norm}. In the near boundary 
region of the boundary $t =i \infty$, we can 
approximate $\|F\|^2$ by $\|F\|_{\rm pol}^2$, which amount to dropping exponentially suppressed correction $\cO(e^{2\pi i t})$.
This latter norm was defined in \eqref{Weil-pol} and arises from the nilpotent orbit approximation. We have 
argued in the proceeding sections that $\|F\|_{\rm pol}^2$ can equally be derived by using the solutions to the 
bulk theory that match the boundary data specified in section \ref{boundary_theory}. 

The crucial outcome of the analysis of section \ref{bulk_boundary} was the construction of 
a bulk solution matching the boundary data. This solution \eqref{simple-h-expansion} relates the 
decomposition of forms on the boundary $H^{p,q}_\infty$, to the one relevant 
in the near boundary region. In particular, it provides us with an explicit expression 
of the Hodge star near the boundary as we will see in the following. 
Recall that we have denoted this near boundary operator by 
$C_{\rm pol}$ in \eqref{Weil-pol}. Concretely we find 
\bea \label{Cpol_approx}
   C_{\rm pol}(t,\bar t) &=& \hat h\,  e^{-\zeta}  C_\infty e^{\zeta}\,  \hat h^{-1} \ , \nn \\
                                   &=& e^{x N^-} h \,  e^{-\zeta}  C_\infty e^{\zeta}\,  h^{-1} e^{-x N^-} \ ,
\eea
Recall that from \eqref{simple-h-expansion} with \eqref{tildeN0fixing}, \eqref{ginfty-properties} the solution $h(y)$ admits the expansion 
\beq \label{hy-solutionlater}
   h(y)  = e^{\zeta} \Big( 1 + \frac{g_1}{y} + \frac{g_2}{y^2} + \ldots \Big) e^{-\zeta}  y^{-\frac{1}{2} N^0} e^{\zeta} \ .
\eeq
Note that $h(y) \sim y^{-\frac{1}{2} N^0} e^{\zeta} $ when we consider very large $y$ 
since all $\frac{g_j}{y^j} \rightarrow 0$.  
Inserted into \eqref{Cpol_approx} we thus find that 
\beq \label{Cpot_approx}
C_{\rm pol}(t,\bar t)\ \sim \ C_{\rm s}(t,\bar t) \equiv e^{x N^-} y^{-\frac{1}{2} N^0}   C_\infty y^{\frac{1}{2} N^0}  e^{-x N^-}\ . %=  e^{x N^-} C_\infty y^{-N^0}  e^{-x N^-}\ . 
\eeq  
We denote the  norm corresponding to $C_{\rm s}$ by $\|\cdot \|_{\rm s }$.
Hence, we find that  very close to the boundary $y\rightarrow \infty$  the Hodge norm is well approximated by \cite{Schmid} \footnote{More exactly, we can show that $\| \cdot \|$ and $\|\cdot \|_{\rm s}$ are mutually bounded, i.e.~there exist positive $a_1,\, a_2$
such that $a_1 \|v \|_{\rm s} \leq  \|v\| \leq a_ 2 \|v \|_{\rm s} $.}
\beq \label{growththeorem}
   \boxed{ \rule[-.25cm]{.0cm}{.8cm}  \quad  \| F \|^2\ \sim\   \| F \|^2_{\rm s} \equiv \big\|  y^{\frac{1}{2}N^0}    e^{-x N^-} F  \big\| ^2_\infty  =  \sum_l y^{l}  \|    \rho_l \|^2_\infty  \ ,
   \quad }
\eeq
where in the second equality we have abbreviated $\rho(x) \equiv e^{-x N^-} F $ and performed an eigendecomposition with respect to $N^0$ via 
\beq \label{rho-decompN0}
   \rho(x) = \sum_{l} \rho_l \ , \qquad   N^0 \rho_l  = l \rho_l\ . 
\eeq
Note that after decomposing $\rho(x)$ we can use the orthogonality of the $N^0$ eigenspaces discussed in section~\ref{boundary_theory}, equation \eqref{orthogonality}, to get the result \eqref{growththeorem}. Considering a bounded $x$, we also infer from \eqref{growththeorem}
that 
\beq \label{growth2}
    \| F \|^2\ \sim\   \sum_l y^{l}  \|    F_l \|^2_\infty \ , \qquad N^0 F_l  = l F_l\ , 
\eeq 
where one uses that $N^-$ acts as a lowering operator and can thus only decrease the growth. Hence, one can not lower the 
leading growth by tuning the field $x$.
The fact that the location of $F$ in the boundary splitting of $\cH=H^{D}(Y_D,\bbC)$ determine the 
leading growth of the Hodge norm is a well-known result of asymptotic Hodge theory \cite{Schmid,CKS}.

\subsection{Proving the finiteness of the flux landscape} \label{Finiteness_proof}

It is an important open problem in the study of flux compactifications 
to show that the number of `well-defined' flux vacua is finite \cite{Grana:2005jc,Douglas:2006es}. Even in the best studied settings, namely 
Type IIB compactifications with three-form flux and their  F-theory and M-theory generalizations, finiteness has 
not been fully established, even though there is compelling evidence from the analysis of the flux density
 \cite{Ashok:2003gk,Denef:2004ze,Eguchi:2005eh,Douglas:2006zj,Lu:2009aw} and individual examples \cite{Braun:2011av}.  
 Concretely, let us consider M-theory or F-theory on a Calabi-Yau fourfold $Y_4$, 
and switch on some background flux $G_4$. It is well-known that these fluxes are constrained by the tadpole 
condition \cite{Duff:1995wd,Sethi:1996es}. Furthermore, one finds that consistency of the vacuum requires, in the absence of any 
non-perturbative corrections,
that $G_4$ satisfies a self-duality condition. Together, these two conditions read  \cite{Becker:1996gj,Dasgupta:1999ss} \footnote{In principle, one can 
also allow for half-quantized fluxes \cite{Witten:1996md}. This does, however, not change the discussion of finiteness. Furthermore, 
one generally has to include a warp-factor in the dimensional reduction yielding a corrected effective action \cite{Grimm:2014efa,Grimm:2015mua}. These corrections do not change the arguments made here.}
\beq  \label{to-show}
  G_4 \in H^{4}(Y_4,\bbZ): \qquad \langle G_4, G_4 \rangle < K\ , \qquad  G_4 = * G_4\ ,
 \eeq
 where $K$ is a positive constant and we recall the definition \eqref{def-S}.
    The finiteness of the $G_4$ flux landscape thus requires, as a necessary condition, 
that the tadpole condition and the self-duality condition have only finitely many solutions $(z_{\rm vac},G_4)$ for a fixed $Y_4$.
Here $z^I_{\rm vac}$ are choices for the complex structure moduli such that the self-duality condition is satisfied and 
we count the connected components in $\cM$ parametrized by $z^I_{\rm vac}$, since not necessarily all $z^I$ might be fixed for a 
given $G_4$. The non-trivial part in answering this question lies entirely in controlling the Hodge star $*$ in \eqref{to-show}
in the boundary regions of the moduli space, where it potentially blows up or decays. In the bulk of the moduli space 
$\cM$ the Hodge norm $\| \cdot \|$ introduced in \eqref{Hodge-norm} is bounded and hence there only finitely many solutions to
\eqref{to-show} as is apparent from $\langle G_4 , G_4 \rangle = \| G_4 \|^2 < K$ and the discreteness of the flux. Hence, the problem of showing finiteness amounts to controlling 
infinite tails of flux vacua. Whether or not such tails exist in certain Type IIA flux compactifications \cite{Grimm:2004ua,Derendinger:2004jn,DeWolfe:2005uu} is an ongoing debate, see e.g.~\cite{Junghans:2020acz,Buratti:2020kda,Marchesano:2020qvg}.  

\noindent
\textbf{Finiteness of supersymmetric vacua and the Hodge conjecture.}
Let us highlight the non-triviality of the finiteness statement. In fact, we might ask the 
slightly less general question if the number of supersymmetric four-form fluxes are finite.  
Evaluating the F-term conditions for the complex structure moduli implies that such $G_4$ fluxes have to be of type $(2,2)$ in 
the Hodge decomposition. Furthermore, demanding that the F-terms for the K\"ahler moduli vanish implies the primitivity 
of $G_4$. Hence, \eqref{to-show} reduces to  \cite{Gukov:1999ya}
\beq  \label{to-show-result}
  G_4 \in H^{4}(Y_4,\bbZ) \cap H^{2,2}_{\rm prim}(Y_4,\bbC): \qquad \int_{Y_4} G_4 \wedge G_4 < K \ ,
 \eeq
 and we can ask for finiteness of pairs $(z^I_{\rm vac},G_4)$ satisfying these conditions. In fact, \eqref{to-show-result} 
 is equivalent to the statement that $G_4$ is a Hodge class with bounded product. It is 
a famous result of Cattani, Deligne, and Kaplan \cite{CDK} that the 
locus in complex structure moduli space at which primitive integral forms are of type $(p,p)$ is a countable 
union of algebraic varieties. Furthermore, they show that, if one also imposes a bound on the wedge-product as in \eqref{to-show-result}, 
the number of connected components at which the fluxes are $(p,p)$ is actually finite. 
Applied to our situation their statement implies that the supersymmetric locus can be given by a finite 
number of complex algebraic equations and hence that there are only finitely many supersymmetric flux vacua. 
The theorems of \cite{CDK} rely crucially on a clever application of the Sl(2)-orbit theorem of \cite{Schmid,CKS}, 
which is also the basis of the bulk-boundary construction presented in this work. 
The mathematical significance of the theorems of \cite{CDK} becomes eminent, when noting that the same conclusion can be obtained by applying 
the Hodge conjecture, which is a famously difficult problem in algebraic geometry \cite{DeligneHodge}.  Indeed, the Hodge conjecture can also be applied to our compactifications, since the introduced supersymmetric $G_4$ fluxes describe Hodge classes 
and simple connected Calabi-Yau manifolds are projective. The result of \cite{CDK} is widely viewed as one of the 
strongest evidences for the Hodge conjecture. 

\noindent
\textbf{Finiteness of self-dual vacua in one parameter limits.} The one-parameter Sl(2) orbit theorem can equally be applied to show the finiteness statement summarize \eqref{to-show} near any one-parameter limit \cite{Schnellletter,GrimmSchnell}.
In fact, since the bulk-boundary construction presented here mimics the proof of the Sl(2)-orbit theorem, the finiteness follows 
from the existence of a boundary theory with the properties described in section \ref{boundary_theory}. More precisely, in showing that \eqref{to-show}
never leads to infinite tails of vacua, one has to control the Hodge star $*$ including its subleading coefficients. In the following, we will sketch 
the proof and highlight how the results of section \ref{bulk_boundary} are central in the argument. The mathematical details can be found in a letter by Schnell~\cite{Schnellletter} and an upcoming work \cite{GrimmSchnell}. 
In the following 
we will consider a series $(z_{\rm vac}^I(n), G_4(n))$ of solutions to \eqref{to-show-result} such that one of the complex structure deformations among the $z_{\rm vac}^I(n)$ approaches the boundary. As before we denote this field by $t$ and index the series of vacua by $n=1,...,\infty$, i.e.~we write 
\beq  
     t_n = x_n + i y_n \ , \qquad y_n \rightarrow \infty \ \ \text{for} \ n\rightarrow \infty\ . 
\eeq
while we keep $x_n$ bounded. We also assume that the remaining $z^I_{\rm vac}(n)$ are bounded. These fields play no role in our 
discussion and will be suppressed in the following. To justify this we note that infinite tails of vacua can only be picked up 
if the Hodge star diverges in a direction \cite{Ashok:2003gk} as indicated above. This means that for a one-parameter limit $t_n\rightarrow i \infty$ to 
a co-dimension one boundary, vacua can only accumulate in this direction.

The first step in showing this result is to consider the tadpole bound \eqref{to-show} and use the self-duality 
of $G_4$ to write it using the Hodge norm 
\beq \label{norm-bound}
    K > \langle G_4(n) , G_4(n) \rangle  = \| G_4(n) \|^2  \ .
\eeq
If we are sufficiently close to the boundary the Hodge star becomes increasingly well approximated $C_s$
introduced in \eqref{Cpot_approx}. We thus find that \eqref{norm-bound} leads to the bound $\| G_4(n) \|^2_{\rm s} < K'$
for some $K'$. As above in \eqref{growththeorem} it turns out to be 
convenient to introduce $\rho(n)=  e^{-x_n N^-} G_4(n)$ and to perform the decomposition $ \rho(n) = \sum_{l} \rho_l (n)$ 
as in \eqref{rho-decompN0}.
The bound then reads 
\beq
   K'  >   \sum_l   y^{l}_n \| \rho_l(n) \|_{\infty}^2\ , 
\eeq
which is a sum of positive terms and implies that all summands are bounded, i.e.~we have 
\beq \label{Rl-bounded}
  K'>  \| \tilde R_l (n) \|_{\infty}^2 \ ,\qquad  \tilde R_l(n) = y_n^{\frac{l}{2}} \rho_l(n)\ .  
\eeq
Here we have introduced the shorthand notation 
\beq \label{def-tildeR}
    \tilde R(n) = \sum_l R_l(n)= y_n^{\frac{1}{2}N^0} \rho(n)\ ,
\eeq
which will be useful later on. 
Since by assumption also the axions $x_n$ are bounded, we conclude that $y^{l}_n \| G_l(n) \|_{\infty}^2$ is bounded, where 
$G_l$ are the components of $G_4$ in the $N^0$-decomposition. 
We can now apply the fact that the fluxes are on a lattice and hence cannot become arbitrarily small. Since $y_n \rightarrow \infty$ for $n\rightarrow \infty $ this means 
that starting at some $n'$ the $G_l(n)$ with $l>0$ have to vanish. In other words we  
have shown that 
\beq \label{G4exp_l}
  n\geq n':  \qquad G_4(n) = \sum_{l\leq 0 } G_l (n) \ , \qquad \rho(n) = \sum_{l\leq 0 } \rho_l (n)\ ,  \qquad \tilde R(n) = \sum_{l\leq 0 } \tilde R_l (n)\ , 
\eeq
where in the last two expressions we have used that $N^-$ lowers the $N^0$-eigenvalue. In the remainder of this subsection 
we will show that the series  $\|\rho(n)\|_\infty$ is bounded by using the self-duality condition in \eqref{to-show}. 
If $\|\rho(n)\|_\infty$ is bounded then we conclude from the boundedness of $x_n$ that also $\|G_4(n)\|_\infty$ is bounded, by using 
the same reasoning leading to \eqref{growth2}.
Recall that the norm $\| \cdot \|_\infty$ does not degenerate and hence we can make general statements about the boundedness 
of $G_4(n)$.\footnote{
The norm $\| \cdot \|_\infty$ can degenerate further if we hit another boundary, i.e.~consider a two-parameter limit. This more general 
situation will not be considered here.}  
Together with the fact that $G_4(n)$ takes values on a lattice is then enough to ensure that $G_4(n)$ can 
only take on finitely many values. Recalling that $y_n$ was an arbitrary path towards the boundary we conclude that 
there are no infinite tails towards any codimension-one boundary as we wanted to show. 

\noindent
\textbf{Boundedness of $G_0(n)$.} Let us first check the most straightforward case and show that 
the component $G_0(n)$ in \eqref{G4exp_l} is bounded. This flux has a $y_n$-independent leading term when evaluating $\| G_0(n)\|$. 
We have argued above that all $y^{l}_n \| G_l(n) \|_{\infty}^2$ are bounded 
and hence conclude that $\| G_0(n)\|_\infty$ is bounded. This highlights again that the crucial point is to control the 
degeneration of the Hodge norm $\|\cdot \|$.

\noindent
\textbf{Boundedness in the strict asymptotic limit.} It remains to show that also the $G_l(n)$ with $l<0$ in \eqref{G4exp_l} are bounded.  
Before doing this generally, we will first focus on the situation in which we simply replace $C_{\rm nil}$ with $C_{\rm s}$.
This approximation was called strict asymptotic 
limit in \cite{Grimm:2019ixq} and the following finiteness result was anticipated in \cite{Grimm:2019ixq}. 
The self-duality condition in the leading approximation \eqref{Cpot_approx} then reads
\beq \label{strict-SD}
    C_\infty  \tilde R(n)  = \tilde R(n)  \ ,
\eeq
where we have used the notation \eqref{def-tildeR}.
The $N^0$ components of $\tilde R(n)$ have already be introduced in \eqref{Rl-bounded}.
Using $C_\infty N^0 = - N^0 C_\infty$, which follows from $(N^0)^\dagger = - C^{-1}_\infty N^0 C_\infty= N^0 $ given in \eqref{N0-dagger}, 
we have 
\beq \label{self-l_cond}
    \tilde R_{-l} (n) = \tilde R_l(n)\ ,\qquad  \tilde R_l(n) \equiv  y_n^{ \frac{1}{2} l} \rho_l(n)\ ,
\eeq
where the second equality is a consequence of $N^0 \rho_l = l\, \rho_l$. This fact can now be combined 
with our general statement \eqref{G4exp_l} that for $n\geq n'$ we have $\rho_{l}(n) = 0$ with
$l>0$. This implies that also $\rho_{l} =0$ with $l<0$ if $n\geq n'$. Hence, we have shown
that all $\rho_{l}(n),\ n\geq n'$ vanish unless $l=0$. This implies that $\rho_{l}(n)$ is bounded for all values 
of $l$ and hence that $G_l(n)$ is bounded for all values of $l$.

\noindent 
\textbf{Boundedness for the full expansion.} Let us now turn to the general situation in which the Hodge star near the boundary 
is given by \eqref{Cpol_approx} with \eqref{hy-solutionlater}, when dropping exponentially suppressed corrections. We follow the 
argument of \cite{Schnellletter}.  
In this case we have to control the corrections appearing in the full expansion of $h(y)$ and $h(y)^{-1}$ in \eqref{Cpol_approx}. 
As in the strict asymptotic case we write the self-duality condition as 
\beq \label{selfRn}
   C_\infty R(n) = R(n)\ .
\eeq 
We now have to determine $R(n)$ from the full Weil operator $C_{\rm pol}$. Using \eqref{Cpol_approx} with 
\eqref{hy-solutionlater} we find 
\bea  
   R(n) &=& y_n^{\frac{1}{2} N^0} \Big(1+ \frac{\tilde f_1}{y_n} + \frac{\tilde f_2}{y_n^2} + ...\Big)  \rho(n)  \label{Rn-rho}\\
   &=& y_n^{\frac{1}{2} N^0} \Big(1+ \frac{\tilde f_1}{y_n} + \frac{\tilde f_2}{y_n^2} + ...\Big)  y_n^{-\frac{1}{2} N^0}  \tilde R(n)\ ,
   \label{Rnexp}
\eea
where we have defined $\tilde g_i = e^\zeta g_i e^{-\zeta}$ and $\tilde f_i = e^\zeta f_i e^{-\zeta}$ and used the definition \eqref{def-tildeR} of $\tilde R(n)$. Note that now $\tilde R(n)$ does not have to satisfy \eqref{strict-SD}, since this latter condition is replaced by \eqref{selfRn}.

Now recall from \eqref{def-hatg-hatf} that $\hat f_i = \rho \tilde f_i \rho^{-1}$ and that we  
have argued in section \ref{expand-matter}, equation \eqref{gf-in_Lam}, that these coefficients have the special property that 
$ \hat f_k \in \bigoplus_{p \leq k-1,\, q} \Lambda^{(p)}_q$. This implies that 
when expanding $\hat f_k$ into $L_0$ eigenvectors with $[L_0,\hat f_k^{(l)}] = l\, \hat f_k^{(l)}$ the decomposition reads
$ \hat f_k = \sum_{l\leq k-1} \hat f_k^{(l)}$. 
Rotated back to the real basis and recalling that $L_0 = \rho N^0 \rho^{-1}$, 
we find that $\tilde f_k$  has an expansion   
\beq \label{tildefcond}
   \tilde f_k = \sum_{l\leq k-1} \tilde f_k^{(l)}\ , \qquad [N^0,\tilde f_k^{(l)}] = l\, \tilde f_k^{(l)}\ .
\eeq 
These conditions come into play when evaluating \eqref{Rnexp}. In fact, using \eqref{tildefcond} together with $\text{Ad}_{e^X} = e^{\text{ad} X}$ and a simple re-summation, we find  
\beq
   R(n) = \bigg(1+ \sum_{k\geq 0} \sum_{j=1}^{k+1} \frac{\tilde f_j^{(2j -k- 2)} }{y^{1+\frac{1}{2} k}_n} \bigg)  \tilde R(n)\ .
\eeq
Let us now consider a $N^0$-component $R_l(n)$ with $l>0$ and use \eqref{G4exp_l} to impose that $\tilde R_l(n )= 0,\ l>0$ for 
sufficiently large $n \geq n'$. Hence, we have 
\beq
R_l (n) =  \sum_{k \geq l} \sum_{j=1}^{k+1} \frac{\tilde f_j^{(2j -k- 2)} }{y^{1+\frac{1}{2} k}_n}   \tilde R_{l-(2j -k- 2)} (n)\ .
\eeq
Note that we have used here that $\tilde R_r(n )= 0,\ r>0$ can be used to see that it suffices to consider $l-(2j -k- 2) \leq 0$ in the sum, which then sets the lower bound on the first sum. We can thus extract an overall factor $y^{-1-\frac{1}{2} l}_n$ and use the boundedness \eqref{Rl-bounded} of $\tilde R_l(n)$
to infer the bound 
\beq \label{boundRl}
  \| R_l (n) \|_\infty \leq B\, y^{-1-\frac{1}{2} l}_n \ , 
\eeq
for a sufficiently large $B \geq 0$. In order to control the 
coefficients $R_l$ with $l<0$, we use the self-duality condition \eqref{selfRn}. Using the same argument as for \eqref{self-l_cond}
we know that $R_l = R_{-l}$, and conclude that for all $l \neq 0$ we have the bound $ \| R_l (n) \|_\infty \leq B\, y^{-1-\frac{1}{2}| l| }_n$.
This implies that
\beq \label{boundRl-gen}
  \| y_n^{-\frac{1}{2} N^0} R (n) \|_\infty \leq B' \ ,
\eeq
where all $N^0$-components with $l\neq 0$ fall of with $y^{-1}$, while $R_0(n)$ can contain a constant term. 
With this bound at hand we can use \eqref{Rn-rho} to infer that $\rho(n)$ is also bounded. Indeed, multiplying \eqref{Rn-rho} 
with $y_n^{-\frac{1}{2}N^0}$ the left-hand side is bounded, while the right-hand side contains a factor $1+\sum_i \tilde f_i/y^i_n$ that 
converges to $1$ in the limit $n\rightarrow \infty$. Hence $\rho(n)$ must be bounded as was required to established the above finiteness statement.

\subsection{Comments on the distance conjecture}  \label{distance_comments}

The result \eqref{growththeorem} for the leading asymptotic of $\| F\|$ has been used intensively in \cite{GPV,Grimm:2018cpv,Corvilain:2018lgw} in the study of the distance conjecture. Phrased from the holographic perspective, we can think of $F \in \cH$ as 
defining a state in the boundary theory, that arises at the limit of the considered asymptotic region of moduli space.  $F$ does not evolve with 
time, while $\rho(x)$ has some well-defined `time' dependence. 
Crucially, one finds that the underlying $\mathfrak{sl}(2)$-structure, 
in particular the fact $\cH$ is spanned by complete representations of $\mathfrak{sl}(2)$, 
dictates which states arise at any limit in moduli space. The boundary 
is at infinite distance, if $\cH$ contains a state 
\beq \label{existenced'>0}
    \Omega_\infty  = |\hat d,\hat d;D \rangle\!\rangle \quad \text{with}\quad \hat d>0\ ,
\eeq
within the splitting \eqref{states_threelables}.
To see this we use \eqref{H-identical} to infer that there is only one boundary state $\Omega_\infty$ corresponding to the 
limit of the $(D,0)$-form $\Omega$. This state is the, up to complex rescalings, unique state with 
charge $D$ under the charge operator $Q_\infty$ as inferred from \eqref{Qinfty_onstates}. The integer
$\hat d$ is dependent on the limit that is considered and determined by the 
principle type I, II, III, IV, ... introduced in section \ref{classification} when discussing the classification of 
boundary theories. The condition $\hat d>0$ then ensures that the limit is at infinite distance, as discussed after
\eqref{asymp_WP}.

Following the arguments of \cite{GPV,Grimm:2018cpv} one can now use $\mathfrak{sl}(2)$-representation theory, 
the growth behaviour \eqref{growththeorem}, and the existence of the state \eqref{existenced'>0} to 
identify candidate D-brane charges relevant to satisfy the distance conjecture. 
$L_{1}$, $L_{-1}$ or $N^+$, $N^-$ 
can be used as creation and annihilation operators and raise or lower the growth by one, as inferred from the 
condition \eqref{growththeorem} and the commutation relations \eqref{sl2R-algebra} and \eqref{commutationrelationsL}. In fact, we realize that for any state having 
growth $y^{l}$, with $l>0$ there  always exists a state with growth $y^{-l}$. Intuitively, we can compare this to the fact  
that in string compactifications on a circle there always exist momentum and winding states.\footnote{Applying the 
above construction to $Y_{1} = T^2$, this interpretation can be made concrete.} The distance conjecture was motivated 
in \cite{Ooguri:2006in} by the existence of momentum and winding states in circle compactifications. The constructions 
of \cite{GPV,Grimm:2018cpv} indicate that it is the underlying $\mathfrak{sl}(2)$ structure that persists in any string compactification 
at infinite distance points. The holographic perspective  
attributes the asymptotic behaviour of the field space metric and the masses of states to the existence of a boundary theory with $\mathfrak{sl}(2)$-symmetry. 

Let us stress that \eqref{growththeorem} does not nearly exploit the complete
information about the solution \eqref{hy-solutionlater} that we gathered in sections \ref{boundary_theory} and \ref{bulk_boundary}. 
Specifically, we can include the subleading corrections involving the $g_i$. Inferring their properties \eqref{gn-property} and explicit form 
in terms of the boundary data was central in section \ref{bulk_boundary}. It is thus possible to expand $F$ into 
a general basis $| d, l;q\rangle\!\rangle$ introduced in \eqref{states_threelables} and determine the complete behaviour of $\|F \|$ in 
the near boundary region in terms of the boundary data. Furthermore, we can also give the complete asymptotic 
expansion of the near boundary $(p,q)$-forms, such as $(D,0)$-form $\Omega$. This allows one, for example, 
to derive general expressions for the central charge and hence extend the analysis of \cite{Grimm:2019ixq} beyond leading 
order. It would be interesting to do this in the future.

%%%%%%%%%%%%%%%%%%%%%%%%%%%%%%%%%%%%%%%%%%%%%%%
\section{Conclusions and further discussions}
\label{sec:con}
%%%%%%%%%%%%%%%%%%%%%%%%%%%%%%%%%%%%%%%%%%%%%%%

Motivated by the recent advances in uncovering quantum gravity constraints 
on effective theories, we argued for a holographic approach to study the field spaces and vacua 
of valid effective theories. Several of the swampland conjectures, such as the distance conjecture, 
are constraining the behaviour of effective theories when moving to the asymptotic regions in the scalar 
field space. In string theory compactifications the complex 
structure moduli space of Calabi-Yau manifolds provides a very general example 
of a field space arising in consistent effective theories. Furthermore, it is known, 
that string dualities can relate the asymptotic regions of the complex structure moduli space to 
other field spaces arising, of example, at large volumes of the compactification space 
or at weak string coupling. Our strategy was therefore to extract the general structures arising 
in the asymptotic regime of the complex structure moduli space and view them as 
universal building blocks that should be considered abstractly and independently of their geometric realization
in string compactifications. This was further motivated by the fact that asymptotic Hodge theory 
provides a universal and rich structure that is independent of specific geometric realizations. 

In developing the holographic perspective we have first discussed several 
aspects of a candidate bulk theory living on the moduli space. The matter fields
on the moduli space are real, group valued fields $\hat h$ that act on a Hilbert space that 
is obtained as a complexificantion of a lattice associated to the effective theory. The latter 
can be the charge lattice or lattice of quantized background fluxes when considering a 
string compactification. We have determined the dynamics of $\hat h$ as being given by a set of
field equations and shown that they partly arise from an action principle. We have discussed 
this coupling of $\hat h$ to gravity on the moduli space with the aim to obtain as solutions to the 
gravity-matter system the geometric results arising in Calabi-Yau compactifications. For 
a real two-dimensional moduli space, this requires to go beyond Einstein gravity and 
we have discussed a some of the relevant field equations for the gravity-coupled 
matter system. Furthermore, we have made some first steps in the construction of an action principle. 
Our construction ensured that a particular set of solutions to the bulk theory corresponds 
to the nilpotent orbits and the Weil-Petersson metric after imposing appropriate boundary conditions.
Nilpotent orbits are known to arise at every boundary in the Calabi-Yau moduli space, while the 
the Weil-Petersson metric is known to be the relevant metric in string compactifications on these 
spaces. 

It was a central task of this work to specify boundary conditions that lead to a the set 
of `physical' bulk solutions. Again we have motivated these conditions 
using asymptotic Hodge theory which ensures that such solutions 
can arise from actual geometric compactification, for example, 
on Calabi-Yau manifolds. The solutions admit a constrained asymptotic behaviour of 
the matter fields $\hat h$ and the metric. In particular, the metric has an asymptotic 
$\mathfrak{sl}(2,\bbR)$ isometry, which becomes an $\mathfrak{sl}(2,\bbC)$ symmetry 
acting on a boundary Hilbert space. The boundary Hilbert space splits under this 
symmetry and admits a compatible norm induced by a charge operator $Q_\infty$. 
The operator $Q_\infty$ induces the analog of a standard Hodge decomposition. It is non-trivial 
that such a decomposition can be constructed on the boundary of moduli space, since in the geometric 
setting the associated geometry would be wildly singular. This boundary data allowed us 
to specify three complex commuting operators: $L^2$, $L_0$, from the $\mathfrak{sl}(2,\bbC)$, 
and $Q_\infty$. We used these to introduce quantum numbers for all states and operators. 
 In addition we have specified an 
operator $\hat \delta$, which we termed phase operator, that encodes how the asymptotic isometry group 
of the metric embeds into the  $\mathfrak{sl}(2,\bbC)$  on the boundary. This operator is 
the essential part of the data which is required to perform a matching of a general bulk 
solution to the boundary as we discussed in section \ref{bulk_boundary}. It turns out 
that there is a single matrix condition \eqref{fixall} that fixes the bulk solution uniquely. 
All coefficients in the near boundary expansion of $\hat h$ are then fixed by universal 
non-commutative polynomials in $\hat \delta$ and the $\mathfrak{sl}(2,\bbC)$ generators.  
The properties of these coefficients are constrained and we have shown that they 
are key in several applications. 

It should be stressed that the precise information about the near boundary expansion 
makes the $Sl(2)$ orbit theorem of \cite{Schmid,CKS} so powerful. One of the aims of our presentation was to 
present the crucial parts of it proof as being part of a holographic correspondence. Furthermore, 
we have suggested to study string compactifications more abstractly, by extracting formal algebraic 
structures common to all geometric settings. This adds a new powerful way to infer general properties of the arising 
effective theories without the need to consider specific examples. We stress, however, that there are numerous 
open questions in developing the holographic correspondence further. Firstly, it would be desirable to 
find a complete action principle for the bulk theory. Secondly, we expect that the discussion of section \ref{boundary_theory}
is only part of a more involved story about the construction of a boundary theory. Most striking would be 
to find a boundary theory dynamically encoding attainable values for the phase operator.  
Thirdly, even on the level studied here one might wonder if the holographic perspective can be generalized to 
higher-dimensional moduli spaces with intersecting boundary components. In mathematics this is part of the multi-variable 
$Sl(2)$-orbit theorem which comes with several additional complications. Eventually one might hope to formulate 
a theory globally on the boundary. Within such a theory many aspects of the bulk physics should have boundary counterparts
and one might hope for developing a dictionary for a complete bulk reconstruction in simple examples.

In the final part of this work we have discussed two finiteness results that use the existence of an 
$Sl(2)$-structure on the boundary and the corresponding asymptotic form of the bulk solutions. 
Firstly, we have pointed out that a famous theorem of Cattani, Deligne, and Kaplan \cite{CDK} implies 
the finiteness of supersymmetric flux vacua in the intensively studied F-theory compactifications with 
$G_4$ fluxes and their Type IIB analogues. The crucial task achieved in \cite{CDK} is to show 
that the tadpole constraint ensures that there 
are no infinite tails of vacua near any boundary of any co-dimension.  
It is well-known that a similar statement can be shown by using the Hodge conjecture, which 
makes the study of the results of \cite{CDK} into an active field of mathematical research. 
Secondly, we have then shown that finiteness persists, at least near co-dimension one 
boundaries, when considering fluxes that are self-dual and not necessarily supersymmetric. 
The argument uses the properties of the coefficients in the $1/y$-expansion of the near boundary 
solution in order to constrain the behaviour of the Hodge star.  We expect that this proof can 
be extended to all co-dimensions \cite{GrimmSchnell}, hence showing finiteness of self-dual flux vacua in full generality. 
It should be clear, however, that several new difficulties have 
to be overcome, as it was done in the general proof of \cite{CDK}, which are connected to a having a much wider 
range of possibilities to pick a path towards the boundary. 

The discussed finiteness results are of physical 
importance in judging the predictive power of string theory. A direct application includes the 
recent constructions of \cite{Demirtas:2019sip,Demirtas:2020ffz,Blumenhagen:2020ire}, in which self-dual fluxes inducing an exponentially 
small superpotential were introduced. Mathematically, such vacua describe certain extended loci of Hodge classes \cite{schnell2014extended}
and such extended loci were shown to be finite. Our arguments show that this finiteness persists even further, e.g.~when the 
fluxes are self-dual but their induced superpotential is not exponentially small. It should be noted that the insights from the 
poofs of the finiteness results seem even more useful than the final statement. In particular, one learns new methods to 
control certain feature of scalar potentials along all possible paths in field space. Furthermore, one concretely sees 
why arbitrary fine-tuning is structurally prohibited. This might help, for example, to give evidence for the conjectures on moduli 
stabilization recently put forward in \cite{Bena:2020xrh}.

Let us close with a further speculation on how the findings of this work might 
yield a deeper understanding of the landscape of effective theories consistent with quantum gravity. 
It is natural to formulate a swampland criterium that states that every
effective theory containing scalar fields admits a sector that can be described holographically with 
a boundary theory based on the described $Sl(2)$-data. 
Such a proposal will then imply the following statements:
\vspace{-.3cm}
\begin{itemize} 
\item Considering any path towards an infinite distance boundary, 
a continuous global symmetry becomes approximately exact that stems from a unipotent monodromy symmetry
and hence can be encoded by a nilpotent $N^-$.
\vspace{-.1cm}

\item Associated to each limit there exists a lattice $\cL$
and a Hilbert space $\cH = \bbC \otimes \cL$ with an action of $N^-$ completed into $\mathfrak{sl}(2,\bbR)$.
The definition of $\cH$ and $\mathfrak{sl}(2,\bbR)$ are such that
(1) the positivity constraints on the effective couplings can be encoded using the norm on 
$\cH$, and (2) the growth of the 
effective couplings in the fields sent to the limit are dictated by the $\mathfrak{sl}(2,\bbR)$ weights. 
\end{itemize}
\vspace{-.3cm}
Note that these statements essentially manifest the observation that there are universal constraints 
from positivity and the existence of global symmetry. The proposal thus claims that the structures discussed in 
this work are universally present. It appears to be consistent with the recent conjectures  
put forward in \cite{GPV}, \cite{Lee:2019wij}, \cite{Lanza:2020qmt,Lanzatoappear} and the observations made in \cite{Andriolo:2020lul}. 
In particular, refs.~\cite{Lanza:2020qmt,Lanzatoappear} view infinite distance limits as RG flows of 
strings, which seems nicely compatible with the holographic perspective outlined here.  
%Two comments are in order to clarify the status of the above proposal further.  
%Firstly, we have discussed how monodromy symmetries 
%act on scalars, but taking a more general viewpoint these symmetries can arise from an action on  
%the various form fields in the effective theory, see e.g.~\cite{Corvilain:2018lgw,Lanza:2020qmt}. Secondly, note that in this work 
%and in the above proposal we consider the 
%splitting of a complex vector space $\cH$. Considering monodromy and splittings of a discrete space is, 
%at least mathematically, significantly more involved. 
It would be exciting if one could develop this holographic view on the string theory landscape of 
effective theories further and show that many its constraining properties manifest themselves on 
its boundaries.

\subsubsection*{Acknowledgments}

It is a great pleasure to thank Tarek Anous, Brice Bastian, Chris Couzens, Umut G\"ursoy, Damian van de Heisteeg, Chongchuo Li, Jeroen Monnee, Miguel Montero, Eran Palti, Erik Plauschinn, Colleen Robles, Christian Schnell, 
Cumrun Vafa, and Irene Valenzuela for very useful discussions and correspondence. 
I am particularly grateful to Christian Schnell for letting me report some upcoming mathematical results on finiteness. My research 
is partly supported by the Dutch Research Council (NWO) via a Start-Up grant and a VICI grant. 
Some early parts of this work were completed at the KITP, Santa Barbara, and therefore supported in part by 
the National Science Foundation under Grant No.~NSF PHY-1748958.

\appendix

\section{Computing the phase operator for a nilpotent orbit} \label{computedelta}

In the following we will describe how to determine the phase operator $\delta$ 
for a given one-parameter nilpotent orbit $F_{\rm pol}^p = e^{t N^-} F_0^p$. 
In contrast to the rest of the paper we will introduce in the following also the 
monodromy weight filtration $W_i$ induced by $N^-$.
In fact, each nilpotent matrix $N^-$ acting on $H^{D}(Y_D,\bbR)$ defines a unique set of real 
vector spaces $W_k(N^-)$ of weight $D$ with 
\beq
  0 \subs W_0 \subs W_1 \subs \cdots \subs W_{2D}= H^{D}(Y_D,\bbR)\ ,
\eeq
such that for all $k$ one has
\beq
  N^- W_k      \    \subs                           \ W_{k - 2}\ ,\qquad (N^-)^k: \quad Gr_{2D + k} \  \cong\   Gr_{2D - k}\ ,
\eeq
where $Gr_{k}= \frac{W_{k} }{W_{k-1} } $. The symbol $\cong$ indicates that $N^k$ is an isomorphism. 

The set of vector spaces $W^\bbC_k = W_k \otimes \bbC $ together with $F_0^p$ 
can now be used to define the so-called Deligne splitting 
by setting 
\beq \label{def-Ipq}
   I^{p,q} = F^p_0 \cap W^\bbC_{p+q} \cap \Big( \bar F^{q}_0 \cap W^\bbC_{p+q} + \sum_{j \geq 1} \bar F^{q-j}_0 \cap W^\bbC_{p+q-j-1}  \Big)\ .
\eeq
The $I^{p,q}$ define the unique splitting satisfying  
\beq  \label{Fp-Wi_split} 
  F^{p}_0  = \bigoplus_{r\geq p} \bigoplus_s I^{r,s} \ ,\qquad W^\bbC_{l} = \bigoplus_{p+q \leq l} I^{p,q} \ ,  \qquad
     \overline{I^{p,q}} = I^{q,p} \ \text{mod}\ \bigoplus_{r < q,s<p} I^{r,s}\ .
\eeq
The most crucial point here, is that in general one does \textit{not} find that 
$ \overline{I^{p,q}} = I^{q,p} $. We can now define  vector spaces $V_l$ and a semisimple grading operator
\beq
  V_l^\bbC  = \bigoplus_{p + q = l} I^{p, q}\ , \qquad T v_l = l v_l \ \ \text{for}\ \ v_l \in V_l^\bbC\ .
\eeq
The operator $T$ should be compared with the operator $N^0$ introduced in section \ref{boundary_theory}. However,
while $\bar N^0=N^0$, this is not necessarily the case for $T$, since in general $ \bar V_l^\bbC  \neq  V_l^\bbC $.
In other words, there is in general no real slice in $V^\bbC_l$ on which $T$ acts as $l$ and 
which yields the space $V_l^\bbC$ as complexification. This is in stark contrast to the vector spaces \eqref{V-decomposition-c}
introduced on the boundary. We will now describe that there how to construct the unique rotation of $T$, such that 
a real split exists. 

To being with, let $\conj{T}$ be the complex conjugate of the grading operator $T$ defined by  $\conj{T}(v) := \conj{T(\conj{v})}$, 
for all $v \in H^{D}(Y_D,\bbC)$. One can now show \cite{CKS} that $\conj{T}$ and $T$ are related by 
\beq \label{Tconj}
  \conj{T} = e^{-2\im\delta} T e^{2\im\delta},
\eeq
where the real operator $\delta$ acts on $I^{p, q}$ by decreasing $p,q$, i.e.~
\beq \label{location-delta}
  \delta(I^{p, q}) \subs \bigoplus_{\substack{r < p\\s < q}} I^{r, s}\ , 
\eeq
holding for all $p, q$. Requiring that $\delta \in \mathfrak{g}_{\bbR}$ and that $[N^-,\delta]=0$, one shows that there is 
a unique operator $\delta$ satisfying \eqref{Tconj} and \eqref{location-delta} (see Proposition 2.20 of \cite{CKS} for details). 
For any given nilpotent orbit we can thus compute the unique charge operator $\delta$. Simple example for such a computation 
can be found e.g.~in \cite{Grimm:2018cpv}.

\bibliographystyle{jhep}
\bibliography{Holo_Bib}

\providecommand{\href}[2]{#2}\begingroup\raggedright\begin{thebibliography}{10}

\bibitem{Palti:2019pca}
E.~Palti, \emph{{The Swampland: Introduction and Review}},
  \href{https://doi.org/10.1002/prop.201900037}{\emph{Fortsch. Phys.}
  {\bfseries 67} (2019) 1900037}
  [\href{https://arxiv.org/abs/1903.06239}{{\ttfamily 1903.06239}}].

\bibitem{Ooguri:2006in}
H.~Ooguri and C.~Vafa, \emph{{On the Geometry of the String Landscape and the
  Swampland}},
  \href{https://doi.org/10.1016/j.nuclphysb.2006.10.033}{\emph{Nucl. Phys.}
  {\bfseries B766} (2007) 21}
  [\href{https://arxiv.org/abs/hep-th/0605264}{{\ttfamily hep-th/0605264}}].

\bibitem{GPV}
T.~W. Grimm, E.~Palti and I.~Valenzuela, \emph{{Infinite Distances in Field
  Space and Massless Towers of States}},
  \href{https://doi.org/10.1007/JHEP08(2018)143}{\emph{JHEP} {\bfseries 08}
  (2018) 143} [\href{https://arxiv.org/abs/1802.08264}{{\ttfamily
  1802.08264}}].

\bibitem{Lee:2018urn}
S.-J. Lee, W.~Lerche and T.~Weigand, \emph{{Tensionless Strings and the Weak
  Gravity Conjecture}},
  \href{https://doi.org/10.1007/JHEP10(2018)164}{\emph{JHEP} {\bfseries 10}
  (2018) 164} [\href{https://arxiv.org/abs/1808.05958}{{\ttfamily
  1808.05958}}].

\bibitem{Lee:2018spm}
S.-J. Lee, W.~Lerche and T.~Weigand, \emph{{A Stringy Test of the Scalar Weak
  Gravity Conjecture}},
  \href{https://doi.org/10.1016/j.nuclphysb.2018.11.001}{\emph{Nucl. Phys.}
  {\bfseries B938} (2019) 321}
  [\href{https://arxiv.org/abs/1810.05169}{{\ttfamily 1810.05169}}].

\bibitem{Grimm:2018cpv}
T.~W. Grimm, C.~Li and E.~Palti, \emph{{Infinite Distance Networks in Field
  Space and Charge Orbits}},
  \href{https://doi.org/10.1007/JHEP03(2019)016}{\emph{JHEP} {\bfseries 03}
  (2019) 016} [\href{https://arxiv.org/abs/1811.02571}{{\ttfamily
  1811.02571}}].

\bibitem{Corvilain:2018lgw}
P.~Corvilain, T.~W. Grimm and I.~Valenzuela, \emph{{The Swampland Distance
  Conjecture for K{\"a}hler moduli}},
  \href{https://doi.org/10.1007/JHEP08(2019)075}{\emph{JHEP} {\bfseries 08}
  (2019) 075} [\href{https://arxiv.org/abs/1812.07548}{{\ttfamily
  1812.07548}}].

\bibitem{Font:2019cxq}
A.~Font, A.~Herr{\'a}ez and L.~E. Ib{\'a}{\~n}ez, \emph{{The Swampland Distance
  Conjecture and Towers of Tensionless Branes}},
  \href{https://doi.org/10.1007/JHEP08(2019)044}{\emph{JHEP} {\bfseries 08}
  (2019) 044} [\href{https://arxiv.org/abs/1904.05379}{{\ttfamily
  1904.05379}}].

\bibitem{Lee:2019xtm}
S.-J. Lee, W.~Lerche and T.~Weigand, \emph{{Emergent Strings, Duality and Weak
  Coupling Limits for Two-Form Fields}},
  \href{https://arxiv.org/abs/1904.06344}{{\ttfamily 1904.06344}}.

\bibitem{Marchesano:2019ifh}
F.~Marchesano and M.~Wiesner, \emph{{Instantons and infinite distances}},
  \href{https://doi.org/10.1007/JHEP08(2019)088}{\emph{JHEP} {\bfseries 08}
  (2019) 088} [\href{https://arxiv.org/abs/1904.04848}{{\ttfamily
  1904.04848}}].

\bibitem{Grimm:2019wtx}
T.~W. Grimm and D.~Van De~Heisteeg, \emph{{Infinite Distances and the Axion
  Weak Gravity Conjecture}},
  \href{https://doi.org/10.1007/JHEP03(2020)020}{\emph{JHEP} {\bfseries 03}
  (2020) 020} [\href{https://arxiv.org/abs/1905.00901}{{\ttfamily
  1905.00901}}].

\bibitem{Lee:2019wij}
S.-J. Lee, W.~Lerche and T.~Weigand, \emph{{Emergent Strings from Infinite
  Distance Limits}},  \href{https://arxiv.org/abs/1910.01135}{{\ttfamily
  1910.01135}}.

\bibitem{Baume:2019sry}
F.~Baume, F.~Marchesano and M.~Wiesner, \emph{{Instanton Corrections and
  Emergent Strings}},
  \href{https://doi.org/10.1007/JHEP04(2020)174}{\emph{JHEP} {\bfseries 04}
  (2020) 174} [\href{https://arxiv.org/abs/1912.02218}{{\ttfamily
  1912.02218}}].

\bibitem{Enriquez-Rojo:2020pqm}
M.~Enr\'\i{}quez~Rojo and E.~Plauschinn, \emph{{Swampland conjectures for type
  IIB orientifolds with closed-string U(1)s}},
  \href{https://doi.org/10.1007/JHEP07(2020)026}{\emph{JHEP} {\bfseries 07}
  (2020) 026} [\href{https://arxiv.org/abs/2002.04050}{{\ttfamily
  2002.04050}}].

\bibitem{Gendler:2020dfp}
N.~Gendler and I.~Valenzuela, \emph{{Merging the Weak Gravity and Distance
  Conjectures Using BPS Extremal Black Holes}},
  \href{https://arxiv.org/abs/2004.10768}{{\ttfamily 2004.10768}}.

\bibitem{Heidenreich:2020ptx}
B.~Heidenreich and T.~Rudelius, \emph{{Infinite Distance and Zero Gauge
  Coupling in 5d Supergravity}},
  \href{https://arxiv.org/abs/2007.07892}{{\ttfamily 2007.07892}}.

\bibitem{Banks:1988yz}
T.~Banks and L.~J. Dixon, \emph{{Constraints on String Vacua with Space-Time
  Supersymmetry}},
  \href{https://doi.org/10.1016/0550-3213(88)90523-8}{\emph{Nucl. Phys. B}
  {\bfseries 307} (1988) 93}.

\bibitem{Banks:2010zn}
T.~Banks and N.~Seiberg, \emph{{Symmetries and Strings in Field Theory and
  Gravity}}, \href{https://doi.org/10.1103/PhysRevD.83.084019}{\emph{Phys. Rev.
  D} {\bfseries 83} (2011) 084019}
  [\href{https://arxiv.org/abs/1011.5120}{{\ttfamily 1011.5120}}].

\bibitem{Grimm:2019ixq}
T.~W. Grimm, C.~Li and I.~Valenzuela, \emph{{Asymptotic Flux Compactifications
  and the Swampland}},
  \href{https://doi.org/10.1007/JHEP06(2020)009}{\emph{JHEP} {\bfseries 06}
  (2020) 009} [\href{https://arxiv.org/abs/1910.09549}{{\ttfamily
  1910.09549}}].

\bibitem{AST_1989__179-180__67_0}
E.~Cattani and A.~Kaplan, \emph{{Degenerating variations of Hodge structure}},
  in \emph{Th\'eorie de Hodge - Luminy, Juin 1987} (B.~D., E.~H., E.~F.,
  V.~Jean-Louis and V.~E., eds.), no.~179-180 in Ast\'erisque, pp.~67--96.
\newblock Soci\'et\'e math\'ematique de France, 1989.

\bibitem{HodgeTheoryMN49}
E.~Cattani, F.~E. Zein, P.~A. Griffiths and L.~D. Tr\'ang, \emph{Hodge Theory
  (MN-49)}. Princeton University Press, Princeton, 2014.

\bibitem{Hori:2003ic}
K.~Hori, S.~Katz, A.~Klemm, R.~Pandharipande, R.~Thomas, C.~Vafa et~al.,
  \emph{{Mirror symmetry}}, vol.~1 of \emph{Clay mathematics monographs}. AMS,
  Providence, USA, 2003.

\bibitem{Kerr2019HodgeTO}
M.~Kerr and R.~Laza, \emph{Hodge theory of degenerations, (i): Consequences of
  the decomposition theorem},
  \href{https://arxiv.org/abs/1901.01896}{{\ttfamily 1901.01896}}.

\bibitem{Kerr2020HodgeTO}
M.~Kerr and R.~Laza, \emph{Hodge theory of degenerations, (ii): vanishing
  cohomology and geometric applications.},
  \href{https://arxiv.org/abs/2006.03953}{{\ttfamily 2006.03953}}.

\bibitem{wang1}
C.-L. Wang, \emph{{On the incompleteness of the Weil-Petersson metric along
  degenerations of Calabi-Yau manifolds}}, {\emph{Mathematical Research
  Letters} {\bfseries 4} (1997) 157}.

\bibitem{Schmid}
W.~Schmid, \emph{{Variation of Hodge structure: the singularities of the period
  mapping}}, \href{https://doi.org/10.1007/BF01389674}{\emph{Invent. Math. ,
  22:211--319, 1973} }.

\bibitem{CKS}
E.~Cattani, A.~Kaplan and W.~Schmid, \emph{{Degeneration of Hodge Structures}},
  \href{https://doi.org/10.2307/1971333}{\emph{Annals of Mathematics}
  {\bfseries 123} (1986) 457}.

\bibitem{donaldson1984}
S.~K. Donaldson, \emph{Nahm's equations and the classification of monopoles},
  {\emph{Comm. Math. Phys.} {\bfseries 96} (1984) 387}.

\bibitem{Cecotti:1991me}
S.~Cecotti and C.~Vafa, \emph{{Topological antitopological fusion}},
  \href{https://doi.org/10.1016/0550-3213(91)90021-O}{\emph{Nucl. Phys. B}
  {\bfseries 367} (1991) 359}.

\bibitem{Cecotti:2020rjq}
S.~Cecotti, \emph{{Special Geometry and the Swampland}},
  \href{https://arxiv.org/abs/2004.06929}{{\ttfamily 2004.06929}}.

\bibitem{Cecotti:2020uek}
S.~Cecotti, \emph{{Moduli spaces of Calabi-Yau $d$-folds as
  gravitational-chiral instantons}},
  \href{https://arxiv.org/abs/2007.09992}{{\ttfamily 2007.09992}}.

\bibitem{DeligneRHS}
P.~Deligne, \emph{{Structures de Hodge mixtes r\'eelles}}, {\emph{Proc. Sympos.
  Pure Math.} {\bfseries 55} (1994) 509}.

\bibitem{Lanza:2020qmt}
S.~Lanza, F.~Marchesano, L.~Martucci and I.~Valenzuela, \emph{{Swampland
  Conjectures for Strings and Membranes}},
  \href{https://arxiv.org/abs/2006.15154}{{\ttfamily 2006.15154}}.

\bibitem{Lanzatoappear}
S.~Lanza, F.~Marchesano, L.~Martucci and I.~Valenzuela, \emph{{to appear}},
  {\emph{2020} }.

\bibitem{Grana:2005jc}
M.~Gra{\~n}a, \emph{{Flux compactifications in string theory: A Comprehensive
  review}}, \href{https://doi.org/10.1016/j.physrep.2005.10.008}{\emph{Phys.
  Rept.} {\bfseries 423} (2006) 91}
  [\href{https://arxiv.org/abs/hep-th/0509003}{{\ttfamily hep-th/0509003}}].

\bibitem{Douglas:2006es}
M.~R. Douglas and S.~Kachru, \emph{{Flux compactification}},
  \href{https://doi.org/10.1103/RevModPhys.79.733}{\emph{Rev. Mod. Phys.}
  {\bfseries 79} (2007) 733}
  [\href{https://arxiv.org/abs/hep-th/0610102}{{\ttfamily hep-th/0610102}}].

\bibitem{Ashok:2003gk}
S.~Ashok and M.~R. Douglas, \emph{{Counting flux vacua}},
  \href{https://doi.org/10.1088/1126-6708/2004/01/060}{\emph{JHEP} {\bfseries
  01} (2004) 060} [\href{https://arxiv.org/abs/hep-th/0307049}{{\ttfamily
  hep-th/0307049}}].

\bibitem{Denef:2004ze}
F.~Denef and M.~R. Douglas, \emph{{Distributions of flux vacua}},
  \href{https://doi.org/10.1088/1126-6708/2004/05/072}{\emph{JHEP} {\bfseries
  05} (2004) 072} [\href{https://arxiv.org/abs/hep-th/0404116}{{\ttfamily
  hep-th/0404116}}].

\bibitem{Schnellletter}
C.~Schnell, \emph{{Letter to T.~Grimm}}, {\emph{2020} }.

\bibitem{GrimmSchnell}
T.~W. Grimm and C.~Schnell, \emph{{in preparation}}, {\emph{2021} }.

\bibitem{CDK}
E.~Cattani, P.~Deligne and A.~Kaplan, \emph{{On the locus of Hodge classes}},
  \href{https://doi.org/10.1090/S0894-0347-1995-1273413-2}{\emph{Jour. A.M.S.}
  {\bfseries 8-2} (1995) 483}
  [\href{https://arxiv.org/abs/alg-geom/9402009}{{\ttfamily
  alg-geom/9402009}}].

\bibitem{Hironaka}
H.~Hironaka, \emph{{Resolution of Singularities of an Algebraic Variety Over a
  Field of Characteristic Zero: I}}, {\emph{Ann. of Math.} {\bfseries 79}
  (1964) 109}.

\bibitem{Viehweg}
E.~Viehweg, \emph{Quasi-projective Moduli for Polarized Manifolds}, Ergebnisse
  der Mathematik und ihrer Grenzgebiete. 3. Folge / A Series of Modern Surveys
  in Mathematics. Springer Berlin Heidelberg, 1995.

\bibitem{Palti:2020qlc}
E.~Palti, C.~Vafa and T.~Weigand, \emph{{Supersymmetric Protection and the
  Swampland}}, \href{https://doi.org/10.1007/JHEP06(2020)168}{\emph{JHEP}
  {\bfseries 06} (2020) 168}
  [\href{https://arxiv.org/abs/2003.10452}{{\ttfamily 2003.10452}}].

\bibitem{BastianGrimmHeisteeg}
B.~Bastian, T.~W. Grimm and D.~v.~d. Heisteeg, \emph{{in preparation}},
  {\emph{2021} }.

\bibitem{robles_2016}
C.~Robles, \emph{Classification of horizontal $\text{SL}(2)$s},
  {\emph{Compositio Mathematica} {\bfseries 152} (2016) 918}
  [\href{https://arxiv.org/abs/arXiv:1405.3163}{{\ttfamily arXiv:1405.3163}}].

\bibitem{Kerr2017}
M.~{Kerr}, G.~J. {Pearlstein} and C.~{Robles}, \emph{{Polarized relations on
  horizontal \(\operatorname{SL}(2)\)'s.}},
  \href{https://doi.org/10.25537/dm.2019v24.1295-1360}{\emph{{Doc. Math.}}
  {\bfseries 24} (2019) 1295}.

\bibitem{Lu99}
Z.~Lu, \emph{{On the geometry of classifying spaces and horizontal slices}},
  {\emph{Amer. J. Math.} {\bfseries 121} (1999) 177 }.

\bibitem{Lu01}
Z.~Lu, \emph{{On the Hodge metric of the universal deformation space of
  Calabi-Yau threefolds}}, {\emph{J. Geom. Anal.} {\bfseries 11} (2001) 103}.

\bibitem{lu_sun_2004}
Z.~Lu and X.~Sun, \emph{{Weil-Petersson geometry on moduli space of polarized
  Calabi-Yau manifolds}}, {\emph{J. Inst. of Math. Jussieu} {\bfseries 3}
  (2004) 185 }.

\bibitem{LuFang05}
H.~Fang and Z.~Lu, \emph{{Generalized Hodge metrics and BCOV torsion on
  Calabi-Yau moduli}}, {\emph{J. reine und angewandte Math.} {\bfseries 588}
  (2005) 49}.

\bibitem{Douglas:2006zj}
M.~Douglas and Z.~Lu, \emph{{On the geometry of moduli space of polarized
  Calabi-Yau manifolds}},  \href{https://arxiv.org/abs/math/0603414}{{\ttfamily
  math/0603414}}.

\bibitem{Lu:2009aw}
Z.~Lu and M.~R. Douglas, \emph{{Gauss-Bonnet-Chern theorem on moduli space}},
  \href{https://doi.org/10.1007/s00208-013-0907-4}{\emph{Math. Ann.} {\bfseries
  357} (2013) 469} [\href{https://arxiv.org/abs/0902.3839}{{\ttfamily
  0902.3839}}].

\bibitem{PearlsteinPeters}
C.~Peters and G.~Pearlstein, \emph{Differential geometry of the mixed hodge
  metric}, {\emph{Communications in Analysis and Geometry} (2017) }
  [\href{https://arxiv.org/abs/1407.4082}{{\ttfamily 1407.4082}}].

\bibitem{Bershadsky:1993cx}
M.~Bershadsky, S.~Cecotti, H.~Ooguri and C.~Vafa, \emph{{Kodaira-Spencer theory
  of gravity and exact results for quantum string amplitudes}},
  \href{https://doi.org/10.1007/BF02099774}{\emph{Commun. Math. Phys.}
  {\bfseries 165} (1994) 311}
  [\href{https://arxiv.org/abs/hep-th/9309140}{{\ttfamily hep-th/9309140}}].

\bibitem{lu2005hodge}
Z.~Lu, \emph{{On the Hodge Metric of the Universal Deformation Space of
  Calabi-Yau Threefolds}}, {\emph{J. Geom. Anal.} {\bfseries 11} (2005) 103}
  [\href{https://arxiv.org/abs/math/0505582}{{\ttfamily math/0505582}}].

\bibitem{GrimmHeisteegMonee}
T.~W. Grimm, D.~v.~d. Heisteeg and J.~Monnee, \emph{Bulk reconstruction in
  moduli space holography}, {\emph{2021} }.

\bibitem{hitchin1983}
N.~J. Hitchin, \emph{On the construction of monopoles}, {\emph{Comm. Math.
  Phys.} {\bfseries 89} (1983) 145}.

\bibitem{Teitelboim:1983ux}
C.~Teitelboim, \emph{{Gravitation and Hamiltonian Structure in Two Space-Time
  Dimensions}}, \href{https://doi.org/10.1016/0370-2693(83)90012-6}{\emph{Phys.
  Lett. B} {\bfseries 126} (1983) 41}.

\bibitem{Jackiw:1984je}
R.~Jackiw, \emph{{Lower Dimensional Gravity}},
  \href{https://doi.org/10.1016/0550-3213(85)90448-1}{\emph{Nucl. Phys. B}
  {\bfseries 252} (1985) 343}.

\bibitem{Grumiller:2002nm}
D.~Grumiller, W.~Kummer and D.~Vassilevich, \emph{{Dilaton gravity in
  two-dimensions}},
  \href{https://doi.org/10.1016/S0370-1573(02)00267-3}{\emph{Phys. Rept.}
  {\bfseries 369} (2002) 327}
  [\href{https://arxiv.org/abs/hep-th/0204253}{{\ttfamily hep-th/0204253}}].

\bibitem{Grumiller:2006rc}
D.~Grumiller and R.~Meyer, \emph{{Ramifications of lineland}}, {\emph{Turk. J.
  Phys.} {\bfseries 30} (2006) 349}
  [\href{https://arxiv.org/abs/hep-th/0604049}{{\ttfamily hep-th/0604049}}].

\bibitem{deAlfaro:1976vlx}
V.~de~Alfaro, S.~Fubini and G.~Furlan, \emph{{Conformal Invariance in Quantum
  Mechanics}}, \href{https://doi.org/10.1007/BF02785666}{\emph{Nuovo Cim. A}
  {\bfseries 34} (1976) 569}.

\bibitem{Chamon:2011xk}
C.~Chamon, R.~Jackiw, S.-Y. Pi and L.~Santos, \emph{{Conformal quantum
  mechanics as the CFT$_1$ dual to AdS$_2$}},
  \href{https://doi.org/10.1016/j.physletb.2011.06.023}{\emph{Phys. Lett. B}
  {\bfseries 701} (2011) 503}
  [\href{https://arxiv.org/abs/1106.0726}{{\ttfamily 1106.0726}}].

\bibitem{Anninos:2019oka}
D.~Anninos, D.~M. Hofman and J.~Kruthoff, \emph{{Charged Quantum Fields in
  AdS$_2$}}, \href{https://doi.org/10.21468/SciPostPhys.7.4.054}{\emph{SciPost
  Phys.} {\bfseries 7} (2019) 054}
  [\href{https://arxiv.org/abs/1906.00924}{{\ttfamily 1906.00924}}].

\bibitem{Grimm:2019bey}
T.~W. Grimm, F.~Ruehle and D.~van~de Heisteeg, \emph{{Classifying Calabi-Yau
  threefolds using infinite distance limits}},
  \href{https://arxiv.org/abs/1910.02963}{{\ttfamily 1910.02963}}.

\bibitem{Skenderis:2002wp}
K.~Skenderis, \emph{{Lecture notes on holographic renormalization}},
  \href{https://doi.org/10.1088/0264-9381/19/22/306}{\emph{Class. Quant. Grav.}
  {\bfseries 19} (2002) 5849}
  [\href{https://arxiv.org/abs/hep-th/0209067}{{\ttfamily hep-th/0209067}}].

\bibitem{DeJonckheere:2017qkk}
T.~De~Jonckheere, \emph{{Modave lectures on bulk reconstruction in AdS/CFT}},
  \href{https://doi.org/10.22323/1.323.0005}{\emph{PoS} {\bfseries Modave2017}
  (2018) 005} [\href{https://arxiv.org/abs/1711.07787}{{\ttfamily
  1711.07787}}].

\bibitem{Harlow:2018fse}
D.~Harlow, \emph{{TASI Lectures on the Emergence of Bulk Physics in AdS/CFT}},
  \href{https://doi.org/10.22323/1.305.0002}{\emph{PoS} {\bfseries TASI2017}
  (2018) 002} [\href{https://arxiv.org/abs/1802.01040}{{\ttfamily
  1802.01040}}].

\bibitem{Eguchi:2005eh}
T.~Eguchi and Y.~Tachikawa, \emph{{Distribution of flux vacua around singular
  points in Calabi-Yau moduli space}},
  \href{https://doi.org/10.1088/1126-6708/2006/01/100}{\emph{JHEP} {\bfseries
  01} (2006) 100} [\href{https://arxiv.org/abs/hep-th/0510061}{{\ttfamily
  hep-th/0510061}}].

\bibitem{Braun:2011av}
A.~P. Braun, N.~Johansson, M.~Larfors and N.-O. Walliser, \emph{{Restrictions
  on infinite sequences of type IIB vacua}},
  \href{https://doi.org/10.1007/JHEP10(2011)091}{\emph{JHEP} {\bfseries 10}
  (2011) 091} [\href{https://arxiv.org/abs/1108.1394}{{\ttfamily 1108.1394}}].

\bibitem{Duff:1995wd}
M.~J. Duff, J.~T. Liu and R.~Minasian, \emph{{Eleven-dimensional origin of
  string-string duality: A One loop test}},
  \href{https://doi.org/10.1016/0550-3213(95)00368-3}{\emph{Nucl. Phys.}
  {\bfseries B452} (1995) 261}
  [\href{https://arxiv.org/abs/hep-th/9506126}{{\ttfamily hep-th/9506126}}].

\bibitem{Sethi:1996es}
S.~Sethi, C.~Vafa and E.~Witten, \emph{{Constraints on low dimensional string
  compactifications}},
  \href{https://doi.org/10.1016/S0550-3213(96)00483-X}{\emph{Nucl. Phys.}
  {\bfseries B480} (1996) 213}
  [\href{https://arxiv.org/abs/hep-th/9606122}{{\ttfamily hep-th/9606122}}].

\bibitem{Becker:1996gj}
K.~Becker and M.~Becker, \emph{{M-theory on eight manifolds}},
  \href{https://doi.org/10.1016/0550-3213(96)00367-7}{\emph{Nucl. Phys.}
  {\bfseries B477} (1996) 155}
  [\href{https://arxiv.org/abs/hep-th/9605053}{{\ttfamily hep-th/9605053}}].

\bibitem{Dasgupta:1999ss}
K.~Dasgupta, G.~Rajesh and S.~Sethi, \emph{{M-theory, orientifolds and
  G-flux}}, \href{https://doi.org/10.1088/1126-6708/1999/08/023}{\emph{JHEP}
  {\bfseries 08} (1999) 023}
  [\href{https://arxiv.org/abs/hep-th/9908088}{{\ttfamily hep-th/9908088}}].

\bibitem{Witten:1996md}
E.~Witten, \emph{{On flux quantization in M theory and the effective action}},
  \href{https://doi.org/10.1016/S0393-0440(96)00042-3}{\emph{J. Geom. Phys.}
  {\bfseries 22} (1997) 1}
  [\href{https://arxiv.org/abs/hep-th/9609122}{{\ttfamily hep-th/9609122}}].

\bibitem{Grimm:2014efa}
T.~W. Grimm, T.~G. Pugh and M.~Weissenbacher, \emph{{The effective action of
  warped M-theory reductions with higher derivative terms --- part I}},
  \href{https://doi.org/10.1007/JHEP01(2016)142}{\emph{JHEP} {\bfseries 01}
  (2016) 142} [\href{https://arxiv.org/abs/1412.5073}{{\ttfamily 1412.5073}}].

\bibitem{Grimm:2015mua}
T.~W. Grimm, T.~G. Pugh and M.~Weissenbacher, \emph{{The effective action of
  warped M-theory reductions with higher-derivative terms - Part II}},
  \href{https://doi.org/10.1007/JHEP12(2015)117}{\emph{JHEP} {\bfseries 12}
  (2015) 117} [\href{https://arxiv.org/abs/1507.00343}{{\ttfamily
  1507.00343}}].

\bibitem{Grimm:2004ua}
T.~W. Grimm and J.~Louis, \emph{{The Effective action of type IIA Calabi-Yau
  orientifolds}},
  \href{https://doi.org/10.1016/j.nuclphysb.2005.04.007}{\emph{Nucl. Phys.}
  {\bfseries B718} (2005) 153}
  [\href{https://arxiv.org/abs/hep-th/0412277}{{\ttfamily hep-th/0412277}}].

\bibitem{Derendinger:2004jn}
J.-P. Derendinger, C.~Kounnas, P.~Petropoulos and F.~Zwirner,
  \emph{{Superpotentials in IIA compactifications with general fluxes}},
  \href{https://doi.org/10.1016/j.nuclphysb.2005.02.038}{\emph{Nucl. Phys. B}
  {\bfseries 715} (2005) 211}
  [\href{https://arxiv.org/abs/hep-th/0411276}{{\ttfamily hep-th/0411276}}].

\bibitem{DeWolfe:2005uu}
O.~DeWolfe, A.~Giryavets, S.~Kachru and W.~Taylor, \emph{{Type IIA moduli
  stabilization}},
  \href{https://doi.org/10.1088/1126-6708/2005/07/066}{\emph{JHEP} {\bfseries
  07} (2005) 066} [\href{https://arxiv.org/abs/hep-th/0505160}{{\ttfamily
  hep-th/0505160}}].

\bibitem{Junghans:2020acz}
D.~Junghans, \emph{{O-Plane Backreaction and Scale Separation in Type IIA Flux
  Vacua}}, \href{https://doi.org/10.1002/prop.202000040}{\emph{Fortsch. Phys.}
  {\bfseries 68} (2020) 2000040}
  [\href{https://arxiv.org/abs/2003.06274}{{\ttfamily 2003.06274}}].

\bibitem{Buratti:2020kda}
G.~Buratti, J.~Calderon, A.~Mininno and A.~M. Uranga, \emph{{Discrete
  Symmetries, Weak Coupling Conjecture and Scale Separation in AdS Vacua}},
  \href{https://doi.org/10.1007/JHEP06(2020)083}{\emph{JHEP} {\bfseries 06}
  (2020) 083} [\href{https://arxiv.org/abs/2003.09740}{{\ttfamily
  2003.09740}}].

\bibitem{Marchesano:2020qvg}
F.~Marchesano, E.~Palti, J.~Quirant and A.~Tomasiello, \emph{{On supersymmetric
  AdS$_{4}$ orientifold vacua}},
  \href{https://doi.org/10.1007/JHEP08(2020)087}{\emph{JHEP} {\bfseries 08}
  (2020) 087} [\href{https://arxiv.org/abs/2003.13578}{{\ttfamily
  2003.13578}}].

\bibitem{Gukov:1999ya}
S.~Gukov, C.~Vafa and E.~Witten, \emph{{CFT's from Calabi-Yau four folds}},
  \href{https://doi.org/10.1016/S0550-3213(01)00289-9,
  10.1016/S0550-3213(00)00373-4}{\emph{Nucl. Phys.} {\bfseries B584} (2000) 69}
  [\href{https://arxiv.org/abs/hep-th/9906070}{{\ttfamily hep-th/9906070}}].

\bibitem{DeligneHodge}
P.~Deligne, \emph{{The Hodge conjecture}}, {\emph{The Millennium Prize
  Problems, Clay Math. Inst.} (2006) 45 }.

\bibitem{Demirtas:2019sip}
M.~Demirtas, M.~Kim, L.~Mcallister and J.~Moritz, \emph{{Vacua with Small Flux
  Superpotential}},
  \href{https://doi.org/10.1103/PhysRevLett.124.211603}{\emph{Phys. Rev. Lett.}
  {\bfseries 124} (2020) 211603}
  [\href{https://arxiv.org/abs/1912.10047}{{\ttfamily 1912.10047}}].

\bibitem{Demirtas:2020ffz}
M.~Demirtas, M.~Kim, L.~Mcallister and J.~Moritz, \emph{{Conifold Vacua with
  Small Flux Superpotential}},
  \href{https://arxiv.org/abs/2009.03312}{{\ttfamily 2009.03312}}.

\bibitem{Blumenhagen:2020ire}
R.~\'Alvarez-Garc\'\i{}a, R.~Blumenhagen, M.~Brinkmann and L.~Schlechter,
  \emph{{Small Flux Superpotentials for Type IIB Flux Vacua Close to a
  Conifold}},  \href{https://arxiv.org/abs/2009.03325}{{\ttfamily 2009.03325}}.

\bibitem{schnell2014extended}
C.~Schnell, \emph{{The extended locus of Hodge classes}},
  \href{https://arxiv.org/abs/1401.7303}{{\ttfamily 1401.7303}}.

\bibitem{Bena:2020xrh}
I.~Bena, J.~Bl\r{a}b\"ack, M.~Gra\~na and S.~L\"ust, \emph{{The Tadpole
  Problem}},  \href{https://arxiv.org/abs/2010.10519}{{\ttfamily 2010.10519}}.

\bibitem{Andriolo:2020lul}
S.~Andriolo, T.-C. Huang, T.~Noumi, H.~Ooguri and G.~Shiu, \emph{{Duality and
  axionic weak gravity}},
  \href{https://doi.org/10.1103/PhysRevD.102.046008}{\emph{Phys. Rev. D}
  {\bfseries 102} (2020) 046008}
  [\href{https://arxiv.org/abs/2004.13721}{{\ttfamily 2004.13721}}].

\end{thebibliography}\endgroup

\end{document}